\documentclass[twocolumn,secnumarabic,amssymb, nobibnotes, pra]{revtex4-2}

\usepackage[utf8]{inputenc} 

\usepackage{geometry} 
\usepackage{graphicx} 
\usepackage{booktabs} 
\usepackage{array} 
\usepackage{paralist} 
\usepackage{verbatim} 
\usepackage{subcaption}
\usepackage{tikz}

\usetikzlibrary{backgrounds}
\usetikzlibrary{arrows}
\usetikzlibrary{shapes,shapes.geometric,shapes.misc}
\usetikzlibrary{decorations.pathmorphing}
\usetikzlibrary{decorations.pathreplacing}
\usetikzlibrary{decorations.markings}

\tikzstyle{every picture}=[baseline=-0.25em,scale=0.5]

\pgfkeys{/tikz/tikzit fill/.initial=0}
\pgfkeys{/tikz/tikzit draw/.initial=0}
\pgfkeys{/tikz/tikzit shape/.initial=0}
\pgfkeys{/tikz/tikzit category/.initial=0}

\newcommand{\tikzfig}[1]{%
\IfFileExists{#1.tikz}
  {\input{#1.tikz}}
  {%
    \IfFileExists{./figures/#1.tikz}
      {\input{./figures/#1.tikz}}
      {\tikz[baseline=-0.5em]{\node[draw=red,font=\color{red},fill=red!10!white] {\textit{#1}};}}%
  }%
}

\pgfdeclarelayer{edgelayer}
\pgfdeclarelayer{nodelayer}
\pgfsetlayers{background,edgelayer,nodelayer,main}
\tikzstyle{none}=[inner sep=0mm]
\tikzstyle{every loop}=[]
\tikzstyle{mark coordinate}=[inner sep=0pt,outer sep=0pt,minimum size=3pt,fill=black,circle]

\input{zx.tikzdefs}

\tikzstyle{box}=[shape=rectangle, text height=1.5ex, text depth=0.25ex, yshift=0.5mm, fill=white, draw=black, minimum height=5mm, yshift=-0.5mm, minimum width=5mm, font={\small}]

\tikzstyle{Z dot}=[inner sep=0mm, minimum size=2mm, shape=circle, draw=black, fill={rgb,255: red,216; green,248; blue,216}]
\tikzstyle{Z phase dot}=[minimum size=5mm, font={\footnotesize\boldmath}, shape=rectangle, rounded corners=2mm, inner sep=0.2mm, outer sep=-2mm, scale=0.8, tikzit shape=circle, draw=black, fill={rgb,255: red,216; green,248; blue,216}, tikzit draw=blue]
\tikzstyle{X dot}=[Z dot, shape=circle, draw=black, fill={rgb,255: red,255; green,136; blue,136}]
\tikzstyle{X phase dot}=[Z phase dot, tikzit shape=circle, tikzit draw=blue, fill={rgb,255: red,255; green,136; blue,136}, font={\footnotesize\color{black}\boldmath}]

\tikzstyle{hadamard}=[fill=yellow, draw=black, shape=rectangle, inner sep=0.6mm, minimum height=1.5mm, minimum width=1.5mm]
\tikzstyle{vertex}=[inner sep=0mm, minimum size=1mm, shape=circle, draw=black, fill=black]
\tikzstyle{vertex set}=[inner sep=0mm, minimum size=1mm, shape=circle, draw=black, fill=white, font={\footnotesize\boldmath}]

\tikzstyle{hadamard edge}=[-, color=blue, dashed, dash pattern=on 3pt off 1.5pt, thick]
\tikzstyle{brace edge}=[-, tikzit draw=blue, decorate, decoration={brace,amplitude=1mm,raise=-1mm}]
\tikzstyle{diredge}=[->]

\usetikzlibrary{quantikz}
\usepackage{adjustbox}
\usepackage{todonotes}

\usepackage{algorithm}
\usepackage{algpseudocode}

\makeatletter
\newcommand{\algmargin}{\the\ALG@thistlm}
\makeatother
\algnewcommand{\parState}[1]{\State%
  \parbox[t]{\dimexpr\linewidth-\algmargin}{\raggedright\hangindent=6pt \strut #1\strut}}

\usepackage{fancyhdr} 
\usepackage{mathtools}
\usepackage{braket}
\usepackage{stmaryrd}
\DeclarePairedDelimiter\dbracket{\llbracket}{\rrbracket}
\DeclarePairedDelimiter\ceil{\lceil}{\rceil}
\DeclarePairedDelimiter\floor{\lfloor}{\rfloor}

\usepackage{dcolumn}
\newcolumntype{d}[1]{D..{#1}}
\usepackage{bbold}
\usepackage{dsfont}
\usepackage{hyperref}

\hypersetup{
  colorlinks   = true, 
  urlcolor     = blue, 
  linkcolor    = blue, 
  citecolor   = red 
}

\begin{document}

\title{Quantum computation on a 19-qubit wide 2d nearest neighbour qubit array.}
\author{Alexis T. E. Shaw}
\email{alexis@alexisshaw.com}
\affiliation{Centre for Quantum Computation and Communication Technology}
\affiliation{Centre for Quantum Software and Information}
\affiliation{School of Computer Science}
\affiliation{Faculty of Engineering \& Information Technology, University of Technology Sydney, NSW 2007, Australia}
\author{Michael J. Bremner}
\affiliation{Centre for Quantum Computation and Communication Technology}
\affiliation{Centre for Quantum Software and Information}
\affiliation{School of Computer Science}
\affiliation{Faculty of Engineering \& Information Technology, University of Technology Sydney, NSW 2007, Australia}
\author{Alexandru Paler}
\affiliation{Aalto University, 02150 Espoo, Finland}
\author{Daniel Herr}
\affiliation{d-fine GmbH, An der Hauptwache 7, 60213, Frankfurt, Germany.}
\author{Simon J. Devitt}
\affiliation{Centre for Quantum Software and Information}
\affiliation{School of Computer Science}
\affiliation{Faculty of Engineering \& Information Technology, University of Technology Sydney, NSW 2007, Australia}
\date{\today}%

\begin{abstract}
In this paper, we explore the relationship between the width of a qubit lattice constrained in one dimension and physical thresholds for scalable, fault-tolerant quantum computation. To circumvent the traditionally low thresholds of small fixed-width arrays, we deliberately engineer an error bias at the lowest level of encoding using the surface code. We then address this engineered bias at a higher level of encoding using a lattice-surgery surface code bus that exploits this bias, or a repetition code to make logical qubits with unbiased errors out of biased surface code qubits. Arbitrarily low error rates can then be reached by further concatenating with other codes, such as Steane $\dbracket{7,1,3}$ code and the $\dbracket{15,7,3}$ CSS code. This enables a scalable fixed-width quantum computing architecture on a square qubit lattice that is only 19 qubits wide, given physical qubits with an error rate of $8.0\times 10^{-4}$. This potentially eases engineering issues in systems with fine qubit pitches, such as quantum dots in silicon or gallium arsenide.
\end{abstract}

\maketitle

\section{Introduction}
Quantum processor architectures have evolved significantly since their first conception, just over a quarter of a century ago~\cite{lloyd1993, Horsman_2012, Veldhorst2017, Bourassa2021}. They started with early discussions on how to build basic gates with two-level systems~\cite{lloyd1993} and have evolved into recent plans for machines with millions of physical qubits~\cite{Horsman_2012, roetteler2017, Lekitsch2017}. Architecture design must work around the peculiarities of their constituent qubits. Physical limitations on qubit size, gate speed, decoherence rates, temperature, control wiring, and infrastructure all have a major effect on the potential scalability of a given architecture~\cite{Veldhorst2017, Bourassa2021, Lekitsch2017}. 

Quantum dot qubits in silicon and gallium arsenide (GaAs) present interesting constraints when investigating quantum architecture. They have demonstrated relatively low decoherence rates~\cite{Xue2022, Zwerver2022}, high operation temperature~\cite{Yang2020}, and high gate speeds~\cite{He2019}. Significantly, they offer the potential of high qubit density and ease of manufacture for large systems on a single wafer~\cite{Vandersypen2017, Zwerver2022, Ansaloni2020}. Yet, the size and small qubit pitch that could allow for high qubit density come with significant drawbacks. 

Running control wires into gates that have qubit spacings on the order of nanometers in a two-dimensional geometry is extremely challenging. This is especially important because architectures using the highest threshold quantum error correcting codes, such as the surface or honeycomb codes, tend to assume two-dimensional lattices of unrestricted size~\cite{Litinski2019,Litinski2019b,Veldhorst2017}. In response, quantum architectures that utilize three-dimensional fabrication have been proposed~\cite{Veldhorst2017}, even though the expected fabrication complexity is formidable.  One approach to solving this is to restrict the width of the array, which limits the interconnect density because the system grows only in one dimension. 

The idea of minimizing the width of a qubit array is certainly not new. Since the early days of fault tolerance, people have considered fixed or minimal-width arrays of qubits. However, previous approaches have always required undesirable tradeoffs, either increasing qubit requirements or requiring long-distance qubit interactions. For example, CSS codes were leveraged in the paper by Veldhorst et. al.~\cite{Ashley2009}, and the subsequent threshold was extremely small (less than $10^{-5}$) due to the costs involved in interacting non-adjacent qubits in a nearest-neighbour array. Other examples include specific investigations into linear nearest neighbours in silicon~\cite{JonesFogartyMorelloGyureDzurakLadd2018} with a threshold of about $10^{-4}$ and the use of resonators to fold a square lattice into a bi-linear array~\cite{Mohiyaddin2021}. Unfortunately, the latter approach required long resonators for interaction between qubits, and these resonators had unknown manufacturability and performance in spin systems given their length and complexity.

In this paper, we present a method of reducing the array width without requiring a lower error threshold or long-distance qubit interactions. Instead, we propose a coding structure that provides good reductions of width for realistic error rates, with a threshold array width as low as 19 qubits for a physical error rate of $8\times 10^{-4}$. It locally has the same nearest neighbour interactivity and the same threshold error rate as the surface code. Of course, this comes at the cost of significantly more qubits; however, this may be an acceptable trade-off for certain technologies, such as silicon quantum dots, where qubits are hoped to be relatively cheap.



We primarily consider two schemes that are based on combining the surface code with techniques for long-range gate compilation and CSS codes. The basic idea is that the 
surface code can be used to decrease errors enough to allow techniques for longer distance interactions. Subsequently, CSS codes are used in a fixed width array.
In section~\ref{sec:rectsc}, we examine the performance of the surface code for rectangular patches for each type of error. These results are then used in section~\ref{sec:bus}
to evaluate the performance of a lattice surgery scheme for remote qubit interaction using a very narrow width strip of surface code qubits that we call the surface code bus. In section~\ref{sec:parity}, we consider
the performance of the Steane $\dbracket{7,1,3}$ and $\dbracket{15,7,3}$ CSS codes when using the flag qubit compilation approach of Reichardt~\cite{ChaoReighardt2020,Reichardt2020} and a shared long-distance interaction capacity, like that provided by the bus of section~\ref{sec:bus}.
We then combine the results of section~\ref{sec:bus} and \ref{sec:parity} and determine the performance of the concatenated scheme in section~\ref{sec:concat_perf}. Finally, in section~\ref{sec:bias}, we examine the possibility of using
the rectangular patches evaluated in section~\ref{sec:rectsc} to engineer a logical error bias that can then be concatenated with the repetition code to create a memory with unbiassed errors and high enough logical fidelity that a CSS code could
be concatenated above it. This enables us to reduce lattice width further, giving a minimum width of 27 for a threshold physical error rate of $1.5 \times 10^{-3}$, 19 for a threshold error rate of $8 \times 10^{-4}$, and 11 for a threshold error rate of $ 2\times 10^{-4}$.

                 \section{Background}
                 \label{sec:bg}
In this section, for the sake of completeness and in order to guide the reader, we summarise material and definitions and provide references to further resources.
\subsection{Errors}
\label{bg:err}


The threshold and performance of a code are usually quoted in terms of an error rate given an error model. In this work, we assume a standard error model with balanced single-qubit depolarising noise channels. That is, each single-qubit physical operation, including leaving qubits idle, is acted on by the single-qubit depolarising channel 
\begin{equation}
\mathcal{E}_1 = (1-p)\rho + \frac{p}{3}(X\rho X + Y\rho Y + Z\rho Z)
\end{equation}
before each gate and measurement, as well as after each initialisation. Two-qubit gates are assumed to be operated
on by the two-qubit depolarising channel
\begin{equation}
\begin{split}
\mathcal{E}_2(\rho) =&\ (1-p)\rho \ +\\
&\ \frac{p}{15}\sum_{P \in {\{I,X,Y,Z\}}^{\otimes 2} \setminus\{I\otimes I\}} P\rho P^\dagger
\end{split}
\end{equation}
before each gate. We assume that there are no correlated higher-order interactions or correlated events apart from those generated by these channels.

A common assumption for existing architecture designs is that the physical error rate of a quantum computer should be approximately one order of magnitude below the threshold for the surface code, that is $p = 10^{-3}$, which we take as the highest value in our analysis~\cite{OGorman2017, Nagayama_2017}. However, we also provide estimates for lower physical error rates, down to $10^{-4}$, to inform the reader of the trade-offs between reducing physical errors and handling higher interconnect density.

In this paper, the logical error rate is the probability of a logical error when performing a single logical CNOT gate, unless otherwise stated. The maximum tolerable logical error rate for computation of several significant early quantum algorithms has been estimated in the order of $10^{-14} - 10^{-18}$~\cite{OGorman2017, roetteler2017, Litinski2019b}. We use the logical error rate of $10^{-15}$ as a benchmark for each proposed architecture, as it reflects these estimates. An attempt has been made to create a rough equivalence for the difference in quantum operations; however, compilation differences between the examined options mean that this is still a loose comparison.

\subsection{Stabilizer Codes}
\label{bg:stab}
One of the key theoretical improvements toward quantum error correction was the development of the quantum stabilizer code~\cite{Shor1995, Steane1996, BenOr1997, CalderbankShor1996}. By measuring a collection of non-local multi-body 
Pauli operators, it is possible to ``stabilise" a logical space that is not within the span of the space of those operators. In effect, the measurement of those operators forces the remaining state to be within a subspace of the total Hilbert space~\cite{KBLW2001}. If low-weight errors are uniquely identifiable by the stabilising operators, then this provides a way to correct for those errors.

The repetition code is the simplest of the stabiliser codes, and perhaps the simplest error correction code. In the classical variant, $n$ copies
of the message are made and then compared against each other. The quantum bit-flip repetition code encodes the state $\alpha\ket{0} + \beta\ket{1}$ as $\alpha\ket{0\cdots 0} + \beta\ket{1\cdots 1}$~\cite[Chapter~10.1.1]{nielsen_chuang_2010}. A quantum repetition code cannot correct for all possible errors, as it can only correct for either bit-flip or phase-flip errors, but not for both. Even so, quantum repetition codes have the highest possible code capacity, with the ability to correct for at most $\frac{n-1}{2}$ errors of that one type. We use this to improve performance to the required level in Section~\ref{sec:bus} and \ref{sec:bias}.

In order to implement our concatenation scheme efficiently, we need an efficient implementation of a block code that has good qubit density and a relatively high threshold. In order to obtain this, we leverage the recent advances in flagged syndrome extraction~\cite{ChaoReighardt2018, ChaoReighardt2018b, ChaoReighardt2020, Reichardt2020}, discussed in section~\ref{bg:flag_qubits}. We consider the Steane $\dbracket{7,1,3}$ code and the $\dbracket{15,7,3}$ code for use as higher level codes in this work as they appear to have a good mix of performance and density, as well as having flagged extraction circuits.

The Steane $\dbracket{7,1,3}$ code is a self-dual CSS code defined using the $7$-bit hamming code~\cite{Steane1996}, and has stabiliser generators
listed in Table \ref{tab:steane_stab}.
\begin{table}[!htbp]
\begin{tabular}[t]{ l  c}
\toprule
1 & $IIIXXXX$ \\
2 & $IIIZZZZ$ \\
3 & $IXXIIXX$ \\
4 & $IZZIIZZ$ \\
5 & $XIXIXIX$\\
6 & $ZIZIZIZ$ \\
\bottomrule
\end{tabular}
\caption{Stabiliser generators for the Steane $\dbracket{7,1,3}$ code}
\label{tab:steane_stab}
\end{table}
This code can correct for a single error, and is the smallest of the self-dual codes.

The $\dbracket{15,7,3}$ self-dual CSS code is defined using the $15$-bit hamming code~\cite{ChaoReighardt2018b}, and has stabiliser generators listed in
Table~\ref{tab:fifteen_seven_stab}. This code can also correct for a single error, however it encodes logical qubits significantly
more efficiently than the Steane code, at the expense of a greater number of more complicated stabilisers.
\begin{table}[!htbp]
\begin{tabular}[t]{ l  c}
\toprule
1 & $IIIIIIIXXXXXXXX$ \\
2 & $IIIIIIIZZZZZZZZ$ \\
3 & $IIIXXXXIIIIXXXX$ \\
4 & $IIIZZZZIIIIZZZZ$ \\
5 & $IXXIIXXIIXXIIXX$ \\
6 & $IZZIIZZIIZZIIZZ$ \\
7 & $XIXIXIXIXIXIXIX$ \\
8 & $ZIZIZIZIZIZIZIZ$ \\
\bottomrule
\end{tabular}
\caption{Stabiliser generators for the Steane $\dbracket{15,7,3}$ code}
\label{tab:fifteen_seven_stab}
\end{table}

\subsection{Flaged Syndrome Extraction}
\label{bg:flag_qubits}
Flagged syndrome extraction is a newly proposed technique for the fault-tolerant measurement of stabilisers in quantum error correcting codes~\cite{ChaoReighardt2018,ChaoReighardt2018b, Reichardt2020,ChaoReighardt2020,PrabhuReichardt2021}. It enables significantly lower extraction circuit complexity and lowers the number of ancillae required compared to previous techniques for fault-tolerant stabiliser measurement.

Earlier methods of fault-tolerant syndrome extraction assumed that error-amplification caused by stabiliser measurement led to severe degradation of code performance. It was assumed that an error with weight beyond the code capacity was uncorrectable, hence stringent measures were proposed in order to prevent error amplification during syndrome extraction. Shor's proposed method~\cite{Shor1996} was to create a GHZ state the size of the stabilizer to be measured, verify it by interacting each element with one of the qubits supported by the stabiliser, and then measure the GHZ state. This would require both increased circuit depth and a substantially greater number of ancilla qubits than what would be required by a non-fault-tolerant stabiliser measurement circuit. In addition to the increase in resources, this method also leads to a lower threshold for the code, as there is more time and space for potential errors in each round of syndrome measurements.

Flagged syndrome extraction capitalises on a realisation that some temporary amplification of errors can be allowed, so long as any higher weight errors can be uniquely distinguished and corrected given enough rounds of syndrome extraction. This loosening of requirements allows a notable reduction in the number of ancilla qubits required, with only two ancilla qubits required for weight-3 codes. Further, in some cases, it is possible to perform this extraction with zero additional gates over a basic extraction (such as in the circuit described in Figure \ref{fig:steane_circuit}). This is achieved by the careful ordering of otherwise commuting gates, as well as through inserting detection gadgets to disambiguate different error channels. The main downside of this approach is that the circuit and decoder designs are substantially more difficult than in earlier forms of fault-tolerant extraction. This is because the faults identified must be propagated forward through the extraction circuit so that the higher weight errors as well as the extraction of other syndromes may be corrected~\cite{ChaoReighardt2018b}, whilst the earlier methods do not introduce additional data qubit errors during syndrome extraction, and so only need a simple hamming-code decoder.

\subsection{The Surface Code}
\label{bg:sc}
\begin{figure*}
\begin{tikzpicture}
\node[draw=none,fill=none] at (0,0){\includegraphics[width=14cm]{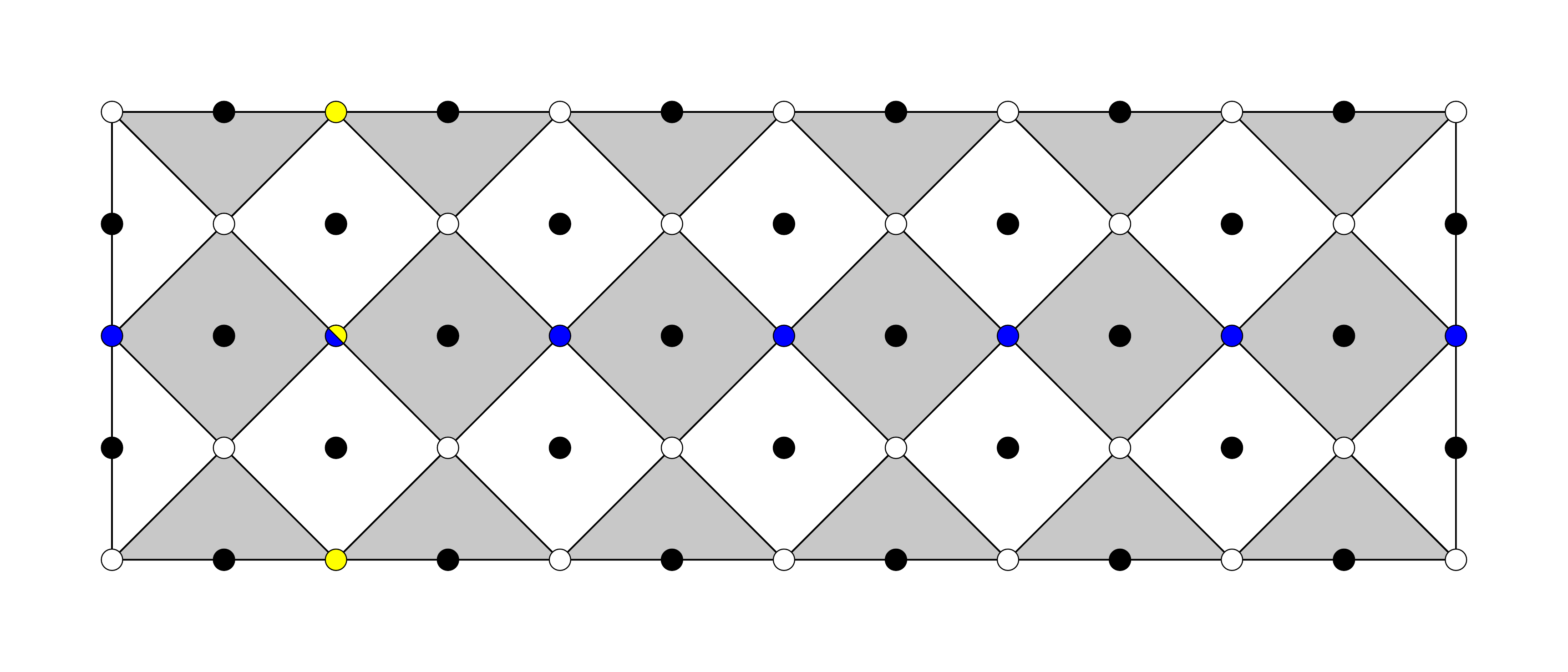}};
\node at (-10,-5.5){Z-Stabiliser};
\node[draw=none,fill=none] at (-10,-9){\includegraphics[width=3cm]{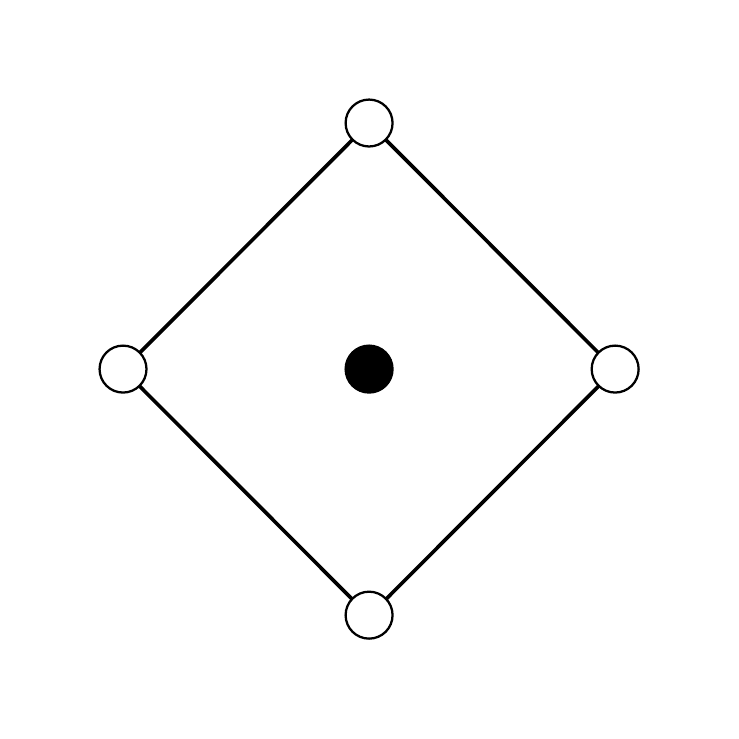}};
\node at (2,-5.5){X-Stabiliser};
\node[draw=none,fill=none] at (2,-9){\includegraphics[width=3cm]{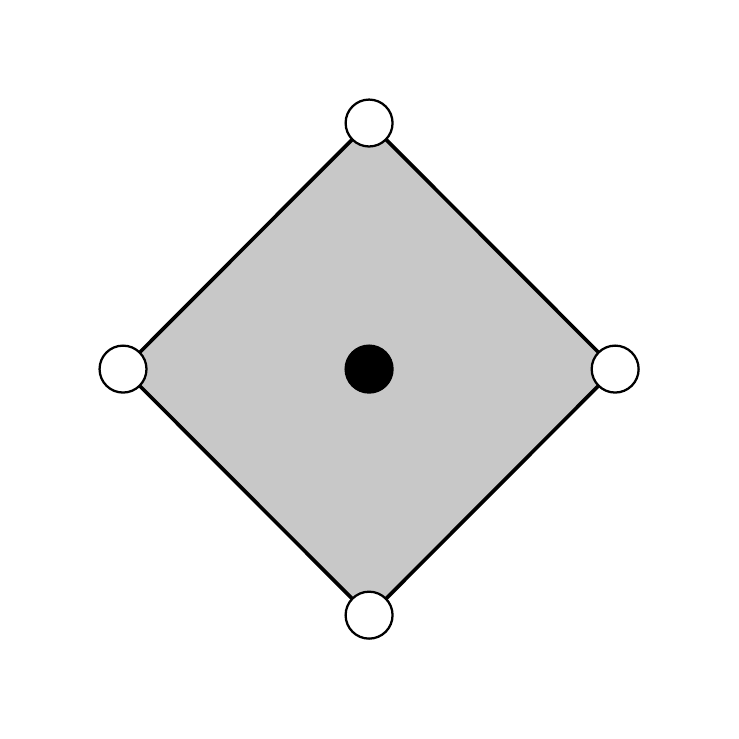}};

    \node[scale=0.5] at (-4,-9){
        \begin{quantikz}
        \lstick{1}        & \ctrl{4} &\qw       &\qw       &\qw      &\qw& \\
        \lstick{2}        & \qw      &\ctrl{3}  &\qw       &\qw      &\qw& \\
        \lstick{3}        & \qw      &\qw       &\ctrl{2}  &\qw      &\qw& \\
        \lstick{4}        & \qw      &\qw       &\qw       &\ctrl{1}  &\qw& \\[0.5 cm]
        \lstick{$\ket{0}$}& \targ{}  &\targ{}   &\targ{}   &\targ{}  &\meter{}&
        \end{quantikz}
    };
    \node[scale=0.5] at (9,-9){
        \begin{quantikz}
        \lstick{1}        & \qw      & \targ{}  &\qw       &\qw       &\qw      &\qw       &\qw& \\
        \lstick{2}        & \qw      & \qw      &\targ{}   &\qw       &\qw      &\qw       &\qw& \\
        \lstick{3}        & \qw      & \qw      &\qw       &\targ{}   &\qw      &\qw       &\qw& \\
        \lstick{4}        & \qw      & \qw      &\qw       &\qw       &\targ{}  &\qw       &\qw& \\[0.5 cm]
        \lstick{$\ket{0}$}&\gate{H}& \ctrl{-4}&\ctrl{-3} &\ctrl{-2} &\ctrl{-1}&\gate{H}&\meter{}&
        \end{quantikz}
    };
\end{tikzpicture}
         \caption{A $d_X = 3$, $d_Z  = 7$ surface code. One $Z$ logical operator is highlighted in blue, and an $X$ operator is highlighted in yellow.}
         \label{fig:rect_sc}
\end{figure*}
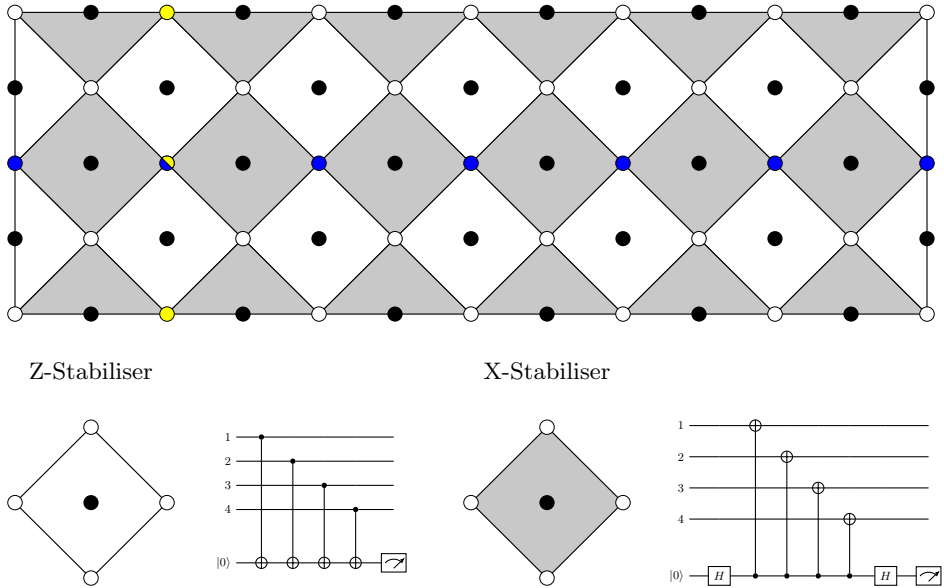

The surface code is one of a family of stabiliser codes that has physically local stabilisers. Here, the qubits that make up the
stabilisers are supported by qubits that are physically close to each other. In a surface code such as in Figure \ref{fig:rect_sc}
the plane is tiled with two sets of square plaquettes~\cite{FowlerMariantoniMartinisCleland2012, Bravyi2018}. The X plaquettes, and the Z plaquettes, are represented here as dark and pale diamonds.
 For historical reasons, the Z plaquettes are also called faces and the X plaquettes are called vertices~\cite{kitaev2003}. However,
the diamond presentation used here shows that they are completely symmetrical, with the exception of the ancilla initialisation and
measurement.

The logical operators in the surface code run from one boundary of a type to the other boundary of that type. The type of the logical operator is determined by the plaquette type on the boundary, with an X-boundary having partial X-plaquettes on the boundary, and similarly for Z-boundaries.  For the same reason used to determine plaquette names, Z-type boundaries are often called rough boundaries, and X-type logical boundaries are called smooth boundaries. To measure a logical operator, you measure qubits in the appropriate basis along that logical operator.

Physical errors are always correctable if the length of the longest error chain is less than half the distance between two boundaries of the same type. That is, the code distance is equal to the smallest number of code qubits between two boundaries of the same type. This makes these codes have a significantly lower density than other stabiliser codes (for example the $\dbracket{7,1,3}$ code needs 9 total qubits for fault tolerance~\cite{ChaoReighardt2018}, but the rotated $d=3$ surface code uses at least 13 qubits, and the non-rotated surface code uses 25~\cite{Tomita2014}, quantum LDPC codes have even higher densities~\cite{Breuckmann2021, Horsman_2012, Breuckmann2021}). However, as a trade-off, their small stabiliser circuit sizes and physical locality means that syndrome extraction is extremely fast and relatively easy to perform. Further, the low circuit depth and locality of errors lead to one of the highest thresholds of any quantum error-correcting code. 

In many ways, the discovery of the surface code is what may make the development of practical digital quantum computers possible. Simulations have put the logical
error rate of a surface code of distance $d$ at approximately 
\begin{equation}
\label{eqn:scfit}
p_{l_{\mathrm{bus}}} \sim 0.3(70p)^{\floor{\frac{d+1}{2}}}
\end{equation}
per round of syndrome extraction~\cite{Devitt2016}, and a threshold of over 1\%~\cite{WangFowlerHollenberg2011}.

\subsection{Lattice Surgery}
\label{bg:lc}
Lattice surgery is the most space efficient manner of fault-tolerant 2-qubit interaction between surface code qubits that is currently known~\cite{Litinski2019b,Horsman_2012}. When combined with faulty state injection and state distillation, it enables the surface code to perform fault-tolerant universal quantum computation~\cite{Litinski2019, Litinski2019b,Horsman_2012}.

The key operations used in lattice surgery are merging two adjacent surface code patches into one larger patch and taking a large patch and then splitting it into smaller ones. 
\begin{figure*}
        \includegraphics[width=\textwidth]{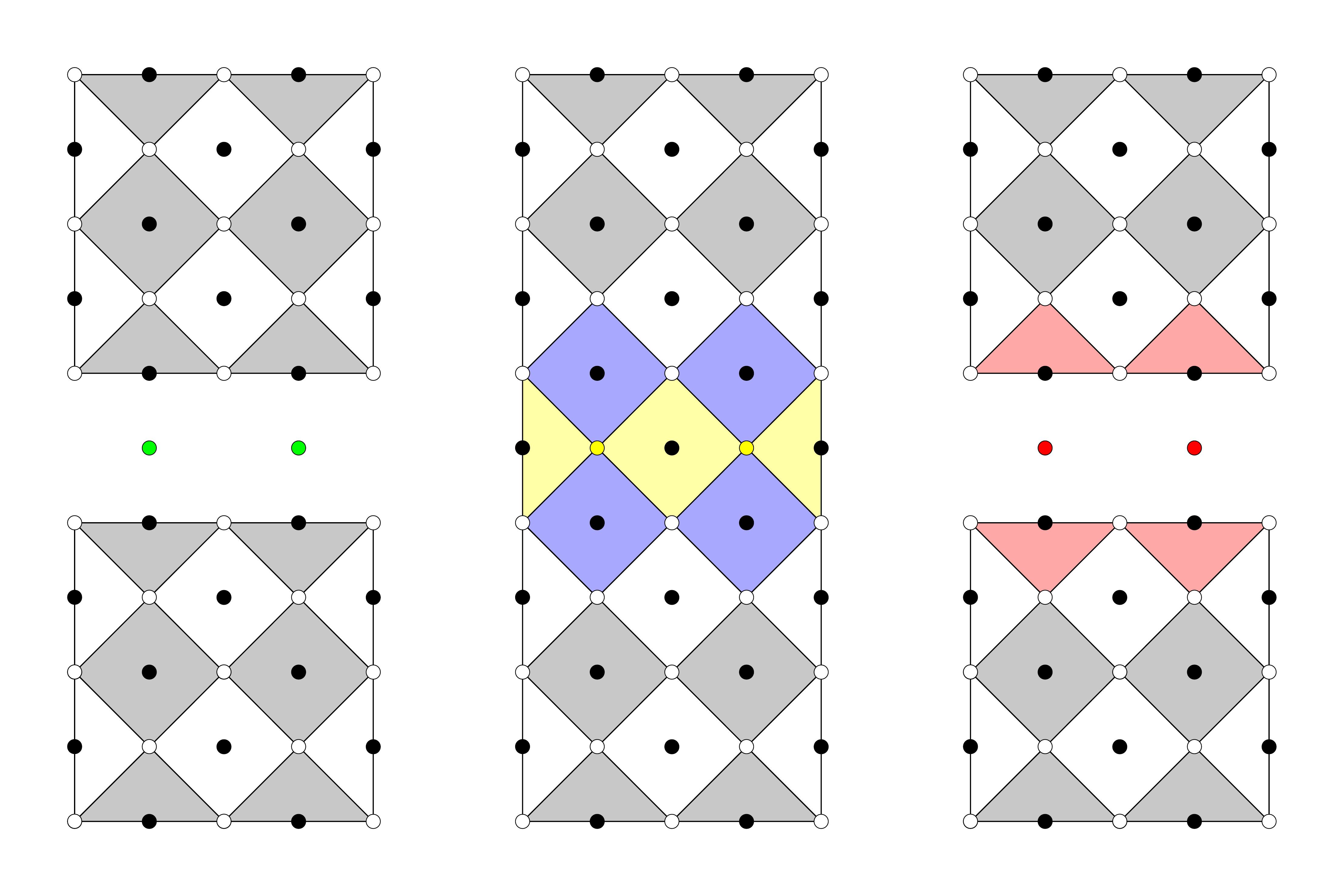}
         \caption{Illistration of a smooth lattice surgery merge followed by a smooth split, implementing an $XX$ parity measurement. The qubits in green are initialised in the $\ket{+}$ state. Then the new stabilisers, shown in yellow, are measured, and the stabilisers in blue are expanded. Finally, the stabilisers in red are reduced, and the qubits in red are measured in the $X$ basis.}
         \label{fig:ls_xx}
\end{figure*}

Patches are merged by measuring new stabilisers between two surface code patches, as shown in the first operation of Figure~\ref{fig:ls_xx}.
The logical state of the new patch is a function of the logical states of the two patches before the merger, the type of boundaries merged, and the parity of the new and modified stabilisers after measurement. The parity of these new measurements encodes a coherent logical parity measurement between the logical qubits. If the parity is zero after this measurement, then the operation encodes the mapping $\ket{0}_L\bra{00}_L + \ket{1}_L\bra{11}_L$. Otherwise, a correction must be made that is equivalent to choosing one of the patches and flipping it before the merge. The correction forces the measurement to be parity zero, and so the mapping for the parity zero measurement can be used.

A patch is split by modifying the stabilisers to separate the patches, as shown in the second operation of Figure~\ref{fig:ls_xx}, and then measuring the data-qubits that are between the patches. This operation creates an entangled state, which is described by the mapping $\ket{00}_L\bra{0}_L + \ket{11}_L\bra{1}_L$.

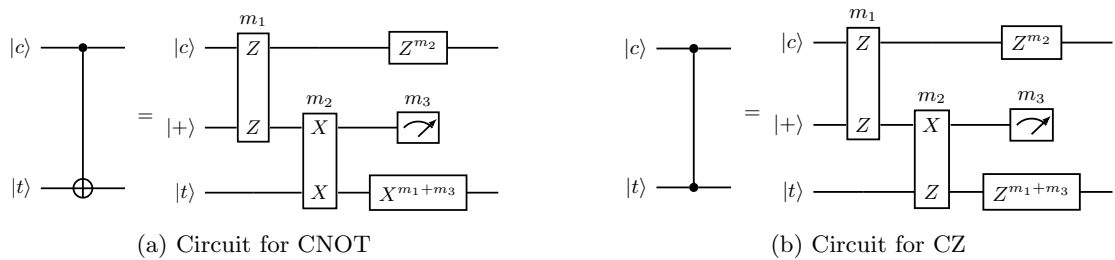
\begin{figure*}
      \begin{subfigure}[b]{0.45\textwidth}
\begin{adjustbox}{width=1\textwidth}
$
\begin{quantikz}
\lstick{$\ket{c}$}& \ctrl{1}&\qw \\[1.5cm]
\lstick{$\ket{t}$} & \targ{}&\qw
\end{quantikz}
 =
\begin{quantikz}
\lstick{$\ket{c}$}& \gate[style={fill=none,draw=none}]{Z} \gategroup[2,steps=1,style={fill=white!20,inner sep=-4pt},background]{$m_1$}& \qw&\gate{Z^{m_2}}&\qw \\
\lstick{$\ket{+}$}& \gate[style={fill=none,draw=none}]{Z} & \gate[style={fill=none,draw=none}]{X}\gategroup[2,steps=1,style={fill=white!20,inner sep=-4pt},background]{$m_2$}&\meter{$m_3$}& \\
\lstick{$\ket{t}$} & \qw  &\gate[style={fill=none,draw=none}]{X}& \gate{X^{m_1 + m_3}}&\qw
\end{quantikz}
$
\end{adjustbox}

         \caption{Circuit for CNOT}
         \label{fig:gt_cnot}
     \end{subfigure}
     \hfill
     \begin{subfigure}[b]{0.45\textwidth}
\begin{adjustbox}{width=1\textwidth}
$
\begin{quantikz}
\lstick{$\ket{c}$}& \ctrl{1}&\qw \\[1.5cm]
\lstick{$\ket{t}$} & \ctrl{}&\qw
\end{quantikz}
 =
\begin{quantikz}
\lstick{$\ket{c}$}& \gate[style={fill=none,draw=none}]{Z} \gategroup[2,steps=1,style={fill=white!20,inner sep=-4pt},background]{$m_1$}& \qw&\gate{Z^{m_2}}&\qw \\
\lstick{$\ket{+}$}& \gate[style={fill=none,draw=none}]{Z} & \gate[style={fill=none,draw=none}]{X}\gategroup[2,steps=1,style={fill=white!20,inner sep=-4pt},background]{$m_2$}&\meter{$m_3$}& \\
\lstick{$\ket{t}$} & \qw  &\gate[style={fill=none,draw=none}]{Z}& \gate{Z^{m_1 + m_3}}&\qw
\end{quantikz}
$
\end{adjustbox}
         \caption{Circuit for CZ}
         \label{fig:bus_fold_b}
     \end{subfigure}
         \caption{CNOT and CZ gates implemented using parity measurements.}
         \label{fig:gt_CNOT_CZ}
\end{figure*}

The combination of these two operations allow circuits to be compiled 
far more space-efficiently than other techniques, such as braiding~\cite{FowlerGidney2018U}. Proposed compilation approaches have only a $50\%$ space penalty when compared to quantum memory alone~\cite{Litinski2019}. The combination of a merge followed by a split and a Pauli-correction results in a parity measurement between the two patches, with the measured operator determined by the types of boundaries merged. By combining two different parity measurements and one ancilla surface code patch, it is possible to implement either a CNOT or CZ gate (see Figure~\ref{fig:gt_CNOT_CZ}).

\subsection{The ZX-Calculus}
\label{bg:ZXcalculus}
The ZX-calculus was developed as a graphical system for reasoning about quantum linear maps. It represents quantum operations with tagged graphs that are manipulated using a collection of simplification rules~\cite{2012.13966, CoeckeDuncan2008}. Every quantum circuit can be efficiently mapped to a ZX diagram, and the graphs can be used to prove the properties of those circuits. 

In this work, we use ZX-calculus to show the correctness of the surface code bus in Section~\ref{bus:folded}. It is especially useful for discussing the logical effect of lattice surgery operations~\cite{deBeaudrap2020}, as the operations of merging and splitting of the surface code map neatly and naturally to the ZX-calculus. Some care is needed, however, to understand the action of joint measurements in the lattice surgery. The ZX-calculus can even be said to be complete, in the sense that the diagram re-writing rules are sufficient to convert any two diagrams that represent the same linear map to the same diagram~\cite{Backens2014, JeandelPendrixVilmart2018}.

Here we reproduce some basic definitions and results that are used in this work. Specifically, we use the fact that state preparation and measurements are represented by degree-one nodes
\[
\ket{+} = 
\begin{tikzpicture}
    \begin{pgfonlayer}{nodelayer}
        \node [style=Z phase dot] (0) at (0.00, 0.00) {};
        \node [style=none] (1) at (2.00, 0.00) {};
    \end{pgfonlayer}
    \begin{pgfonlayer}{edgelayer}
        \draw (0) to (1);
    \end{pgfonlayer}
\end{tikzpicture}\quad , \quad \bra{+}R_z(\alpha) = 
\begin{tikzpicture}
    \begin{pgfonlayer}{nodelayer}
        \node [style=Z phase dot] (0) at (2.00, 0.00) {$\alpha$};
        \node [style=none] (1) at (0.00, 0.00) {};
    \end{pgfonlayer}
    \begin{pgfonlayer}{edgelayer}
        \draw (0) to (1);
    \end{pgfonlayer}
\end{tikzpicture};
\]
that a $Z$-flip, or other $Z$-rotation, is represented by a degree-2 $Z$-node
\[
\ket{0}\bra{0} + e^{i\alpha}\ket{1}\bra{1} = 
\begin{tikzpicture}
    \begin{pgfonlayer}{nodelayer}
        \node [style=none] (0) at (0.00, 0.00) {};
        \node [style=Z phase dot] (1) at (1.50, 0.00) {$\alpha$};
        \node [style=none] (2) at (3.00, 0.00) {};
    \end{pgfonlayer}
    \begin{pgfonlayer}{edgelayer}
        \draw (0) to (1);
        \draw (1) to (2);
    \end{pgfonlayer}
\end{tikzpicture};
\]
that a smooth split is represented by a $Z$-spider
\[
S_{S} = \ket{00}\bra{0} + \ket{11}\bra{1} = 
\begin{tikzpicture}[baseline={([yshift=-.5ex]current bounding box.center)}]
    \begin{pgfonlayer}{nodelayer}
        \node [style=none] (0) at (0.00, 1.00) {};
        \node [style=Z phase dot] (1) at (1.50, 1.00) {};
        \node [style=none] (2) at (3.00, 0.00) {};
        \node [style=none] (3) at (3.00, 2.00) {};
    \end{pgfonlayer}
    \begin{pgfonlayer}{edgelayer}
        \draw (1) to [out=-45, in=180] (2.center);
        \draw (1) to [out=45, in=180] (3.center);
        \draw (0) to (1);
    \end{pgfonlayer}
\end{tikzpicture},
\]
and that a rough merge is an $X$-spider with a correction on one leg that depends on the parity measurement result~\cite{deBeaudrap2020}
\[
M_{S} = 
\begin{tikzpicture}[baseline={([yshift=-.5ex]current bounding box.center)}]
    \begin{pgfonlayer}{nodelayer}
        \node [style=none] (0) at (0.00, 0.00) {};
        \node [style=none] (1) at (0.00, 2.00) {};
        \node [style=Z phase dot] (2) at (1, 1.75) {$b\pi$};
        \node [style=X phase dot] (3) at (2, 1.00) {};
        \node [style=none] (4) at (4.00, 1.00) {};
    \end{pgfonlayer}
    \begin{pgfonlayer}{edgelayer}
        \draw (0.center) to [out=0, in=225] (3);
        \draw (1.center) to [out=0, in=155] (2);
        \draw (2) to (3);
        \draw (3) to (4);
    \end{pgfonlayer}
\end{tikzpicture},
\]
where $b$ is the measurement outcome of the merge. Finally, we note that the CNOT gate can be represented as a $Z$-spider connected to an $X$-spider,
\[
\begin{quantikz}
\lstick{}& \ctrl{1}&\qw \\[0.27cm]
\lstick{} & \targ{}&\qw
\end{quantikz}
=
\begin{tikzpicture}[baseline={([yshift=-.5ex]current bounding box.center)}]
    \begin{pgfonlayer}{nodelayer}
        \node [style=none] (0) at (0.00, 0.00) {};
        \node [style=X phase dot] (1) at (2.00, 0.00) {};
        \node [style=none] (2) at (4.00, 0.00) {};
        \node [style=none] (3) at (0.00, 2.00) {};
        \node [style=Z phase dot] (4) at (2.00, 2.00) {};
        \node [style=none] (5) at (4.00, 2.00) {};
    \end{pgfonlayer}
    \begin{pgfonlayer}{edgelayer}
        \draw (0) to (1);
        \draw (1) to (2);
        \draw (3) to (4);
        \draw (4) to (5);
        \draw (1) to (4);
    \end{pgfonlayer}
\end{tikzpicture}.
\]

In addition to these definitions, we use the spider-merge simplification rule, which states nodes of the same type can be merged as long as their rotations are summed modulo $2\pi$.

\section{Rectangular Surface Codes}
\label{sec:rectsc}

\subsection{Motivation}
\label{rectsc:motive}
In rectangular surface codes with a small $X$ or $Z$ distance, edge effects necessarily have a greater impact than when dimensions get larger. Because of this, the simple scaling rules for taking the well-studied performance of square surface codes~\cite{WangFowlerHollenberg2011,Fowler2012,Nagayama_2017,Hakkaku2021,Russendorf2007,Tomita2014,Cai2019} to derive the performance of a rectangular surface code patch are expected to break down on high aspect ratio codes. Hence. direct analysis and simulations are required to study narrow rectangular surface code patches.

While there has been some recent work in this area~\cite{Utkarsh,Lee2021}, this work does not separately evaluate how rectangular patches bias the rate of logical $X$ and $Z$ errors. In this paper, we manage the biassing effect of these rectangular patches on logical error rates to reduce the minimum required width for a scalable qubit array. This requires us to perform new simulations to understand and characterise these impacts, which is especially important for determining the performance of the surface code bus in Section~\ref{sec:bus}.

\subsection{Simulations of biased surface codes}
\label{rectsc:sims}
We performed simulations of rectangular surface codes
for $d_Z$ surface code syndrome extraction cycles for $d_X \in \{ 3,5,7\}$ with increasing odd values of $d_Z$. We then measured the data qubits. These simulations were performed with the patches initialised in both the $\ket{0}$ and $\ket{+}$ logical states using the high-performance simulator \emph{Stim} developed by Gidney~\cite{Gidney2021}. This tool uses a tableau representation similar to that of the CHP simulator~\cite{AaronsonGottesman2004} of Aaronson et.al. which we re-implemented for our simulations of parity codes in section~\ref{sec:rectsc}. 

In order to determine errors, the Pauli basis was updated for each round of stabiliser using a Minimum Weight Perfect Matching decoder, as described by Fowler~\cite{WangFowlerHollenberg2011}, and then the measurements were interpreted accordingly. Although not an optimal decoder, the Minimum Weight Perfect Matching decoder is significantly
better than a hamming-distance lattice decoder, and it does not require significant extra implementation complexity. In our implementation, the matching graphs created were solved using the Blossom V maximum matching algorithm implementation of Kolmogorov~\cite{Kolmogorov2009}. 

\begin{figure*}
     \begin{subfigure}[b]{0.48\textwidth}
         \includegraphics[width=\textwidth]{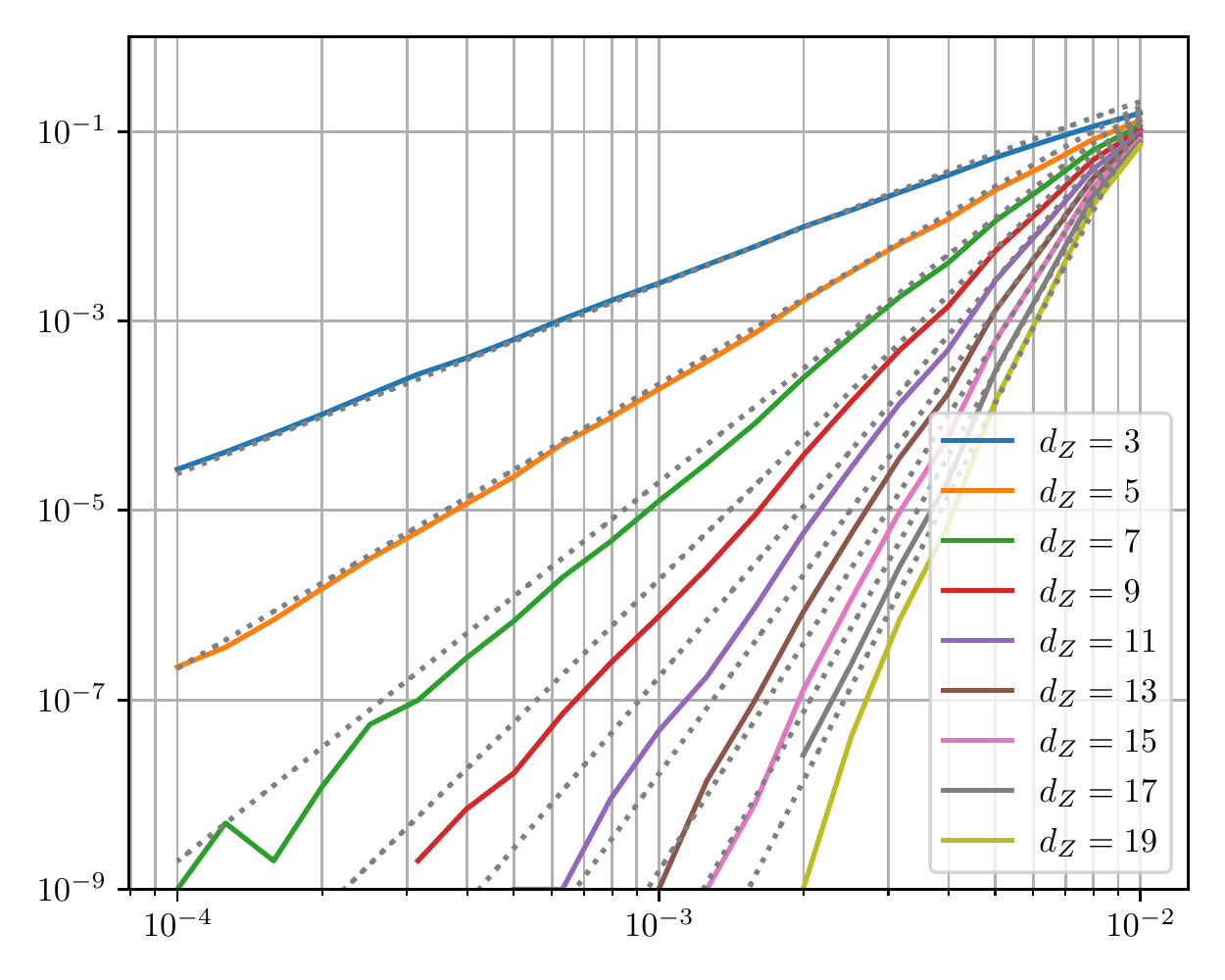}
         \caption{$p_Z$ with $d_X = 3$, $10^{9}$ samples per point.}
         \label{fig:3_long}
     \end{subfigure}
     \hfill
     \begin{subfigure}[b]{0.48\textwidth}
         \includegraphics[width=\textwidth]{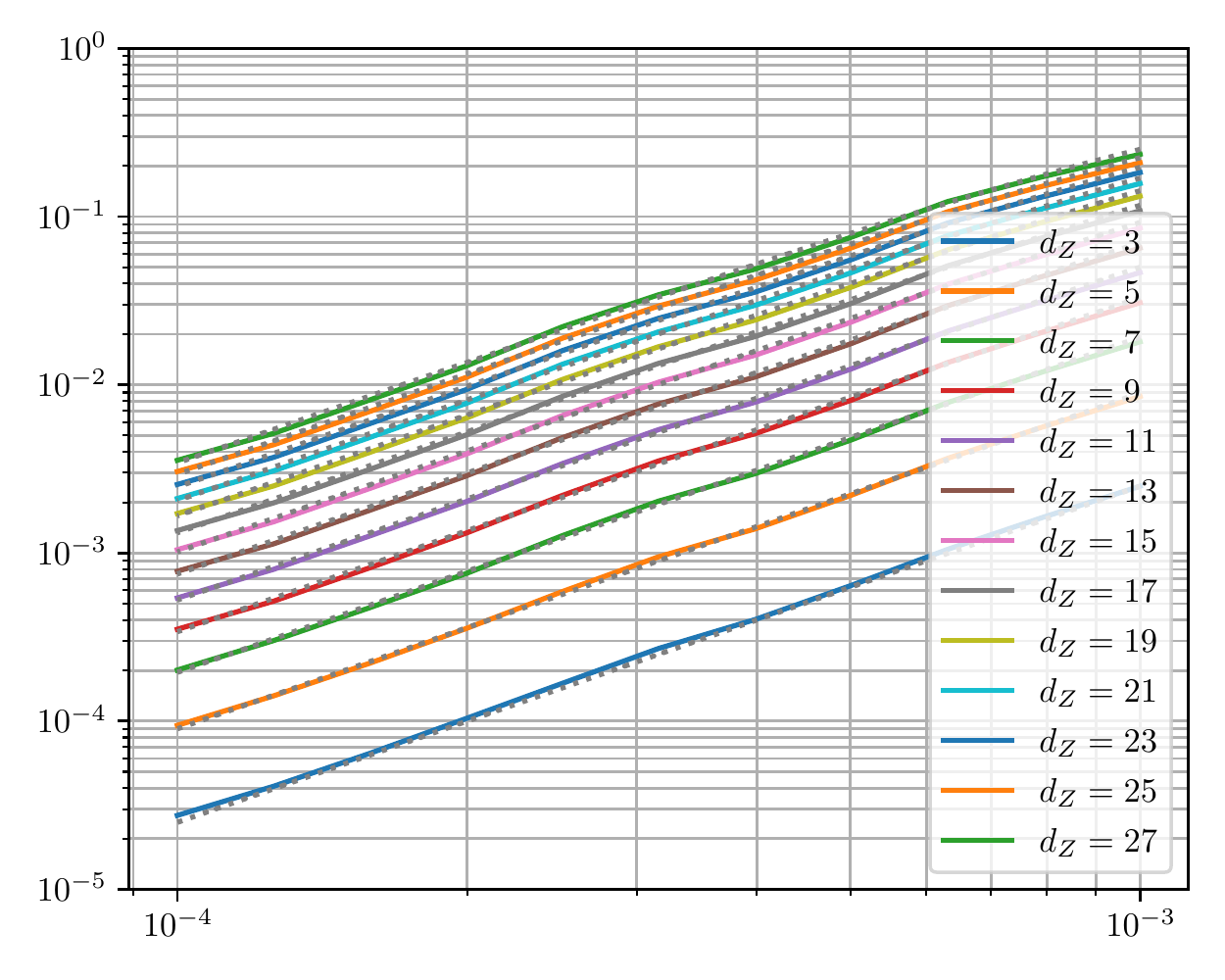}
         \caption{$p_X$ with $d_X = 3$, $10^{8}$ samples per point.}
         \label{fig:3_short}
     \end{subfigure}
     
     \begin{subfigure}[b]{0.48\textwidth}
         \includegraphics[width=\textwidth]{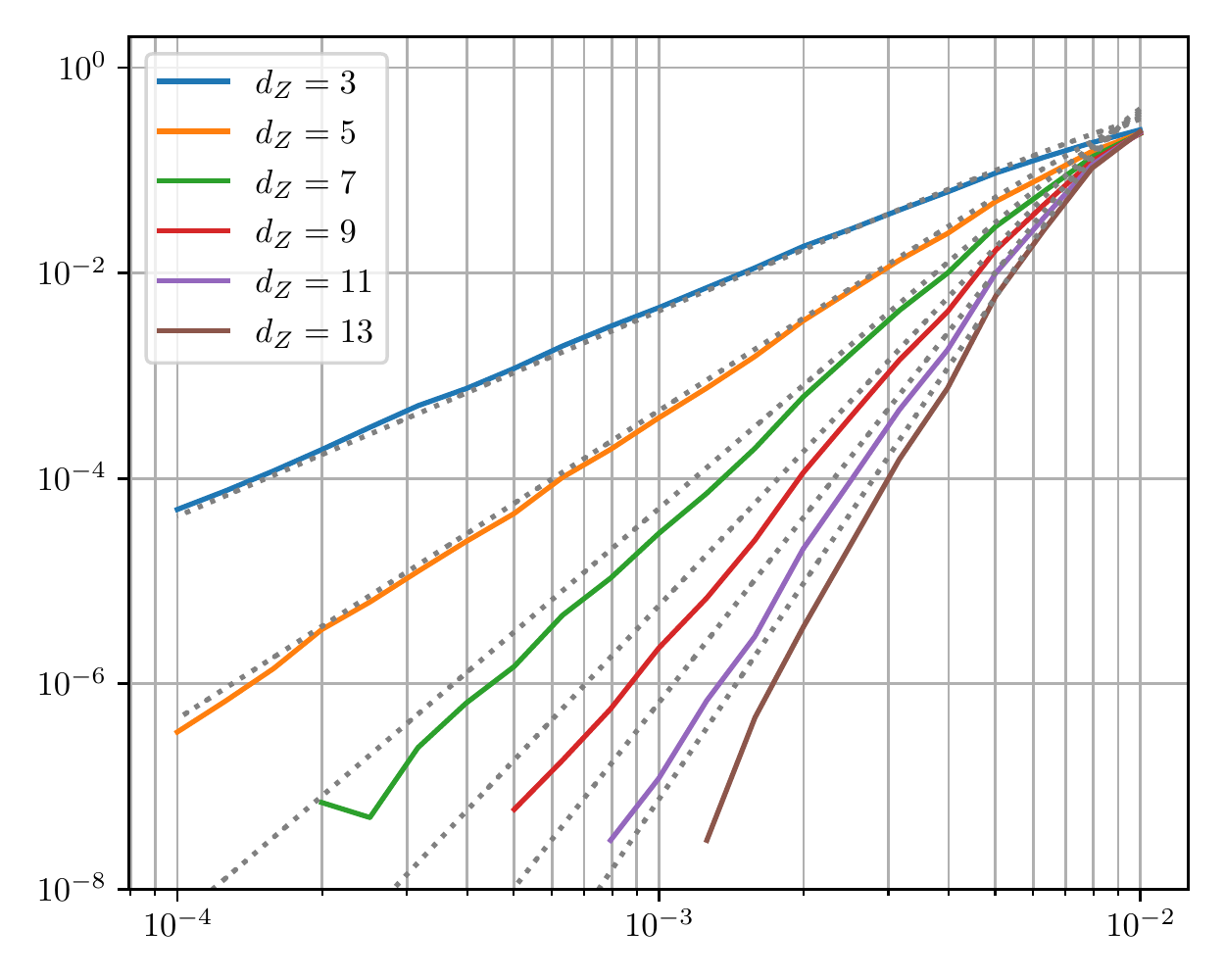}
         \caption{$p_Z$ with $d_X = 5$, $10^{8}$ samples per point.}
         \label{fig:5_long}
     \end{subfigure}
     \hfill
     \begin{subfigure}[b]{0.48\textwidth}
         \includegraphics[width=\textwidth]{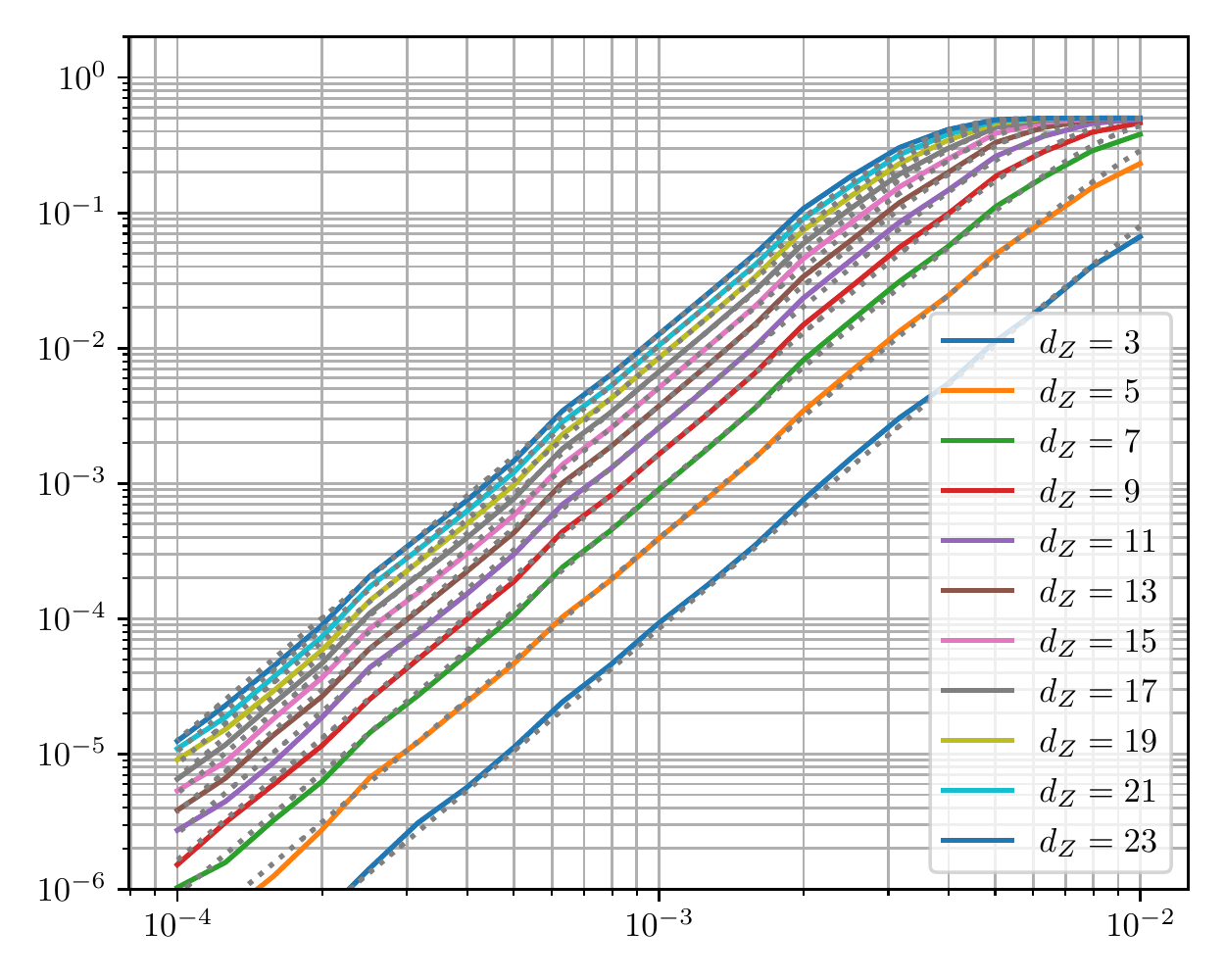}
         \caption{$p_X$ with $d_X = 5$, $10^{6}$ samples per point.}
         \label{fig:5_short}
     \end{subfigure}

     \begin{subfigure}[b]{0.48\textwidth}
         \includegraphics[width=\textwidth]{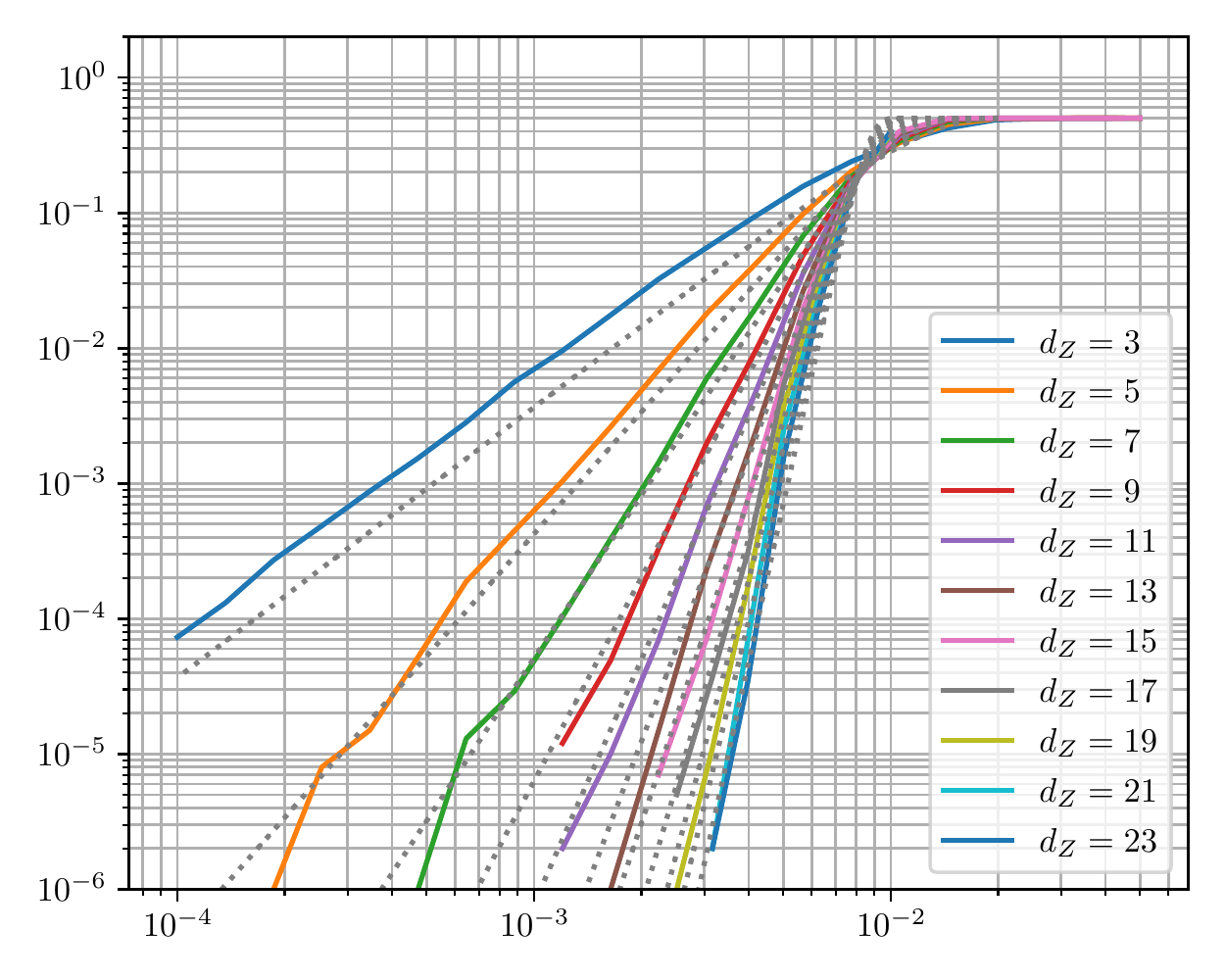}
         \caption{$p_Z$ with $d_X = 7$, $10^{6}$ samples per point.}
         \label{fig:7_long}
     \end{subfigure}
     \hfill
     \begin{subfigure}[b]{0.48\textwidth}
         \includegraphics[width=\textwidth]{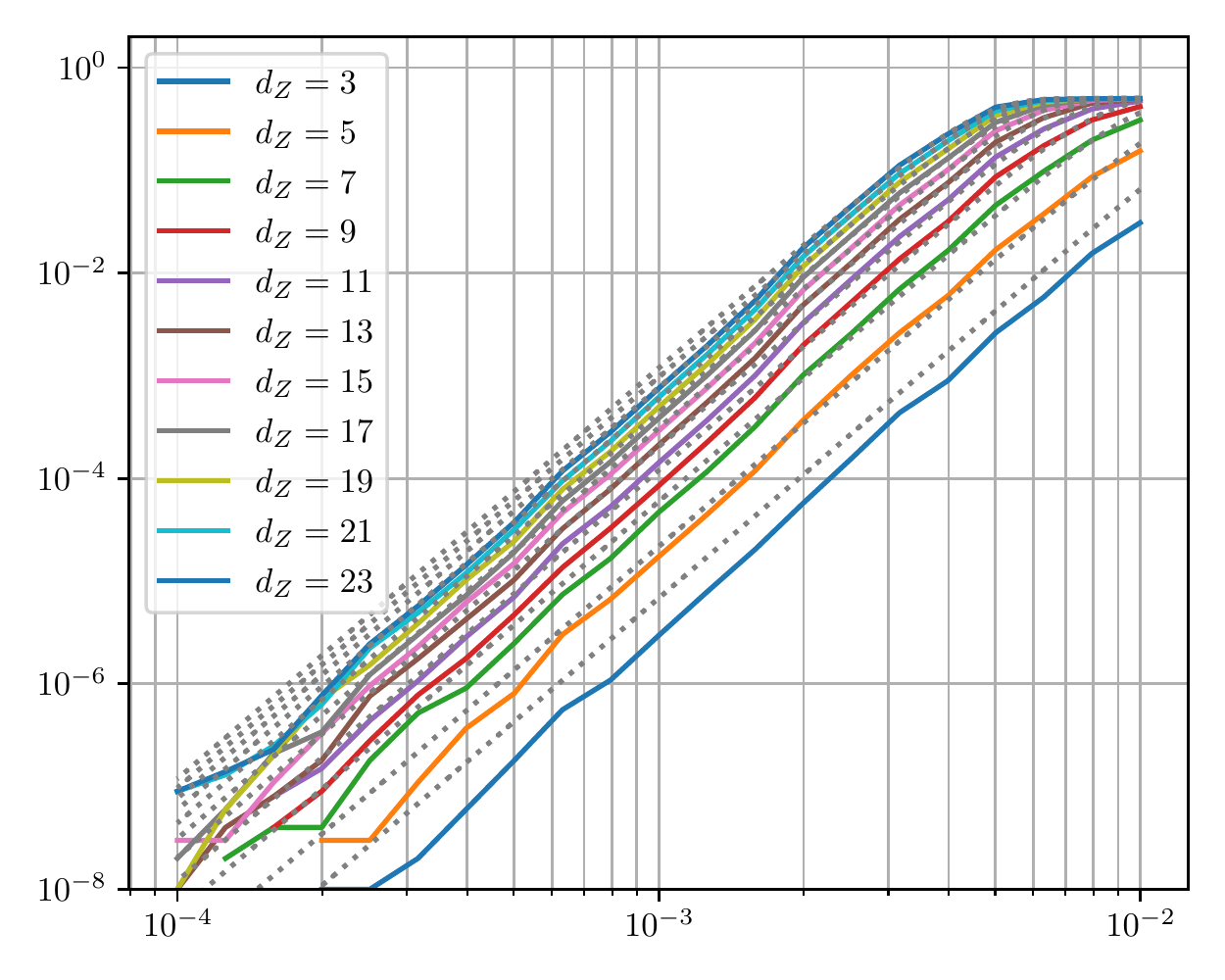}
         \caption{$p_X$ with $d_X = 7$, $10^{6}$ samples per point.}
         \label{fig:7_short}
     \end{subfigure}
        \caption{Plots of the Logical error rates in both the $Z$ and $X$ directions over $d_Z$ rounds of a biased rectangular surface code. 
        }
        \label{fig:biased_graphs}
\end{figure*}

To evaluate the performance of the decoder, simulations were also performed on square lattices and compared with pre-existing results~\cite{WangFowlerHollenberg2011, Fowler2012}. These results were largely the same with only some very minor differences. As such, we don't expect major improvements with better decoders. The results in figure \ref{fig:biased_graphs} demonstrate the upper bound on the performance of the surface code given by this decoder.

\subsection{Evaluation of results}
\label{rectsc:results}
A fitting function was necessary in order to extrapolate the performance of these surface codes and enable further analysis. These fitting functions needed to have a form that made sense given the expected theoretical behaviour of the codes, as well as provide a good fit for the simulation data. The available computing resources limited the number of simulations that could be performed and, in turn, restricted the number of widths (i.e. values of $d_X$) we could analyse. This made it difficult to determine a single fitting function, so we derived separate fits for each of the $p_Z$ and $p_X$ error rates, and for each $d_X$. This gave fits in good agreement with the simulation as shown by the dotted lines in Figure~\ref{fig:biased_graphs}.

Our fits for the $Z$ error rates $p_Z$ are of the form $\alpha\frac{d_X - 0.5}{d_Z - 0.5}d_Z(\beta p )^{\floor{\frac{d_z + 1}{2}}}$. This was chosen by taking the expected $\alpha (\beta p )^{\floor{\frac{d_z + 1}{2}}}$
error per time step, multiplying it by $d_Z$ to account for the number of syndrome extractions required to perform a lattice surgery operation, and then adding a correction factor of $\frac{d_X - 0.5}{d_Z - 0.5}$. The correction factor was empirically found to provide for a better fit. The fitting parameters $\alpha$ and $\beta$ are given in Table~\ref{tab:table2_a} for each $d_X$. 

Fits for the $X$ error rate $p_X$ of the form $f_{d_X}(d_Z) \cdot p^{\floor{\frac{d_X + 1}{2}}}$ were also found, where $f(x)$ is some quadratic polynomial in $d_Z$. A quadratic was chosen because the number of error channels grows approximately proportionally to the number of surface code extractions multiplied by the length of the surface code patch, that is as the square of $d_Z$. A generic quadratic was chosen as this improved the fit over a more restrictive function and seemed to be a physically plausible way to account for edge effects. These polynomial fits were found for each $d_X$, and are given in Table~\ref{tab:table2_b}.

\begin{table*}
\begin{subtable}[t]{0.35\textwidth}
\begin{tabular}[t]{ l  c c }
\toprule
$d_X$ & $\alpha$& $\beta$ \\
\midrule
3 & 0.09 & 95\\
5 & 0.06 & 110\\
7 & 0.03 & 125\\
\bottomrule
\end{tabular}
\caption{Parameterd for the $p_Z$ fit}
\label{tab:table2_a}
\end{subtable}
\begin{subtable}[t]{0.45\textwidth}
\begin{tabular}[t]{l  c }
\toprule
$d_X$ & $f_{d_X}(d_Z)$ \\
\midrule
3 & $f(x) = 500x^2 -700x +250$\\
5 & $f(x) = 26399x^2 - 57295x + 18522$\\
7 & $f(x) = 2.87 \times 10^{6}x^2 -  1.55 \times 10^{7}x + 2.75 \times 10^{7}$\\
\bottomrule
\end{tabular}
\caption{Functions for the $p_X$ fit}
\label{tab:table2_b}
\end{subtable}
\caption{\footnotesize  Parameters for chosen fits.}
\label{tab:table2}
\end{table*}

\section{The Surface Code Bus}
\label{sec:bus}

\subsection{The GHZ state Bus}
\label{bus:ghz}
\begin{figure*}
    \includegraphics[width=\textwidth]{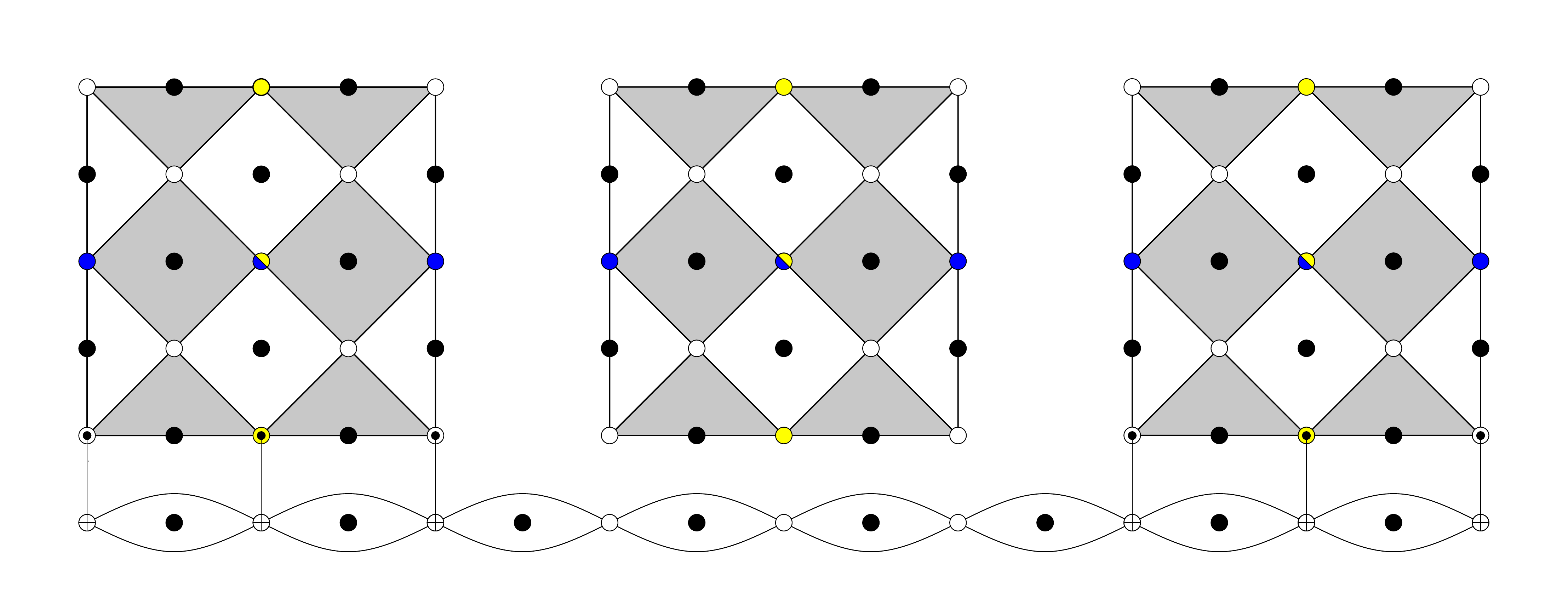}
    \caption{Visual representation of the interaction between the GHZ state, and the two $d=3$ surface code patches in the Unprotected GHZ surface code bus $ZZ$ measurement.}
    \label{fig:OriginalDiagram}
\end{figure*}

The preprint by Herr et. al.~\cite{1902.08117} proposed a method that used a narrow single-qubit wide bus to interconnect two $d\times d$ surface code patches to determine the parity between them, which they called the surface code data bus. This used the procedure we reproduce as Algorithm \ref{alg:unprotected_algorithm}. A visual representation of their proposed bus is presented in Figure \ref{fig:OriginalDiagram}. Sadly, this procedure is not fault-tolerant (however we show a modified version that is fault tolerant in the next section), because at line 9 the parity measured is not protected by the repetition code, meaning any $Z$ error that occurs on any of the data qubits will flip the $X$ parity, taking the GHZ state prepared at line 3 from 
$\ket{0\cdots0} +\ket{1\cdots 1}$  to $\ket{0\cdots0} - \ket{1\cdots 1}$. As both of these have the same values for the $XX$ operators checked in line 5, any such error in the time/space
 volume is uncorrectable. Further, any single measurement error on line 9 will also be undetectable.


\begin{figure*}
\begin{minipage}{\linewidth}
\begin{algorithm}[H]
\caption{\raggedright GHZ Surface Code Bus}
\label{alg:unprotected_algorithm}
\begin{algorithmic}[1]
\Procedure{UnprotectedBus}{$d\times d$ qubit SC patches ${\psi}$, ${\phi}$ indexed as ${\theta}_{[i,j]}$}
    \For{$i\in \{1 \dots d\}$}
      \parState{Create an $2d + m$ quantum register ${b}$ which is initialised to the GHZ state $\frac{1}{\sqrt{2}}(\ket{0\cdots0} +\ket{1\cdots 1})$}
      \For{$j \in \{1 \dots d\}$}
          \parState{Check for errors by measuring the operator $X_{b_{[k]}}X_{b_{[k+1]}}$\\ for $k\in \{1 \dots (2n+m-1)\}$}
     \EndFor
      \State Update Pauli frame to correct errors in GHZ state.
      \parState{Perform a CNOT gate between the GHZ state and SC patch Z boundaries using $$\bigotimes_{k=1}^{d} CNOT_{b_{[k]},\psi_{[1,k]}} \otimes CNOT_{b_{[k]},\phi_{[1,k]}}$$}
      \State Measure $b$ in the $X$ basis and compute the parity $p_i$ of the result.
    \EndFor
    \State Compute the majority vote over all $p_j$, the parity XX over ${\psi}$ and ${\phi}$ and return.
\EndProcedure
\end{algorithmic} 
\end{algorithm}
\end{minipage}
\end{figure*}

The size of the GHZ state depends on the distance between the two qubit patches, as well as their code distance. Together, this means the GHZ state will be $Nd$ qubits wide in the worst
 case, where $N$ is the number of patches along the surface code bus (where the minimum is 2, in the case of adjacent patches), and $d$ is the surface code distance. To detect $X$ errors in the preparation of the
 GHZ state, it is necessary to perform step 5 $d$ times, with each step taking $t \geq 4$ steps. The total time/space volume of the parity check must be then at least $Ntd^2$. This means that $Ntd^2p$ needs to be below $0.5$ for the protocol to be fault-tolerant, as the
 code capacity for the repetition code is $0.5$. This allows for $N\leq 13$ when $p=0.1\%$ and $d=3$; it allows $N\leq 4$ when $d=5$, and only adjacent patches may be interacted fault-tollerantly when $d\geq7$. Further, the number of repetitions in this bus could be substantial depending on how close the system is to the code capacity of the repetition code.  An increase in code distances might be required in order to meet the required error rate for higher-level protocols and algorithms.

\subsection{The Folded Surface Code Bus}
\label{bus:folded}
\begin{figure*}
    \begin{subfigure}[b]{0.47\textwidth}
        \includegraphics[width=\textwidth]{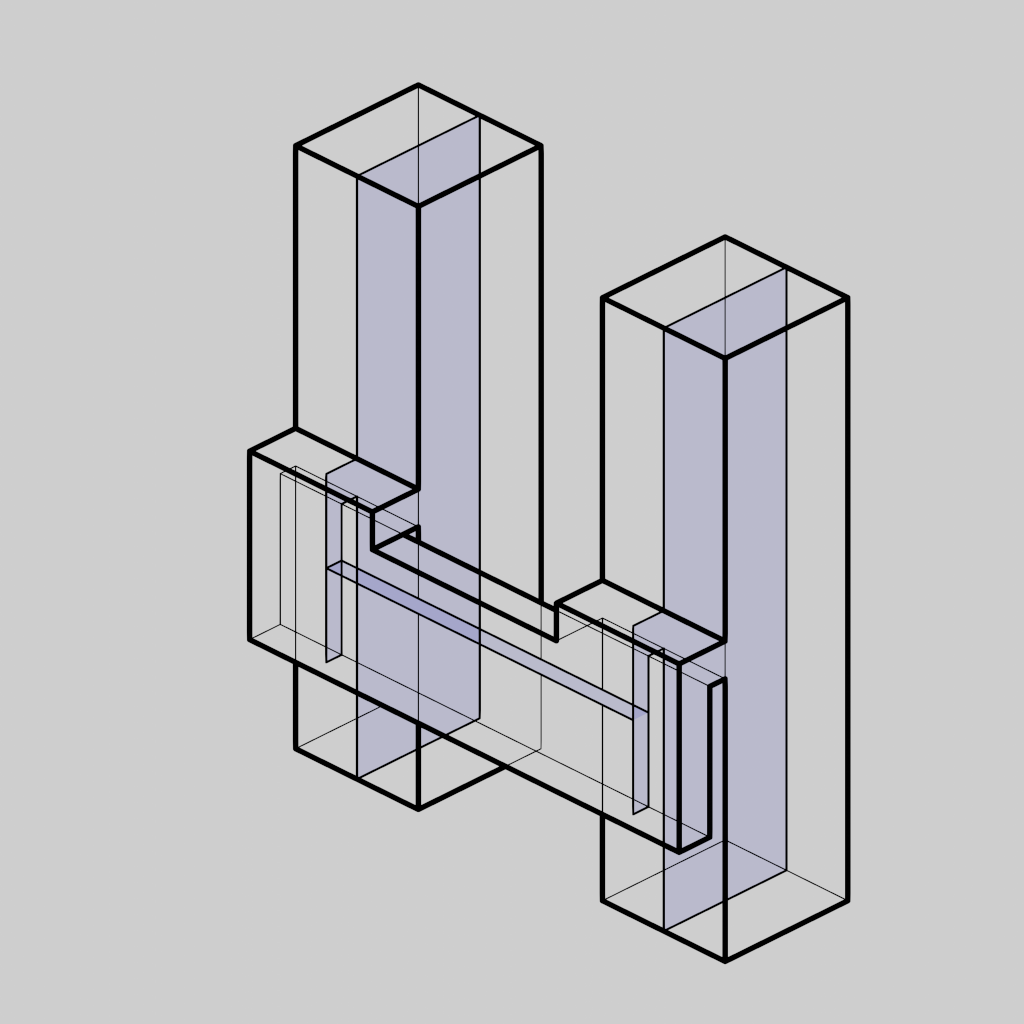}
        \caption{Z Logical Operators}
        \label{fig:FoldedXCorr}
    \end{subfigure}
    \hfill
    \begin{subfigure}[b]{0.47\textwidth}
         \includegraphics[width=\textwidth]{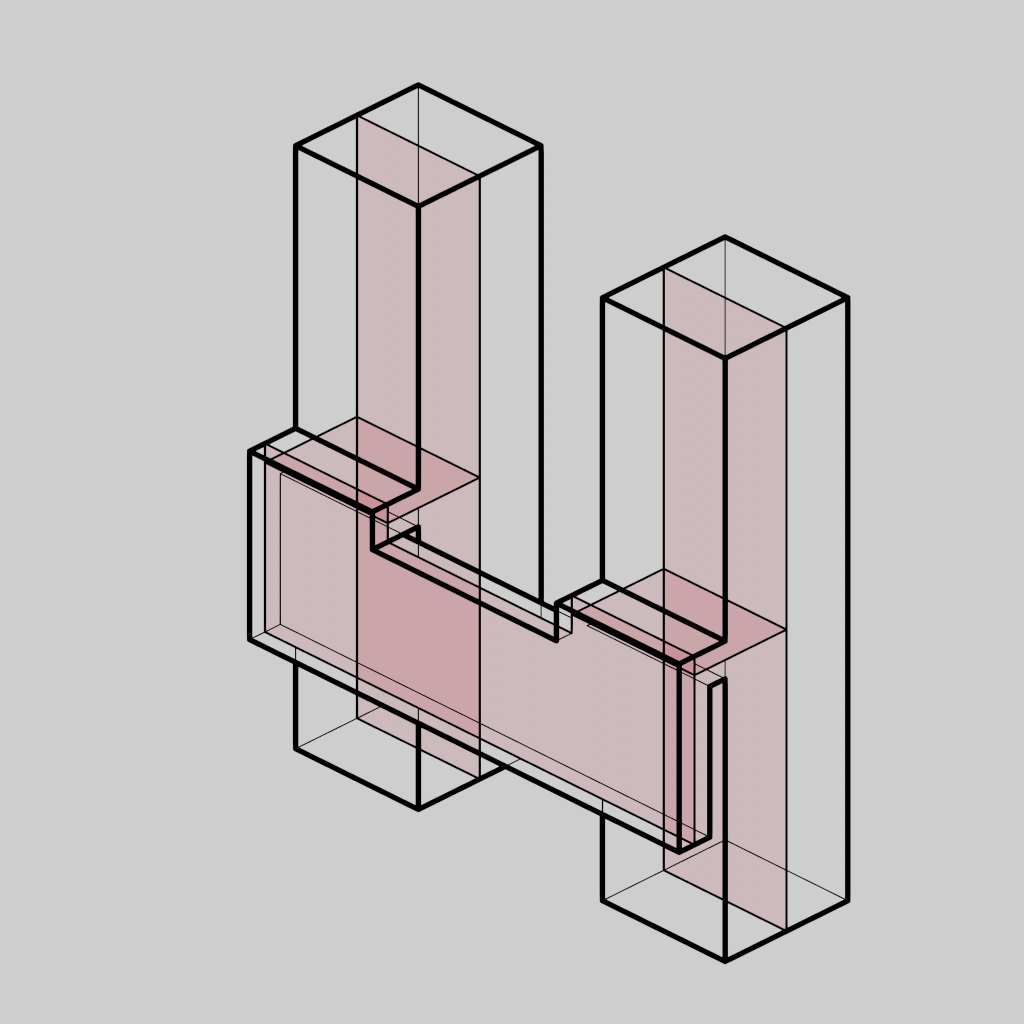}
        \caption{X Logical Operators}
        \label{fig:FoldedZCorr}
    \end{subfigure}
    \caption{Logical operators for the Folded Surface Code Bus for $XX$.  In these diagrams the vertical direction represents
    time, with the cross-section at each point of time representing the surface code at that point in time. The marked surfaces represent
    some of the possible logical operators at each point.}
    \label{fig:FoldedBusRegions}
\end{figure*}

\begin{figure}
    \includegraphics[width=\columnwidth]{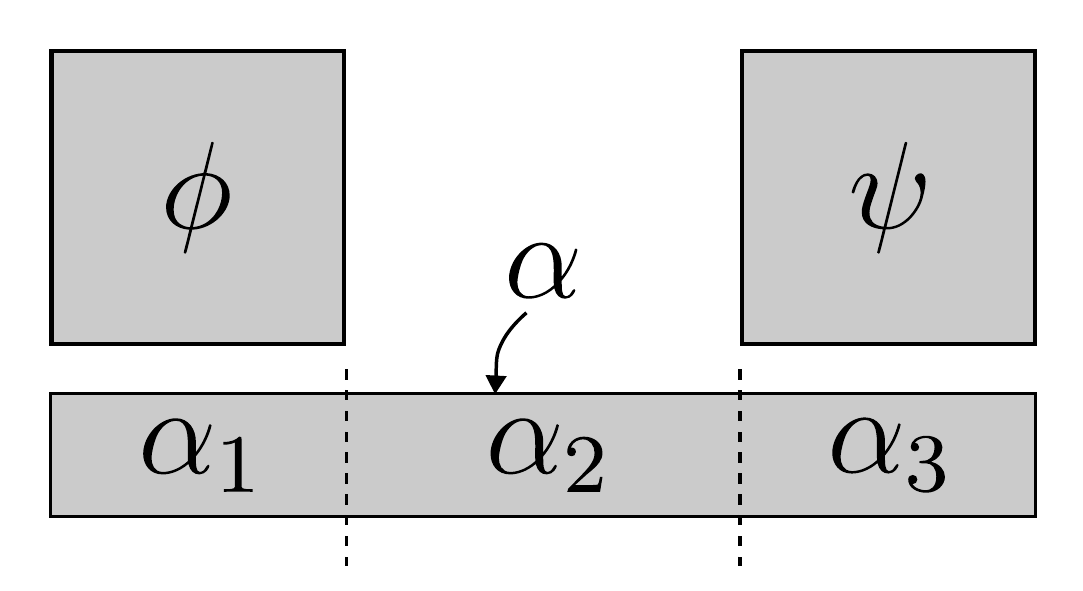}
    \caption{Layout and position of surface code patches for the folded surface code bus.}
    \label{fig:FoldedBus}
\end{figure}

The above technique can be made fault tolerant for any code distance or spacing by using lattice surgery over biassed surface code patches to replace the GHZ state
\footnote{This method was determined in discussions involving the authors of the Herr paper and Craig Gidney, and is presented here with their consent}.
 This modified fault-tolerant procedure is presented in Algorithm~\ref{alg:protected_algorithm} with a visualisation of the logical operators shown in Figure~\ref{fig:FoldedBus}. The visual representation 
is especially useful in understanding the possible error chains and the locations and interactions of various boundaries. Also useful in understanding the algorithm is the schematic view of the surface code regions shown in Figure~\ref{fig:FoldedBusRegions}. We call this the folded surface code bus because it can be understood as folding a temporally thin lattice surgery parity measurement along the time axis so that it is narrow in space.

\begin{figure*}
\begin{minipage}{\linewidth}
\begin{algorithm}[H]
\caption{\raggedright Fault-Tolerant Surface Code Bus $XX$ measurement of width $w$}
\label{alg:protected_algorithm}
\begin{algorithmic}[1]
\Procedure{ProtectedBus-$w$}{$d\times d$ qubit SC patches $\psi$, $\phi$ indexed as $\theta_{[i,j]}$}
    \For{$i\in \{1 \dots d\}$ \label{pa:line:l1}}
        \parState{Create a $d_x = 2d + m$ and $d_z = w$ qubit surface code patch ${\alpha}$, initialised to the logical $\ket{+} = \frac{1}{\sqrt{2}}(\ket{1}+\ket{0})$
                         state. (if $w=1$, this is exactly the GHZ state from Algorithm \ref{alg:unprotected_algorithm}) \label{pa:line:sc}}
        \For{$j \in \{1 \dots d\}$ \label{pa:line:l2}}
            \parState {Measure and record the stabiliser operators of all 3 surface code patches.}
        \EndFor
        \parState{Split the surface code patch ${\alpha}$ into 3 patches with $d_z=w$, 
        in order ${\alpha_1}$ with $d_x=d$, ${\alpha_2}$~with $d_x = m$ and ${\alpha_3}$ with $d_x = d$, recording the measurements from the boundaries. 
        (This leaves us in the logical GHZ state $\frac{1}{\sqrt{2}}(\ket{111} + \ket{000})$). \label{pa:line:s1}}
        \parState{Measure the surface code patch ${\alpha_2}$ in the $Z$ basis leaving the logical state ${\alpha_1\alpha_3}=\frac{1}{\sqrt{2}}(\ket{11}+\ket{00})$\label{pa:line:d1}}
        \parState {Merge Surface code patch ${\psi}$ with surface code patch ${\alpha_1}$ along their rough boundaries creating ${\psi\beta_1\alpha_1}$. Where ${\beta_1}$ 
                          is initialised as $\bigotimes_{k=1}^{d-1}\ket{+}$ and is the boundary. \label{pa:line:m1}}
        \parState {Merge Surface code patch ${\phi}$ with surface code patch ${\alpha_3}$ along their rough boundaries creating ${\phi}{\beta_2}{\alpha_3}$.
                          Where ${\beta_2}$ id initialised as $\bigotimes_{k=1}^{d-1}\ket{+}$ is the boundary. \label{pa:line:m2}}
        \For{$j \in \{1 \dots m\}$ \label{pa:line:l3}}
            \parState {Measure and record the stabiliser operators of both surface code patches ${\psi\alpha_1}$ and ${\psi\alpha_3}$. \\
          and record all results.}
        \EndFor \label{pa:line:endfor}
        \parState{Restore the SC patches ${\alpha}$ and ${\beta}$ to their original size by measuring out the area that was ${\beta_1\alpha_1}$ and ${\beta_2\alpha_3}$.}
        \For{$j \in \{1 \dots d\}$} \Comment{\parbox[t]{.5\linewidth}{(You may merge this loop with that on line~\ref{pa:line:l2} of the following enclosing loop\\ iteration (line~\ref{pa:line:l1}) where it exists.)}}
            \parState {Measure and record the stabiliser operators of both ${\phi}$ and ${\psi}$.}
        \EndFor
        \parState{Update the Pauli frame predictions to correct for the split, and measure in 
         lines~\ref{pa:line:s1}-\ref{pa:line:d1} accounting for detectable errors. \label{pa:line:p1}}
        \parState{Determine the logical parity measurement $p_{i1}$ associated with the merge on line~\ref{pa:line:m1} accounting for detectable errors. \label{pa:line:p2}}
        \parState{Determine the logical parity measurement $p_{i2}$ associated with the merge on line~\ref{pa:line:m2} accounting for detectable errors.\label{pa:line:p3}}
        \State Compute $p_i = p_{i1} \oplus p_{i2}$
    \EndFor
    \parState{Compute the majority vote over all $p_i$, that is the parity XX over ${\psi}$ and ${\phi}$; update the final Pauli Frame, and return.}
\EndProcedure
\end{algorithmic}
\end{algorithm}
\end{minipage}
\end{figure*}

The lattice-surgery operations in Algorithm~\ref{alg:protected_algorithm} create a logical bell pair $$\ket{\alpha_1\alpha_3} = \frac{1}{\sqrt{2}}(\ket{00}+\ket{11})$$
on lines~\ref{pa:line:sc}-\ref{pa:line:d1}. On lines~\ref{pa:line:m1} to \ref{pa:line:endfor}, each half of this bell pair is merged with one of the input logical
qubits, measuring both the parity between $\phi$ and $\alpha_1$ and the parity between $\psi$ and $\alpha_3$. As $\alpha_1$ and $\alpha_3$ encode a bell pair, the parity between these measurements is the parity between the two input logical qubits. In addition to those lattice surgery operations, we have to measure the stabilisers for multiple periods to allow for error detection; compute an updated estimate of the Pauli frame, and correct the observed measurements before the final parity calculations are performed.

The choice of period of stabiliser measurements within the loop at line~\ref{pa:line:l2} ensures that the ancilla patches are in either a $\ket{\Phi^+}$ or a $\ket{\Psi^+}$ bell state, so that no error can propagate to the input patches during the subsequent merges. Further, the choice of period of stabiliser measurements within the loop at line~\ref{pa:line:l3} determines that the uncertainty of the parity measurement due to errors in the merged parity is no greater than that due to the width of the bell preparation ancilla.

We now have to show that the method presented here is, in fact, fault-tolerant. There are two different approaches, and we present
both here. The first is to note that the only surface code errors that can occur with a distance $w$ probability are $X$ errors in the creation of the bell
state and errors in the determination of the parity between each of the biased logical bell-state patches. Figure~\ref{fig:FoldedBus} depicts a single iteration of the loop at Line~\ref{pa:line:l1}, which shows that there are no other possible short error chains. The only short error chains are:
\begin{enumerate}
\item $Z$-error chains in the initialisation of the SC patch ${\alpha}$;
\item $Z$-error chains between initialisation of SC patch ${\alpha}$ and the splitting of the SC patch into ${\alpha_1}$, ${\alpha_2}$, and ${\alpha_3}$;
\item $Z$-error chains when measuring ${\alpha_2}$;
\item parity measurement error-chains along the time axis in performing the rough merges between $\alpha_1$ and $\phi$ and $\alpha_3$ and $\psi$.
\end{enumerate}

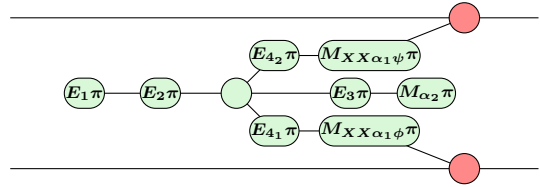
\begin{figure}[H]
    \begin{tikzpicture}
        \begin{pgfonlayer}{nodelayer}
            \node [style=none] (0) at (0.00, 0.00) {};
            \node [style=X phase dot] (1) at (12.00, 0.00) {};
            \node [style=none] (2) at (14.00, 0.00) {};
            \node [style=none] (3) at (0.00, 4.00) {};
            \node [style=X phase dot] (4) at (12.00, 4.00) {};
            \node [style=none] (5) at (14.00, 4.00) {};
            \node [style=Z phase dot] (6) at (2.00, 2.00) {$E_1\pi$};
            \node [style=Z phase dot] (7) at (4.00, 2.00) {$E_2\pi$};
            \node [style=Z phase dot] (8) at (6.00, 2.00) {};
            \node [style=Z phase dot] (9) at (9.00, 2.00) {$E_3\pi$};
            \node [style=Z phase dot] (12) at (11.00, 2.00) {$M_{\alpha_2}\pi$};
            \node [style=Z phase dot] (10) at (7.00, 1.00) {$E_{4_1}\pi$};
            \node [style=Z phase dot] (13) at (9.5, 1.00) {$M_{XX\alpha_1\phi}\pi$};
            \node [style=Z phase dot] (11) at (7.00, 3.00) {$E_{4_2}\pi$};
            \node [style=Z phase dot] (14) at (9.5, 3.00) {$M_{XX\alpha_1\psi}\pi$};
        \end{pgfonlayer}
        \begin{pgfonlayer}{edgelayer}
            \draw (0) to (1);
            \draw (1) to (2);
            \draw (3) to (4);
            \draw (4) to (5);
            \draw (6) to (7);
            \draw (7) to (8);
            \draw (8) to (9);
            \draw (9) to (12);
            \draw (8) to (10);
            \draw (8) to (11);
            \draw (13) to (1);
            \draw (14) to (4);
            \draw (10) to (13);
            \draw (11) to (14);
        \end{pgfonlayer}
    \end{tikzpicture}
    \caption{A ZX-calculus diagram showing the effect of the possible short error chains on the result of the parity measurement. }
    \label{fig:BusZXCalculus}
\end{figure}

The ZX-calculus has been proposed as a language for the description of lattice surgery operations~\cite{deBeaudrap2020}. It can be used to show that this algorithm implements a parity measurement and that all errors appear as errors in the parity measurement, see the
diagram in Figure~\ref{fig:BusZXCalculus}.
This diagram is equivalent to one of the loop iterations of the procedure in Algorithm~\ref{alg:protected_algorithm}, with all short error chain error possibilities included as rotations on the Z-nodes. These errors can be merged into a single $Z$-spider, with weight 
\begin{equation}
\label{eqn:error_rate}
\begin{split}
(E_1 + E_2 + E_3 + E_{4_1} + E_{4_2} +\\
M_{\alpha_2} + M_{XX\alpha_1\phi} + M_{XX\alpha_1\psi})\pi.
\end{split}
\end{equation}

Compare this to the diagram in Figure~\ref{fig:ErrorParityMeasurement}, which shows the ZX-diagram of an $XX$ parity measurement with a possible error in measurement. Both these diagrams are equivalent under the mapping error
$$E = E_1 + E_2 + E_3 + E_{4_1} + E_{4_2} \mod{2}$$ and parity measurement 
$$M = M_{\alpha_2} + M_{XX\alpha_1\phi} + M_{XX\alpha_1\psi} \mod{2}$$
because both can be contracted into a single $Z$-spider.

\begin{figure}[H]
    \begin{minipage}{\columnwidth}
    \[
    \begin{tikzpicture}
    \node[scale=0.6]{
\begin{quantikz}
\lstick{}        & \targ{}  &\qw     &\qw&\qw&\qw \\
\lstick{$\ket{+}$} & \ctrl{-1}&\ctrl{1}&\gate{E}&\meter{$\ket{\pm}$}& \\
\lstick{}        & \qw      &\targ{} &\qw&\qw&\qw
\end{quantikz}
};
\end{tikzpicture}
 = 
    \begin{tikzpicture}[baseline={([yshift=-.5ex]current bounding box.center)}, scale=0.6]
        \begin{pgfonlayer}{nodelayer}
            \node [style=none] (0) at (0.00, 0.00) {};
            \node [style=X phase dot] (1) at (5.00, 0.00) {};
            \node [style=none] (2) at (10.00, 0.00) {};
            \node [style=none] (3) at (0.00, 4.00) {};
            \node [style=X phase dot] (4) at (3.00, 4.00) {};
            \node [style=none] (5) at (10.00, 4.00) {};
            \node [style=Z phase dot] (6) at (1.00, 2.00) {};
            \node [style=Z phase dot] (7) at (3.00, 2.00) {};
            \node [style=Z phase dot] (8) at (5.00, 2.00) {};
            \node [style=Z phase dot] (9) at (7.00, 2.00) {$E\pi$};
            \node [style=Z phase dot] (10) at (9.00, 2.00) {$M\pi$};
        \end{pgfonlayer}
        \begin{pgfonlayer}{edgelayer}
            \draw (0) to (1);
            \draw (1) to (2);
            \draw (3) to (4);
            \draw (4) to (5);
            \draw (6) to (7);
            \draw (7) to (8);
            \draw (8) to (9);
            \draw (9) to (10);
            \draw (1) to (8);
            \draw (4) to (7);
        \end{pgfonlayer}
    \end{tikzpicture}
    \]
    \end{minipage}
    \caption{ZX-calculus diagram of the measurement of the $XX$-parity between two qubits using CNOTs and an ancilla with a possible logical error.}
    \label{fig:ErrorParityMeasurement}
\end{figure}

The probability of an unrecoverable error in the quantum bus in any loop iteration is proportional to the probability of an uncorrectable short-edge error within the bus time/space volume during the course of a single bus cycle. Hence, the error is approximated as the probability of the second and fourth error channels, as the quantum space/time volume of this region is significantly larger than any other short error channel.  Whether this error exceeds the repetition code's code capacity for a certain code distance,
error likelihood or bus length determines the minimum practical width of the bus.

\subsection{Surface code bus performance}
\label{bus:performance}
The performance of the surface code bus is understood through the analysis of the error behaviour of each inner parity measurement. This is done by determining the rates for each term in Equation~\ref{eqn:error_rate}. Terms $E_1, E_2$ and $E_3$ correspond to uncorrectable $Z$-errors in the rectangular surface code patch that spans the width of the bus. Likewise, Terms $E_{4_1}$ and $E_{4_2}$ correspond to uncorrectable errors in smaller rectangular surface code time/space volumes. The sum of the rates of these errors is then the probability of an inner parity measurement failing.

\begin{figure*}
    \includegraphics[width=\textwidth]{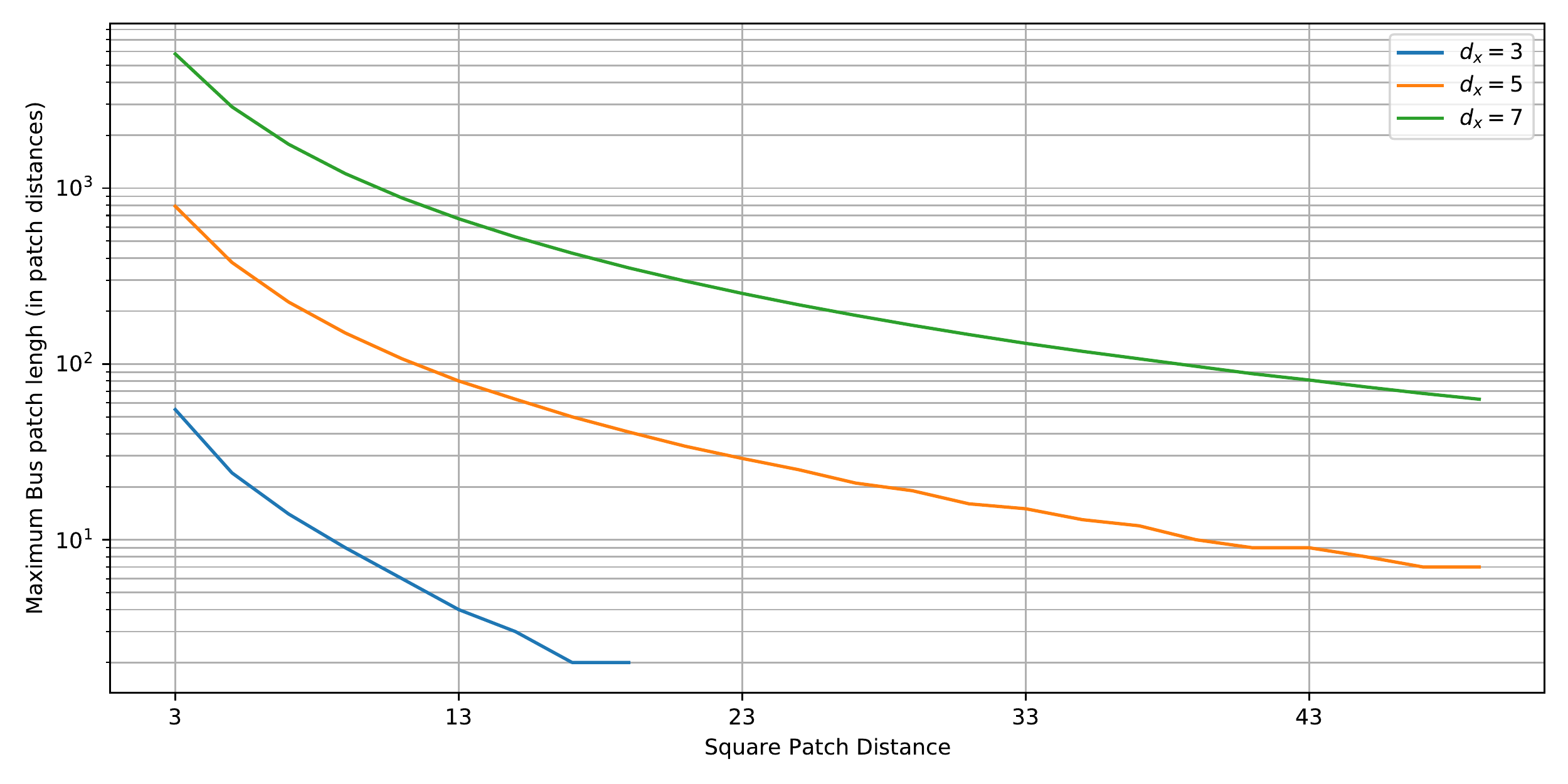}
    \caption{Maximum bus lengths, for differing bus widths, and patch dimensions. Providing space between patches for bus access to each side as in Figure~\ref{fig:sq_sc_bus}.}
    \label{fig:FoldedBusLengths}
\end{figure*}

These error rates are exactly what was simulated in Section~\ref{sec:rectsc}. From these results, the rate of errors that occur in a single folded bus section can be determined. As the repetition code has a threshold of $0.5$, it is trivial to check differing combinations of bus lengths, bus widths and surface code distances to determine if each bus operation is below this threshold. This check was performed, and the results are presented in Figure~\ref{fig:FoldedBusLengths}.

Whilst it is true that the total code presented in Algorithm~\ref{alg:protected_algorithm} will be of distance $d$, the error rates on the bus will strongly depend on the width and the length of the bus. In order to ensure that the rate of errors on the bus is the same or lower than the rate of errors in the data logical qubits, we have evaluated the performance of the surface code bus at differing bus lengths, widths and surface code distances, and then determined the minimum number of repetitions that ensures the probability of an error on the bus is lower than the probability of a memory error. The results of this calculation are shown in Figure~\ref{fig:FoldedBusRates}.

\begin{figure*}
    \begin{subfigure}[b]{0.82\textwidth}
        \includegraphics[width=\textwidth]{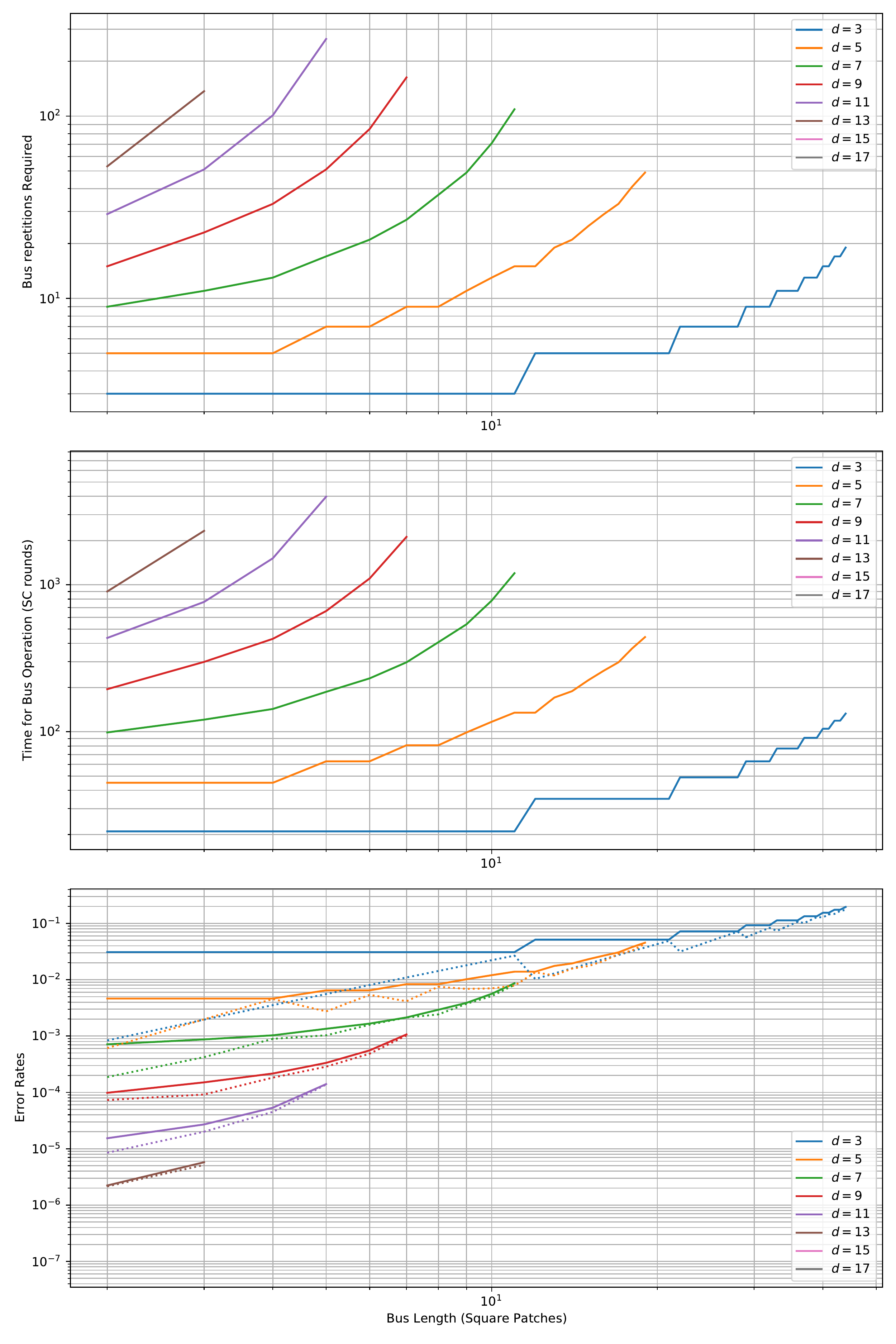}
        \caption{$w = 3$}
        \label{fig:BusRates3}
    \end{subfigure}
    \caption{Performance of the Surface code bus for $w \in \{3,5,7\}$. In the top plot of each group we have the number of repetitions to make the short-edge error equal to the patch error for one bus cycle. In the middle Graph we have the total number of surface code cycles required for a single bus parity measurement. The lower graph shows the effective error rate on the surface code patch (solid), and bus operation(dotted) when using the number of repetitions.}
\end{figure*}
\begin{figure*}\ContinuedFloat
        \begin{subfigure}[b]{0.82\textwidth}
        \includegraphics[width=\textwidth]{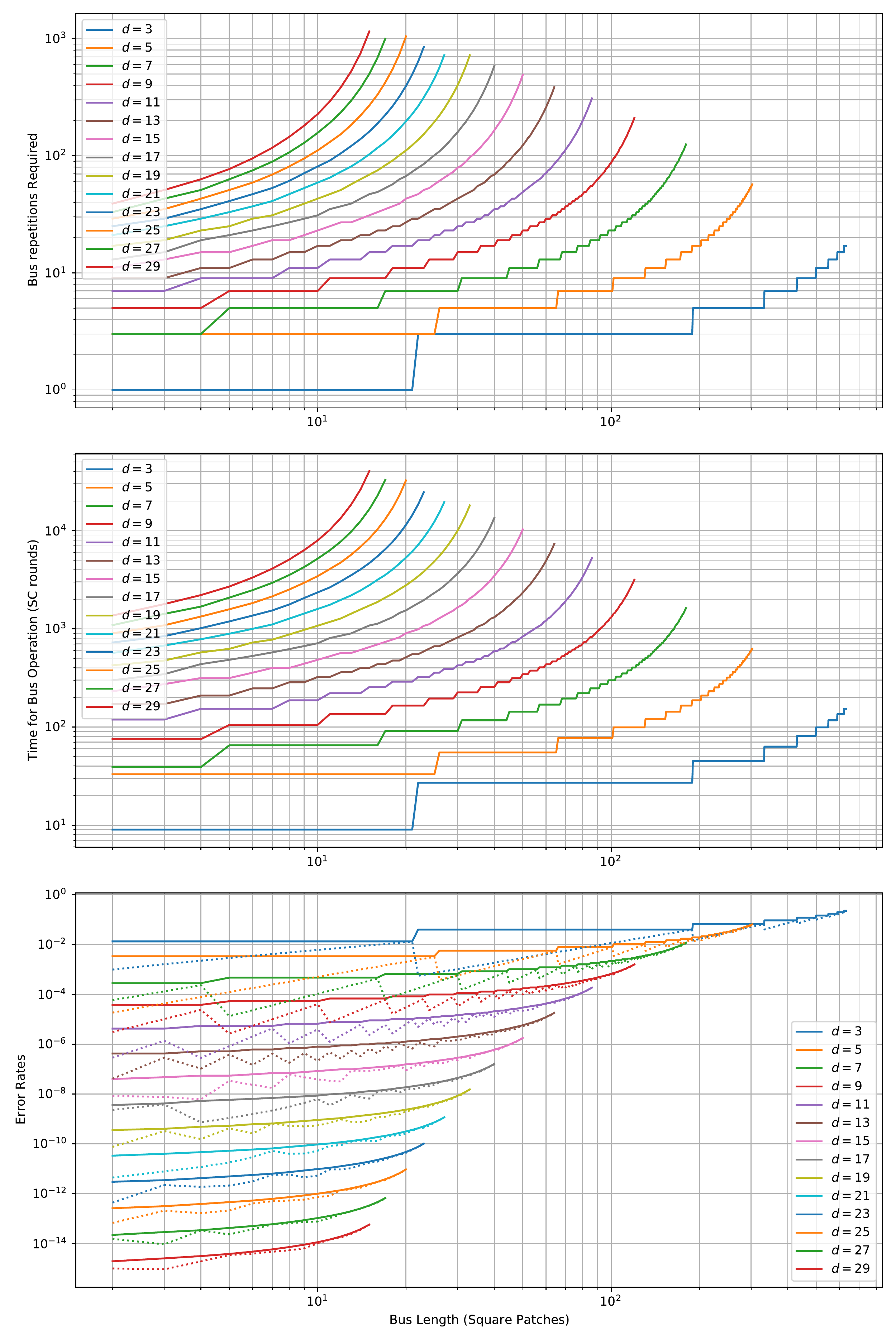}
        \caption{$w = 5$}
        \label{fig:BusRates5}
    \end{subfigure}
    \caption{continued. Performance of the Surface code bus for $w \in \{3,5,7\}$. In the top plot of each group we have the number of repetitions to make the short-edge error equal to the patch error for one bus cycle. In the middle Graph we have the total number of surface code cycles required for a single bus parity measurement. The lower graph shows the effective error rate on the surface code patch (solid), and bus operation(dotted) when using the number of repetitions.}
\end{figure*}
\begin{figure*}\ContinuedFloat
    \begin{subfigure}[b]{0.82\textwidth}
         \includegraphics[width=\textwidth]{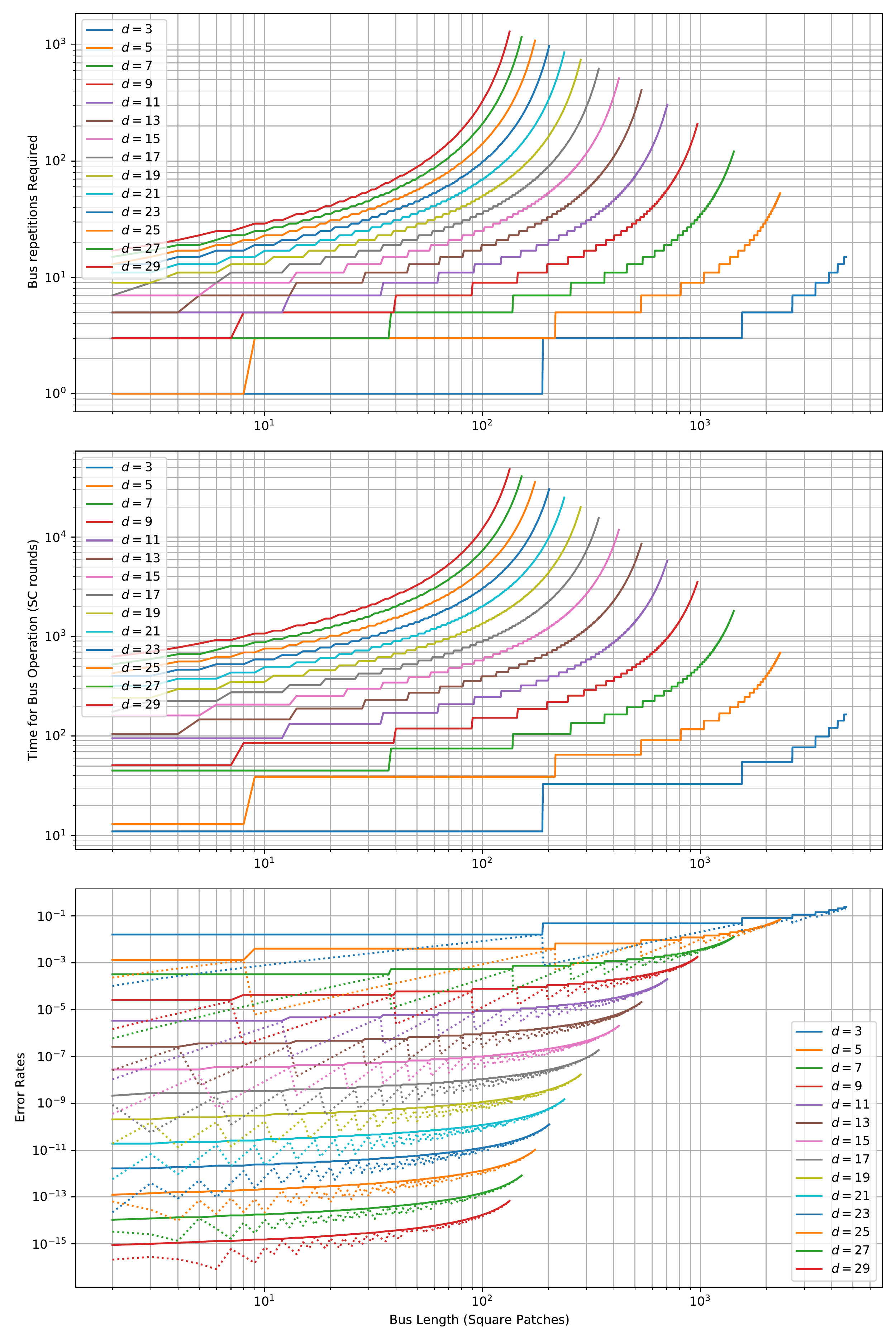}
        \caption{$w = 7$}
        \label{fig:BusRates7}
    \end{subfigure}
    \caption{continued. Performance of the Surface code bus for $w \in \{3,5,7\}$. In the top plot of each group we have the number of repetitions to make the short-edge error equal to the patch error for one bus cycle. In the middle Graph we have the total number of surface code cycles required for a single bus parity measurement. The lower graph shows the effective error rate on the surface code patch (solid), and bus operation(dotted) when using the number of repetitions.}
    \label{fig:FoldedBusRates}
\end{figure*}

         \section{Parity codes for the bus}
         \label{sec:parity}
\begin{figure*}
         \includegraphics[width=\textwidth]{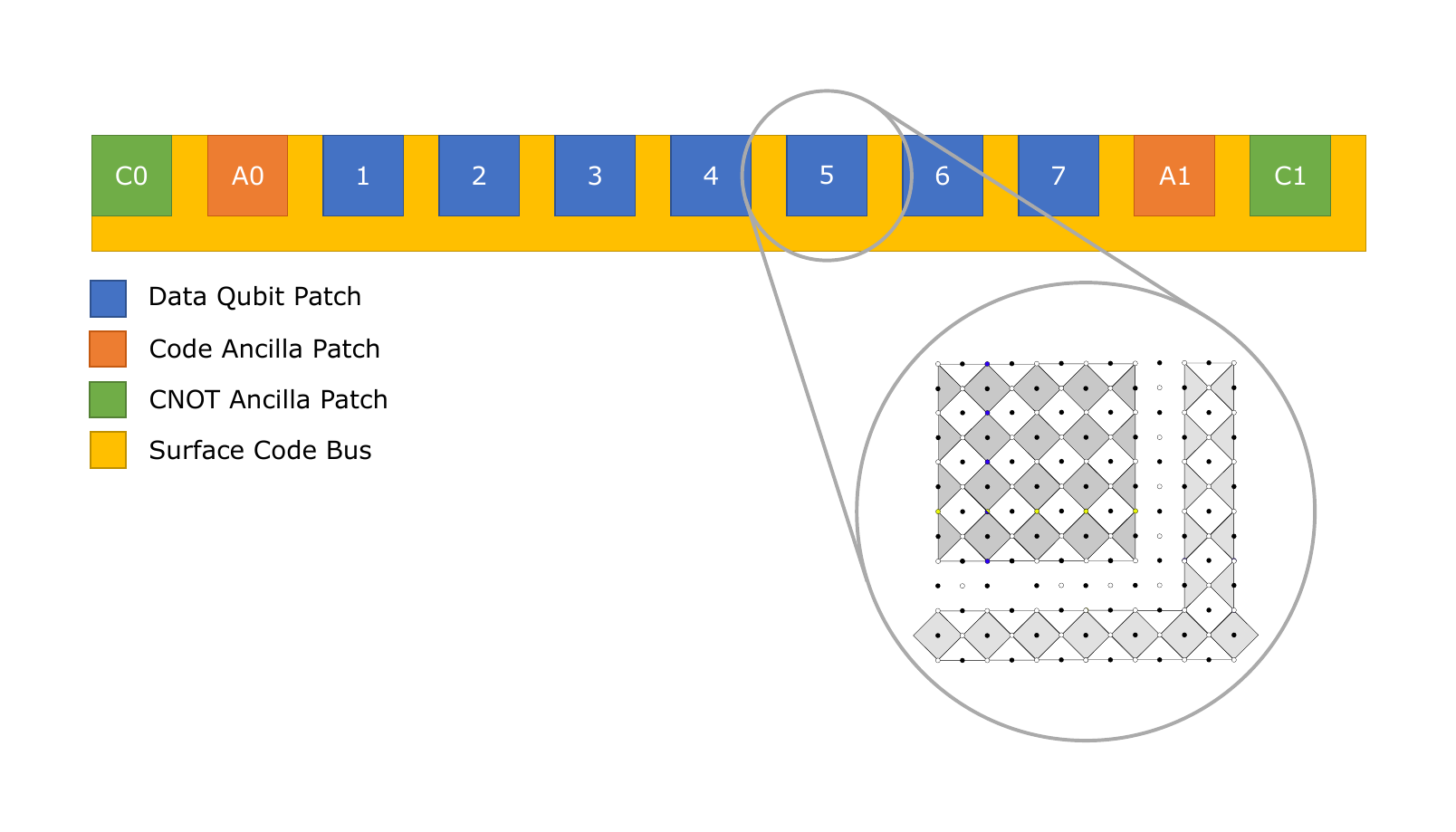}
         \caption{An descriptive illustration of what the Steane $\dbracket{7,1,3}$ code on top of $d=5$ surface code patches connected with a $w=2$ bus might look like.}
         \label{fig:sq_sc_bus}
\end{figure*}

Our first attempt to reduce the width of lattice required to perform universal quantum computing by using the fault-tolerant surface code bus introduced in Section~\ref{sec:bus} to mediate the interactions of logical qubits so that another code may be concatenated above it as illustrated in Figure~\ref{fig:sq_sc_bus}.

With a single bus, such a layout should give that code a
threshold lower than when complete connectivity at the logical layer is possible but higher than what is required if the logical architecture
of the qubits were to be nearest-neighbour linear interactions. This is because while we can interact any two qubits with each other,
each linear section of the bus can only be used for one interaction at a time. 

To determine whether this technique could have applicability, codes were selected to evaluate, for both performance
and ease of implementation. A fault tolerance scheme was chosen for each code, and a decoder was designed as described below. Then the performance of each code was simulated using these choices under the timing constraints of the surface code bus
lower layer. It is worth noting here, that while the bus can implement multi-qubit parity measurements, it is not known whether these can be used fault-tolerantly in this concatenation scheme, so parity measurements on the bus were limited to weight 2 in this consideration.

\subsection{Choice of codes}
\label{par:code_choice}
There were three main considerations in choosing which block codes to test: the expected threshold of the different codes; the ease of implementing the codes, and the amount of scholarly literature that exists on the code performance. The flag-qubit method for fault-tolerance described in the section \ref{bg:flag_qubits} showed significant benefits in terms of reduced qubit count and
circuit simplicity, which enabled a reduced threshold~\cite{ChaoReighardt2018}.
This choice left three codes with flag-qubit circuits within the published literature that could be easily tested: the $\dbracket{5,1,3}$ five qubit code; Steane's $\dbracket{7,1,3}$ code, and the $\dbracket{15,7,3}$ CSS code. As the five-qubit code historically demonstrated a much lower threshold than the other two codes~\cite{CrossDivincenzoTerhal2009}, we analysed the threshold performance of both the $\dbracket{7,1,3}$ code and the $\dbracket{15,7,3}$ code to examine the tradeoff between qubit density and the total width of the processor array, $w$.

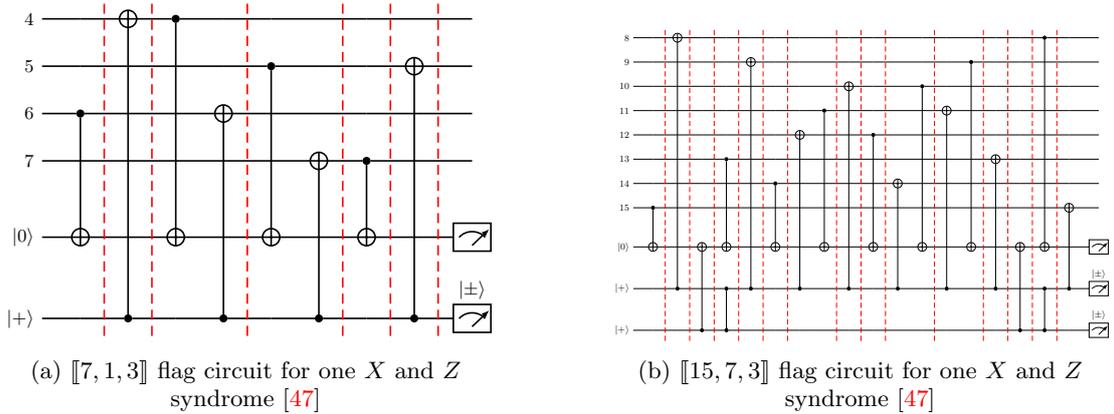
\begin{figure*}
     \begin{subfigure}[b]{0.45\textwidth}
\begin{adjustbox}{width=1\textwidth}
         \begin{quantikz}
\lstick{4}          &\qw       \slice{} &\targ{}   \slice{} &\ctrl{4}   &\qw        \slice{}&\qw        &\qw        \slice{}&\qw  \slice{}      &\qw     \slice{}  &\qw \\
\lstick{5}          &\qw        &\qw        &\qw        &\qw        &\ctrl{3}   &\qw        &\qw        &\targ{}   &\qw \\
\lstick{6}          &\ctrl{2}   &\qw        &\qw        &\targ{}    &\qw        &\qw        &\qw        &\qw       &\qw \\
\lstick{7}          &\qw        &\qw        &\qw        &\qw        &\qw        &\targ{}    &\ctrl{1}   &\qw       &\qw \\ [1 em]
\lstick{$\ket{0}$}  &\targ{}    &\qw        &\targ{}    &\qw        &\targ{}    &\qw        &\targ{}    &\qw       &\meter{}\\
\lstick{$\ket{+}$}  &\qw        &\ctrl{-5}  &\qw        &\ctrl{-3}  &\qw        &\ctrl{-2}  &\qw        &\ctrl{-4}   &\meter{$\ket{\pm}$}
         \end{quantikz}
\end{adjustbox}
         \caption{$\dbracket{7,1,3}$ flag circuit for one $X$ and $Z$ syndrome~\cite{Reichardt2020}}
         \label{fig:steane_circuit}
     \end{subfigure}
     \hfill
     \begin{subfigure}[b]{0.45\textwidth}
        \begin{adjustbox}{width=1\textwidth}
         \begin{quantikz} 
\lstick{8}		&\qw	\slice{}	&\targ{} \slice{}	&\qw	\slice{}	&\qw	\slice{}	&\qw	\slice{}	&\qw	\slice{}	&\qw		&\qw	\slice{}	&\qw	\slice{}	&\qw	\slice{}	&\qw		&\qw	\slice{}	&\qw		&\qw	\slice{}	&\qw	\slice{}	&\qw	\slice{}	&\ctrl{8}\slice{}	&\qw &\qw	\\	
\lstick{9}		&\qw		&\qw		&\qw		&\qw		&\targ{}	&\qw		&\qw		&\qw		&\qw		&\qw		&\qw		&\qw		&\qw		&\ctrl{7}	&\qw		&\qw		&\qw		&\qw		&\qw	\\	
\lstick{10}		&\qw		&\qw		&\qw		&\qw		&\qw		&\qw		&\qw		&\qw		&\targ{}	&\qw		&\qw		&\ctrl{6}	&\qw		&\qw		&\qw		&\qw		&\qw		&\qw		&\qw	\\	
\lstick{11}		&\qw		&\qw		&\qw		&\qw		&\qw		&\qw		&\qw		&\ctrl{5}	&\qw		&\qw		&\qw		&\qw		&\targ{}	&\qw		&\qw		&\qw		&\qw		&\qw		&\qw	\\	
\lstick{12}		&\qw		&\qw		&\qw		&\qw		&\qw		&\qw		&\targ{}	&\qw		&\qw		&\ctrl{4}	&\qw		&\qw		&\qw		&\qw		&\qw		&\qw		&\qw		&\qw		&\qw	\\	
\lstick{13}		&\qw		&\qw		&\qw		&\ctrl{3}	&\qw		&\qw		&\qw		&\qw		&\qw		&\qw		&\qw		&\qw		&\qw		&\qw		&\targ{}	&\qw		&\qw		&\qw		&\qw	\\	
\lstick{14}		&\qw		&\qw		&\qw		&\qw		&\qw		&\ctrl{2}	&\qw		&\qw		&\qw		&\qw		&\targ{}	&\qw		&\qw		&\qw		&\qw		&\qw		&\qw		&\qw		&\qw	\\	
\lstick{15}		&\ctrl{1}	&\qw		&\qw		&\qw		&\qw		&\qw		&\qw		&\qw		&\qw		&\qw		&\qw		&\qw		&\qw		&\qw		&\qw		&\qw		&\qw		&\targ{}	&\qw	\\[1 em]
\lstick{$\ket{0}$}	&\targ{}	&\qw		&\targ{}	&\targ{}	&\qw		&\targ{}	&\qw		&\targ{}	&\qw		&\targ{}	&\qw		&\targ{}	&\qw		&\targ{}	&\qw		&\targ{}	&\targ{}	&\qw		&\meter{}\\	
\lstick{$\ket{+}$}	&\qw		&\ctrl{-9}	&\qw		&\ctrl{1}	&\ctrl{-8}	&\qw		&\ctrl{-5}	&\qw		&\ctrl{-7}	&\qw		&\ctrl{-3}	&\qw		&\ctrl{-6}	&\qw		&\ctrl{-4}	&\qw		&\ctrl{1}	&\ctrl{-2}	&\meter{$\ket{\pm}$}	\\	
\lstick{$\ket{+}$}	&\qw		&\qw		&\ctrl{-2}	&\ctrl{}	&\qw		&\qw		&\qw		&\qw		&\qw		&\qw		&\qw		&\qw		&\qw		&\qw		&\qw		&\ctrl{-2}	&\ctrl{}	&\qw		&\meter{$\ket{\pm}$}
         \end{quantikz}
        \end{adjustbox}
         \caption{$\dbracket{15,7,3}$ flag circuit for one $X$ and $Z$ syndrome~\cite{Reichardt2020}}
         \label{fig:fifteen_seven_circuit}
     \end{subfigure}
        \caption{Flag qubit extraction circuits used for $\dbracket{7,1,3}$  and  $\dbracket{15,7,3}$ codes}
        \label{fig:block_circuits}
\end{figure*}

To choose which syndrome extraction circuit should be used for each code, we evaluated the total bus latency to extract all syndromes for each of the flagged extraction circuits found in the literature~\cite{Reichardt2020, ChaoReighardt2018}. The Steane $\dbracket{7,1,3}$ code circuit in figure \ref{fig:steane_circuit} had the lowest complexity by far, requiring only 18 bus cycles. There were two candidate circuits for the $\dbracket{15,7,3}$. Both had the same bus latency, so the circuit in \ref{fig:fifteen_seven_circuit} was chosen because it was slightly easier to develop its decoder.

\subsection{Decoder design}
\label{par:dec_design}
A decoder had to be designed for each of the top-level block codes after the codes and circuits had been chosen. As each code was distance three, a brute force analysis was feasible. We evaluated all weight-one Pauli errors after initialisation, on idle qubits, and before measurements, and all weight-two Pauli errors before CNOT gates for all gates in a round of fault-tolerant syndrome extraction. The measured error syndrome is recorded for each error, and the errors propagated through the syndrome extraction to create a decoder. The decoder has to provide a unique correction for each error syndrome, or at least leave an error that can be corrected in the next cycle.

For the Steane $\dbracket{7,1,3}$ CSS code, a fault-tolerant round of syndrome extraction consists of either two or three complete sets of syndrome extraction. Each syndrome extraction is made up of three copies of the circuit in figure \ref{fig:steane_circuit}, with the qubits permuted to extract syndromes 1 and 2 first, followed by syndromes 3 and 4, and finally syndromes 5 and 6. If two sets of measurements return identical syndromes, then the procedure is complete. Otherwise, an additional set of measurements is taken to complete the round of fault-tolerant extraction. The decoder for this fault-tolerant syndrome extraction was calculated manually, given the relatively short size of the decision tree required. The correctness of the decoder was checked automatically with an exhaustive search. 

For the $\dbracket{15,7,3}$ CSS code, a fault-tolerant round of syndrome extraction only involves one or two complete sets of syndrome extraction. Each is made up of four copies of the circuit in figure \ref{fig:fifteen_seven_circuit}, with the qubits permuted to extract the syndromes in ordered pairs, first with syndromes 1 and 2, then syndromes 3 and 4, and so on. In this case, a single set of measurements is required if the syndromes all return unchanged values (zero/plus), otherwise two sets of measurements were made. For this code, the decoder was too large to consider optimizing manually, so an automatic routine to extract and verify a decoding table was created.

\subsection{Simulation Design}
\label{par:sim_design}
\begin{figure*}
        \includegraphics[width=\textwidth]{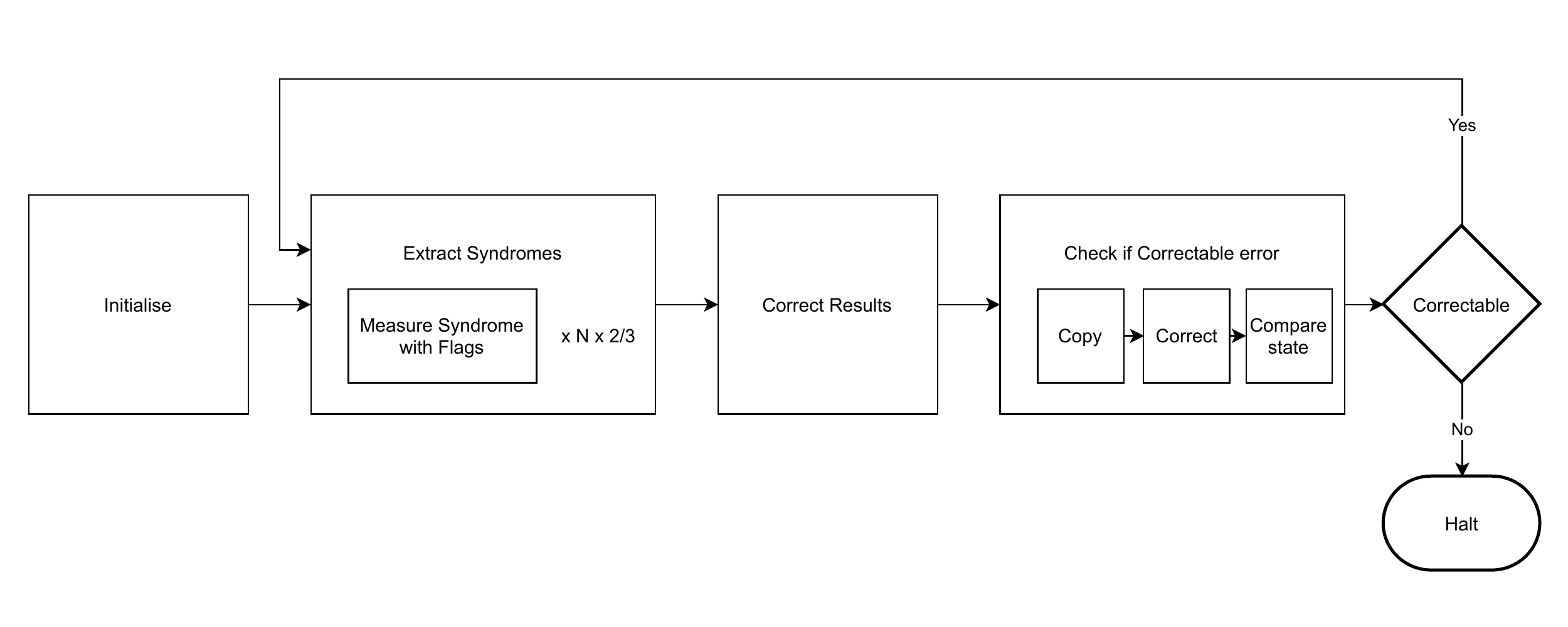}
        \caption{Flow chart for simulation of block code performance simulation.}
        \label{fig:sim_flow}
\end{figure*}
The codes were simulated with the balanced error model described in section~\ref{bg:err} to evaluate their performance. As the bus operations take $d^2$ time steps, the error rate in this simulation assumes that measurement and initialisation of logical ancilla patches are error-free operations, as they effectively take one round of syndrome extraction on the surface code layer. This is because initialization requires $d$ rounds of syndrome extraction in the surface code to be fault-tolerant. However, these can occur simultaneously with the first bus operation, so do not introduce any additional errors or computation time. 

The simulation for these codes was performed with a custom CHP simulator. We ran repeated syndrome extractions with perfect corrections applied between them until an uncorrectable error occurred.  When evaluating the Steane $\dbracket{7,1,3}$ code after each round of syndrome extraction and correction, a copy of the simulated state was placed back into a +1 eigenstate of all the surface code stabilisers by performing syndrome extraction with errors disabled. The parity of the logical operator, in the appropriate basis, was then measured and compared to the initial state. To correct all order one errors, we determined that three rounds of syndrome extraction were required prior to measuring the eigenstate of the logical operator and compared it to the initial value for this implementation of the $\dbracket{15,7,3}$ code. In both cases, the mean and standard deviation of the number rounds before failure were computed and used to find the logical error rate. 

The $\pm 1$ eigenstates of Pauli-$X$, Pauli-$Y$ or Pauli-$Z$ were chosen as our initial encoded states for the $\dbracket{7,1,3}$ code. The states $\ket{0}^{\otimes7}$,$\ket{1}^{\otimes7}$,$\ket{+}^{\otimes7}$,$\ket{-}^{\otimes7}$,$\ket{i+}^{\otimes7}$, and $\ket{i-}^{\otimes7}$ were chosen as our initial encoded states for the $\dbracket{15,7,3}$ code, with the restriction of states here due to limited computing resources.

\subsection{Simulation Results}
\label{par:sim_results}
\begin{figure*}
     \begin{subfigure}[b]{0.7\textwidth}
         \includegraphics[width=\textwidth]{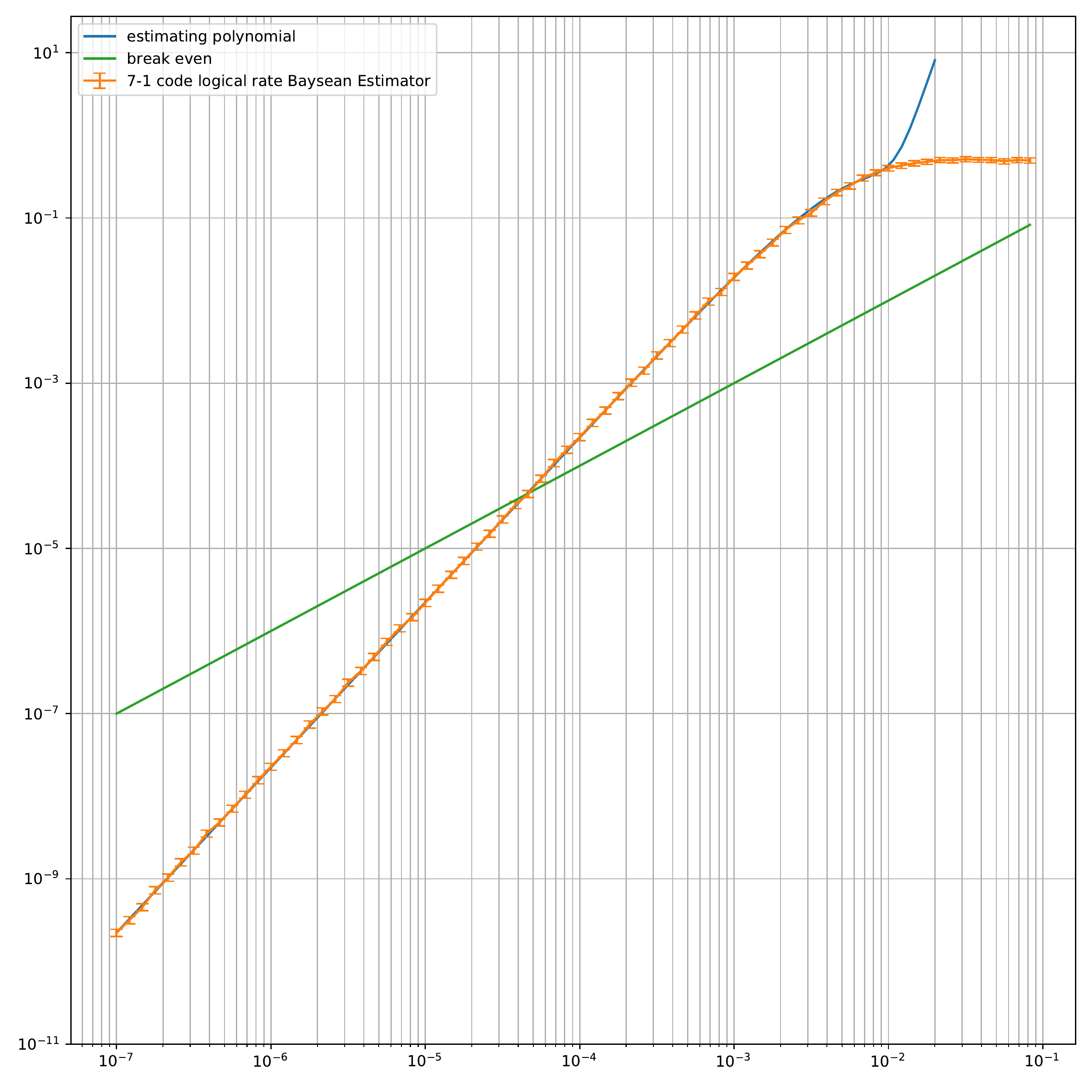}
         \caption{$\dbracket{7,1,3}$  code}
         \label{fig:steane}
     \end{subfigure}
     \bigskip 
     \begin{subfigure}[b]{0.7\textwidth}
         \includegraphics[width=\textwidth]{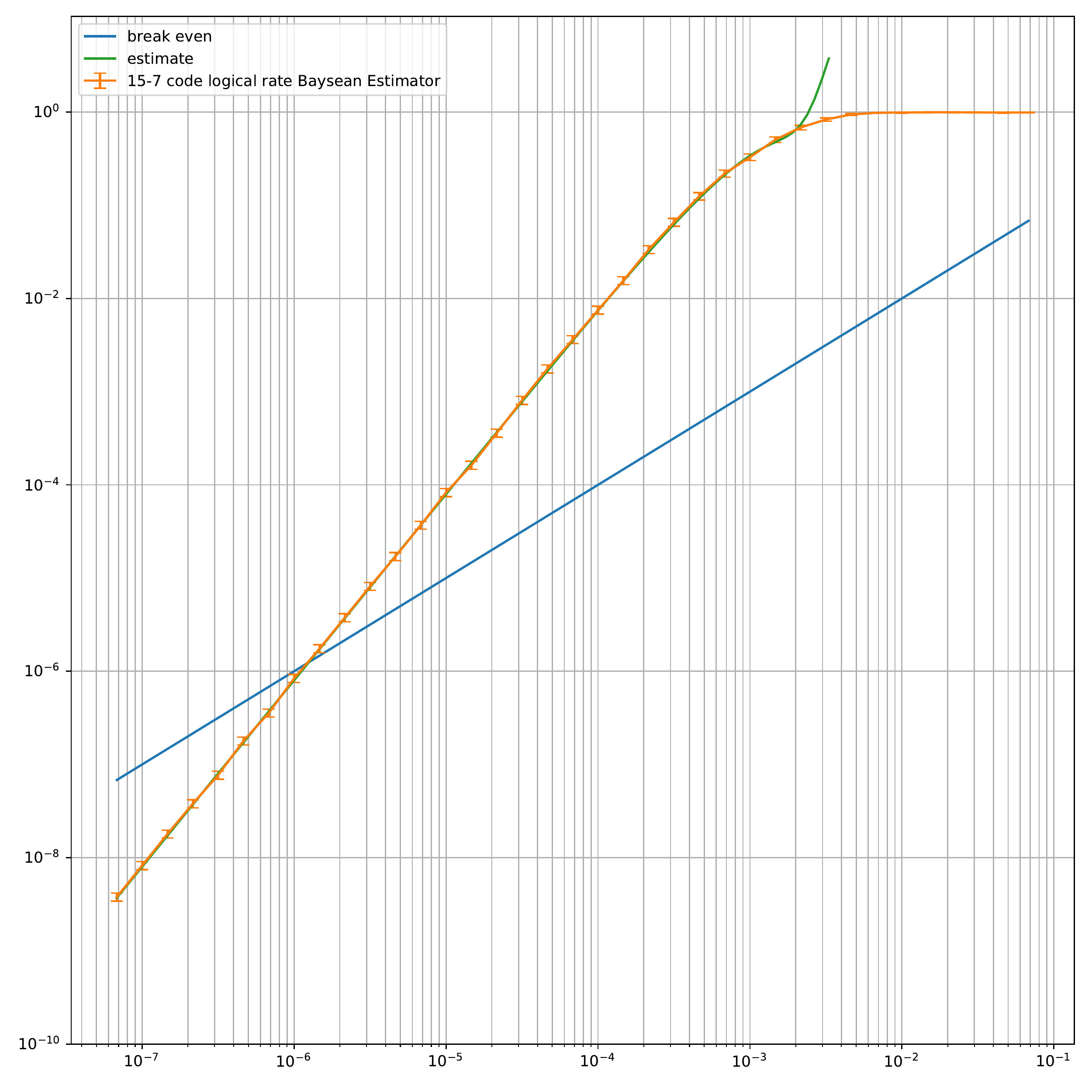}
         \caption{$\dbracket{15,7,3}$ code}
         \label{fig:fifteen_seven_graph}
     \end{subfigure}
        \caption{The logical error rates for the $\dbracket{7,1,3}$  and  $\dbracket{15,7,3}$ codes}
        \label{fig:block_graphs}
\end{figure*}

For each physical error rate, 900 simulations to failure were computed. Then the mean of these gaussian samples was used to compute an estimator for the Bernoulli trials. The maximum likelihood estimator and the Bayesian posterior mean of the posterior distribution were both evaluated, and they produced almost identical results. The posterior mean and $3\sigma$ confidence interval were calculated assuming a uniform prior distribution, with the results presented in Figure~\ref{fig:block_graphs}. As initially expected, the $3\sigma$ confidence interval is approximately $\pm 10\%$ around the estimate. From these parameter estimates, a polynomial fit was manually extracted to degree-4 for each code. These are
\begin{equation}
\begin{aligned}
p_{\dbracket{7,1,3} }(p) =&\quad 2.23\times 10^4 p^2 - 3.5\times 10^6 p^3\\
& + 1.7\times 10^8 p^4
\end{aligned}
\end{equation}
for the $\dbracket{7,1,3}$ code and 
\begin{equation}
\begin{aligned}
p_{\dbracket{15,7,3} }(p) =&\quad 8.00\times 10^5 p^2 - 6.0\times 10^8 p^3\\
&+ 14\times 10^{10} p^4
\end{aligned}
\end{equation}
for the $\dbracket{15,7,3}$ code, which are also depicted in Figure \ref{fig:block_graphs}. 

From these plots, we can see that the Steane $\dbracket{7,1,3}$ code has a pseudo-threshold of approximately $p = 4.52\times 10^{-5}$ when used with the surface code bus. Similarly the pseudo-threshold for the $\dbracket{15,7,3}$ code is approximately $p = 1.25 \times 10^{-6}$.

\section{Performance of the concatenated codes }
\label{sec:concat_perf}
To evaluate the concatenated quantum error correcting codes, we must:
\begin{itemize}
\item Determine the performance of the base surface code layer, and evaluate the performance of two-qubit interactions.
\item Determine a compilation scheme at the first layer for the $\dbracket{7,1,3}$ or $\dbracket{15,7,3}$ codes above the surface code.
\item Determine a scheme to concatenate additional levels of the code above this in a self-similar manner.
\item Calculate the performance of this scheme using these compilation strategies recursively.
\end{itemize}

We must set a target for final logical error performance to make width and area comparisons. We have chosen an error rate of $10^{-15}$ for a logical CNOT or parity measurement operation between two adjacent qubits as our target in this work.

The performance is evaluated at several physical error rates between $10^{-3}$ and $10^{-4}$, which represent both near, and intermediate-term physical error targets. We then determine the required architecture size for these rates. These results are presented in Table~\ref{tab:table1} and are calculated using the procedures described in the remainder of this section.

\begin{table*}[htbp]
\begin{subtable}[t]{\textwidth}
{\setlength{\tabcolsep}{1em}
\begin{tabular}[t]{ l c c c c c c}
\toprule
  &  $d_{sc}$& $d_b$ &  $w$ & $p_l$ & block size & qubit density \\
\midrule
Surface Code - No Bus & 27 & NA & 107 & $5.49 \times 10^{-16}$ & 5778 & 5778\\
Surface Code - Bus    & 31 & 5 & 71 & $ 1.66 \times 10^{-16}$ & 5112  & 5112\\
\midrule                                                  
1L $\dbracket{7,1,3}$ on SC + Bus  & 21 & 7 & 55 & $3.96 \times 10^{-17}$ & $3.39 \times 10^{4}$ & $3.39 \times 10^{4}$\\
2L $\dbracket{7,1,3}$ on SC + Bus  & 15 & 7 & 43 & $1.91 \times 10^{-16}$ & $1.87 \times 10^{5}$ & $1.87 \times 10^{5}$\\
3L $\dbracket{7,1,3}$ on SC + Bus  & 13 & 7 & 39 & $4.64 \times 10^{-16}$ & $1.39 \times 10^{6}$ & $1.39 \times 10^{6}$\\
7L $\dbracket{7,1,3}$ on SC + Bus  & 11 & 7 & 35 & $8.27 \times 10^{-21}$ & $7.37 \times 10^{9}$ & $7.37 \times 10^{9}$\\
\midrule
1L $\dbracket{15,7,3}$ on SC + Bus & 23 & 7 & 59 & $2.74 \times 10^{-17}$ & $7.08 \times 10^{4}$ & $1.01 \times 10^{4}$\\
2L $\dbracket{15,7,3}$ on SC + Bus & 21 & 7 & 55 & $2.50 \times 10^{-20}$ & $1.11 \times 10^{6}$ & $2.26 \times 10^{4}$\\
3L $\dbracket{15,7,3}$ on SC + Bus & 19 & 7 & 51 & $9.73 \times 10^{-21}$ & $1.72 \times 10^{6}$ & $5.01 \times 10^{4}$\\
4L $\dbracket{15,7,3}$ on SC + Bus & 17 & 7 & 47 & $3.84 \times 10^{-17}$ & $2.63 \times 10^{8}$ & $1.10 \times 10^{5}$\\
\bottomrule
\end{tabular}}
\caption{\footnotesize $p = 1.0 \times 10^{-3}$}
\label{tab:table1_a}
\end{subtable}
\begin{subtable}[t]{\textwidth}
{\setlength{\tabcolsep}{1em}
\begin{tabular}[t]{ l c c c c c c}
\toprule
  &  $d_{sc}$& $d_b$ &  $w$ & $p_l$ & block size & qubit density \\
\midrule
Surface Code - No Bus & 21 & NA  & 83 & $3.23 \times 10^{-16}$ & 3652 &3652 \\
Surface Code - Bus    & 23 & 5 & 55 & $3.26\times 10^{-16}$ & 2450  & 2450\\
\midrule                                                  
1L $\dbracket{7,1,3}$ on SC + Bus  & 15 & 5 & 39 & $5.11 \times 10^{-16}$ & $1.72 \times 10^{4}$ & $1.72 \times 10^{4}$\\
2L $\dbracket{7,1,3}$ on SC + Bus  & 11 & 5 & 35 & $3.56 \times 10^{-16}$ & $9.82 \times 10^{4}$ & $9.82 \times 10^{4}$\\
4L $\dbracket{7,1,3}$ on SC + Bus  &  9 & 5 & 27 & $4.31 \times 10^{-19}$ & $6.06 \times 10^{6}$ & $6.06 \times 10^{6}$\\
\midrule
1L $\dbracket{15,7,3}$ on SC + Bus & 17 & 5 & 43 & $6.59 \times 10^{-17}$ & $3.78 \times 10^{4}$ & $5.41 \times 10^{3}$\\
2L $\dbracket{15,7,3}$ on SC + Bus & 15 & 5 & 39 & $2.94 \times 10^{-18}$ & $5.62 \times 10^{5}$ & $1.15 \times 10^{4}$\\
4L $\dbracket{15,7,3}$ on SC + Bus & 13 & 5 & 35 & $4.34 \times 10^{-16}$ & $8.16 \times 10^{6}$ & $2.38 \times 10^{4}$\\
\bottomrule
\end{tabular}}
\caption{\footnotesize  $p = 4.7 \times 10^{-4}$}
\label{tab:table1_b}
\end{subtable}
\begin{subtable}[t]{\textwidth}
{\setlength{\tabcolsep}{1em}
\begin{tabular}[t]{ l c c c c c c}
\toprule
  &  $d_{sc}$& $d_b$ &  $w$ & $p_l$ & block size & qubit density \\
\midrule
Surface Code - No Bus & 17 & NA & 67 & $2.48 \times 10^{-16}$ & 2278 & 2278\\
Surface Code - Bus    & 19 & 3  & 43 & $1.40 \times 10^{-16}$ & 1892 & 1892\\
\midrule                                                  
1L $\dbracket{7,1,3}$ on SC + Bus  & 13 & 3 & 31 & $5.07 \times 10^{-17}$ & $1.09 \times 10^{4}$ & $1.09 \times 10^{4}$\\
2L $\dbracket{7,1,3}$ on SC + Bus  &  9 & 5 & 27 & $6.43 \times 10^{-18}$ & $7.48 \times 10^{4}$ & $7.48 \times 10^{4}$\\
4L $\dbracket{7,1,3}$ on SC + Bus  &  7 & 5 & 23 & $9.82 \times 10^{-22}$ & $4.43 \times 10^{6}$ & $4.43 \times 10^{6}$\\
\midrule
1L $\dbracket{15,7,3}$ on SC + Bus & 13 & 5 & 35 & $1.50 \times 10^{-16}$ & $2.52 \times 10^{4}$ & $3.60 \times 10^{3}$\\
2L $\dbracket{15,7,3}$ on SC + Bus & 13 & 3 & 31 & $1.54 \times 10^{-19}$ & $3.57 \times 10^{5}$ & $7.29 \times 10^{3}$\\
3L $\dbracket{15,7,3}$ on SC + Bus & 11 & 3 & 27 & $3.26 \times 10^{-17}$ & $4.90 \times 10^{6}$ & $1.43 \times 10^{4}$\\
\bottomrule
\end{tabular}}
\caption{\footnotesize  $p = 2.2 \times 10^{-4}$}
\label{tab:table1_c}
\end{subtable}

\begin{subtable}[t]{\textwidth}
{\setlength{\tabcolsep}{1em}
\begin{tabular}[t]{ l c c c c c c}
\toprule
  &  $d_{sc}$& $d_b$ &  $w$ & $p_l$ & block size & qubit density \\
\midrule
Surface Code - No Bus & 15 & NA & 59 & $2.59 \times 10^{-17}$ & 1770 & 1770\\
Surface Code - Bus    & 15 & 3  & 35 & $4.93 \times 10^{-16}$ & 1260 & 1260\\
\midrule                                                  
1L $\dbracket{7,1,3}$ on SC + Bus  & 11 & 3 & 27 & $2.46 \times 10^{-18}$ & $8.32 \times 10^{3}$ & $8.32 \times 10^{3}$\\
2L $\dbracket{7,1,3}$ on SC + Bus  &  7 & 3 & 23 & $6.78 \times 10^{-16}$ & $3.76 \times 10^{4}$ & $3.76 \times 10^{4}$\\
9L $\dbracket{7,1,3}$ on SC + Bus  &  5 & 3 & 19 & $5.10 \times 10^{-22}$ & $1.14 \times 10^{11}$ & $1.14 \times 10^{11}$\\
\midrule
1L $\dbracket{15,7,3}$ on SC + Bus & 11 & 3 & 27 & $1.67 \times 10^{-16}$ & $1.51 \times 10^{4}$ & $2.16 \times 10^{3}$\\
3L $\dbracket{15,7,3}$ on SC + Bus &  9 & 3 & 23 & $1.44 \times 10^{-20}$ & $3.58 \times 10^{5}$ & $1.04 \times 10^{4}$\\
\bottomrule
\end{tabular}}
\caption{\footnotesize  $p = 1.0 \times 10^{-4}$}
\label{tab:table1_d}
\end{subtable}

\caption{\footnotesize Error rates for different code concatenation configurations, parameters, and physical error rates. In this figure $d_{sc}$ is the surface code distance, $d_b$ is the bus width distance, $w$ is the width of the lattice, and $p_l$ is the logical error rate of the concatenated code.}
\label{tab:table1}
\end{table*}

\subsection{Performance of the surface code, and the surface code bus.}
\label{perf:sc_scbus}
The performance of the surface code is determined to provide a base for comparison, which is required when establishing the performance of the concatenated codes. Here, we estimate the error rate of the surface code using the fit in Equation~\ref{eqn:scfit} multiplied by $d+1$, the number of rounds required for a lattice surgery operation. 

We use the procedure described in Section~\ref{bus:performance} to calculate the estimate of the bus error rate using the simplifying assumption that all bus operations have the worst-case length. These bus operations are then used as the underlying physical error rate for concatenated codes, which will result in slightly higher logical error rates than a tighter analysis using exact lengths.

\subsection{Compilation of the Steane $\dbracket{7,1,3}$ code}
\label{perf:steane_comp}
\begin{figure*}
     \begin{subfigure}[b]{\textwidth}
         \includegraphics[width=\textwidth]{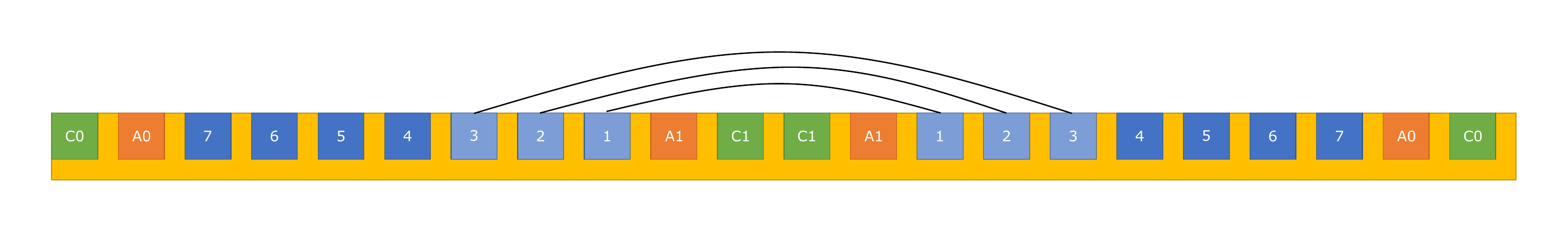}
         \caption{Patch interactions required to perform logical Steane operations using the $XXXIIII$ or $ZZZIIII$ logical operator, between adjacent Steane arrays. }
         \label{fig:concat_patch_1}
     \end{subfigure}
     
     \begin{subfigure}[b]{\textwidth}
         \includegraphics[width=\textwidth]{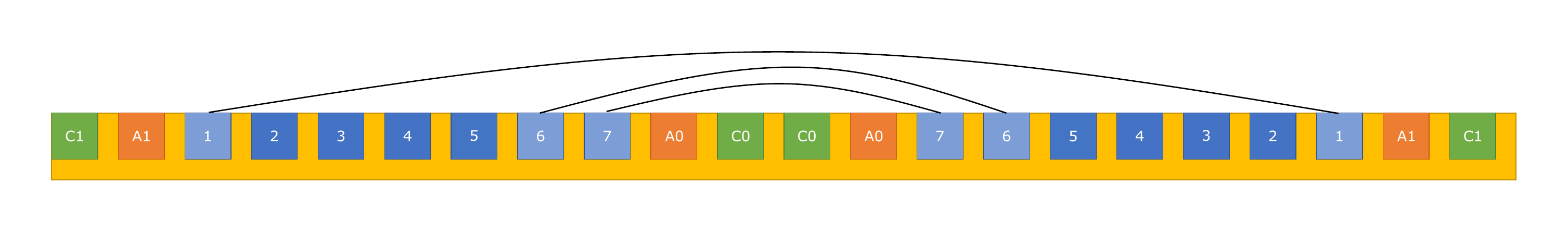}
         \caption{Patch interactions required to perform logical Steane operations using the $XIIIIXX$ or $ZIIIIZZ$ logical operator, between adjacent Steane arrays. }
         \label{fig:concat_patch_2}
     \end{subfigure}
        \caption{Logical operators between order reversed Steane qubits on the bus.}
        \label{fig:concat_sq_patch}
\end{figure*}

A scheme for concatenation is required when concatenating multiple layers of the Steane $\dbracket{7,1,3}$ code with the surface code bus. Logical CNOTs are performed transversally between the two logical blocks at each level of concatenation below the surface code. This requires at least three CNOT gates between qubits in the two logical blocks. There are several ways of doing this including those shown in Figure~\ref{fig:concat_sq_patch}.

\begin{enumerate}
\item For one level (1L) of concatenation of the Steane code above the surface code, we use the bus to perform logical CNOTs between surface code patches. We arrange the qubits as in Figure~\ref{fig:sq_sc_bus}, with CNOT ancillae adjacent to each parity ancilla. This layout enables some minor parallel bus operations. We allocate additional time to enable CNOTS between patches in different logical blocks and perform transversal CNOTs gates between logical patches as in Figure~\ref{fig:concat_sq_patch}. The existing ancillae, both CNOT ancillae and code ancillae, are reused for this purpose.
\item For two levels (2L) of concatenation, logical operations at both levels are performed using the bus. We allocate adequate time in each L1 syndrome extraction to perform a long-distance bus operation between any two of the L1 qubits within the block, as described above. CNOTs between two adjacent L2 CNOT qubits are performed using long bus operations, as are CNOT gates between adjacent 2L qubits. Swaps between 2L qubits are performed by using transversal quantum teleportation of 1L qubits. This ensures that no single bus error can create more than a weight-1 logical error at layer 3 of concatenation.
\item For three levels (3L) of concatenation or higher, a swap network of sufficient length is used to make interacting lower-level qubits adjacent to each other in each concatenated layer. This swap procedure is then applied recursively.
\end{enumerate}

In our analysis, we increase the error rate of the code at each layer of concatenation in proportion to the additional time required to perform operations between blocks. This provides an estimate of the error rate with these interactions. We then apply the equation of fit for the performance of the $\dbracket{7,1,3}$ code. This procedure is performed recursively. To determine the minimum possible combined lattice width that meets the $10^{-15}$ performance target, we iteratively tried each possible bus width and code distance. For each of these, we then determined if it was possible to reach the target and, if so, how many layers of concatenation were required. The narrowest lattices for each number of layers of concatenation that reach our target were then recorded.

\subsection{Compilation of the $\dbracket{15,7,3}$ CSS code}
\label{perf:fifteen_comp}
A concatenation scheme for multiply concatenating the $\dbracket{15,7,3}$ code must be determined in order to concatenate the $\dbracket{15,7,3}$ code above the surface code with bus multiple times. We have chosen the simplest scheme in this work, where we concatenate fifteen (15) level one (L1) code blocks together into a new L2 code block of $7\times 7 = 49$ Level zero L0 logical surface code qubits. This is possible because a transversal CNOT operation of fifteen (15) pairs of qubits between two blocks is equivalent to performing a CNOT between all corresponding logical qubits in each block~\cite{ChaoReighardt2018b}. This work does consider CNOT gates between differing forty-nine (49) L0 logical qubit code blocks on the same qubit. However, we have not considered how to fault-tolerantly implement the internal Clifford group (that is the Clifford group on the seven qubits within a single $\dbracket{15,7,3}$ code block) at any level of concatenation, and the implementation of this may reduce the threshold for this code.

The concatenation scheme is then as follows.
\begin{enumerate}
\item For the first level of concatenation, we perform CNOT operations between code blocks by performing CNOTs in parallel, reusing the ancilla qubits for the syndrome extraction in order to increase bus utilization and perform a CNOT between two L1 code blocks. This can be done in eighteen (18) bus operations. When concatenated again for a higher level code, we perform swaps using the bus to teleport into the ancilla qubits between the blocks, this can be done in thirty-six (36) bus operations. This is done so that no operation can cause a weight 2 error.
\item For the second and higher levels of concatenation, a swap network of sufficient length is then used to make interacting lower-level qubits adjacent in each concatenated layer.  After swapping, the CNOT is applied transversally, and then the qubits are swapped back. This swap procedure is applied recursively at each successive layer.
\end{enumerate}

As with the Steane code, to determine the performance of the concatenated code, the error rate of each layer of concatenation is increased proportionately with the additional time required to perform two qubit interactions before applying the equation of fit. After which, a search was performed to determine the performance of the code.

\subsection{Evaluation of results}
\label{perf:eval}
It can be seen from the results in Table~\ref{tab:table1} that, in most cases, the overall width $w$ can be reduced by about half that required for the surface code with bus, or about a third of the surface code alone. This required the number of qubits per block to increase by about three orders of magnitude when the physical error rate is $10^{-3}$. By increasing the width slightly, to about $70\%$ of that required for the surface code with bus, this penalty is reduced substantially, to only a factor of four over the standard approach. It is possible to have a reduction whilst retaining more density through the use of the $\dbracket{15,7,3}$, at the cost of more complex compilation and a larger smallest unit size. It may be possible to increase density or width further with different choices of block codes.


\section{Repetition codes above biassed surface codes}
\label{sec:bias}
The approach of applying a quantum block code directly above the surface code appears to be insufficient for further reduction of the array width $w$, and so a new approach is required for further improvement. As you increase the length of a surface code patch, one error type is suppressed exponentially, while the other only grows linearly. It is known~\cite{BonillaAtaides2021, Tuckett2019, Lee2021,TuckettBartlettFlamia2018} that highly biased error models can greatly increase the performance of error correction. A further reduction in width may be possible by engineering a highly biased logical error model using a highly asymmetric surface code patch and then concatenating it with a repetition code. These rectangular surface code patches can be much narrower, as the threshold of the repetition code is much higher than that of a code that corrects for all possible quantum errors.

Whilst we might think of looking at more complicated codes than the quantum repetition code because we wish to minimise array widths, we are limited by the complexity of more advanced codes. The most space-efficient method to mediate long-distance interactions between these patches is a single surface code bus. Such a bus must grow in width as interaction distance increases, and each bus can only perform one stabiliser measurement at a time. By measuring the parity of neighbouring qubits, the quantum repetition code can be implemented in constant time for increasing code distance. This reduces the requirements for the performance of the biassed surface code patches in our proposed architecture, allowing the patches to be narrower.

\subsection{Rectangular surface codes and creating biased logical errors}
The repetition code is a classical code, and so only corrects for one type of quantum error---either of type $X$, or of type $Z$. An error of the other type in any classical error-correcting code is never corrected. To be able to use the repetition code as a layer two code, one type of error must be suppressed well below the target rate of the layer three code. This is precisely what we engineer using a highly asymmetric surface code layer, forcing one of the error rates to be many orders of magnitude lower than the other.

This may be achieved using a highly rectangular surface code. The probability of a logical error in a surface code decreases exponentially with the length of the shortest possible uncorrectable error chain. As this length scales proportionally to the length of the shortest logical operator, we can engineer an exponential bias in logical error rates by using rectangular surface patches. Assume, without loss of generality, that the $Z$ logical operator is along the longer edge of a rectangular surface code patch. The effect of increasing this longer patch is to reduce the possibility of a $Z$ logical error exponentially due to the increased number of errors required to form an error chain. The probability of an $X$ logical error only increases linearly though, as the length of the $X$ error chains required to form a logical error does not change, only the number of possible chains does. This means that the probability of an $X$ error is approximately $\frac{d_Z}{d_X} \cdot p_{l_X}(d_X, d_X)$~\cite{Litinski2019b} for a rectangular surface code patch. This does, however, ignore some edge effects, so the simulations of \ref{sec:rectsc} are required for more accurate analysis.

The rectangular surface codes may then be concatenated with the repetition code using lattice surgery. This is a delicate task, however, as it is essential that you never reduce the minimum distance between two $Z$ boundaries when working with rectangular surface code patches, which increases the probability of a logical $Z$ error exponentially. Nonetheless, it is possible to measure the logical $Z$ parity of two adjacent surface code patches within a repetition code. In Appendix~\ref{sec:non_bus_ls} we present a scheme for performing such a parity measurement, and we also offer an analysis of the time needed to perform a parity measurement between two adjacent rectangular surface code qubits in approximately $9d_Z$ time steps. Using the surface code bus, each syndrome measurement takes at least $d_Z(d_Z + 1)  + 1 $ time steps. This results in a total code time of $2(d_Z^2 + d_Z + 1)$, as we require two sets of measurements to implement the repetition code.


\begin{figure*}
     \begin{subfigure}[b]{\textwidth}
         \includegraphics[width=0.5\textwidth]{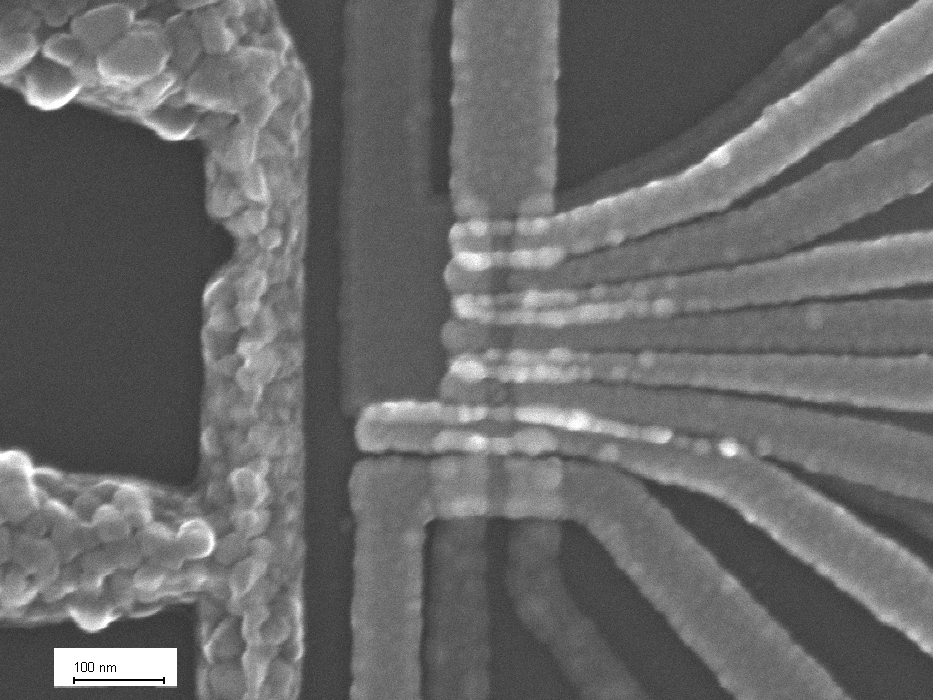}
         \caption{Physical qubits, $p_X = 10^{-4}$, $p_Z = 10^{-4}$, Perhaps silicon quantum dot qubits similar to these~\cite{2201.06679} (Image used with the permission of the authors) }
         \label{fig:concat_qubit}
     \end{subfigure}
     
     \begin{subfigure}[b]{\textwidth}
         \includegraphics[width=\textwidth]{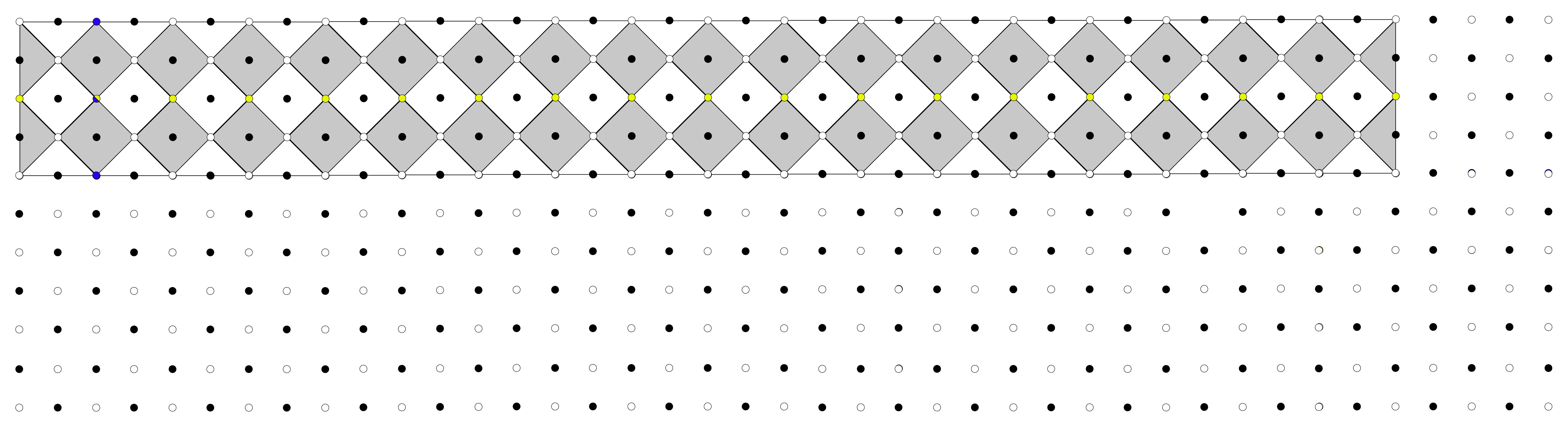}
         \caption{A rectangular surface code patch with $d=2$ bus, $d_X = 3$, $d_Z = 19$, 369 qubits total. With the qubits in Fig \ref{fig:concat_qubit} $p_X < 0.158$, $p_Z < 6.91 \times 10^{-21}$.  }
         \label{fig:concat_patch}
     \end{subfigure}

     \begin{subfigure}[b]{\textwidth}
         \includegraphics[width=\textwidth]{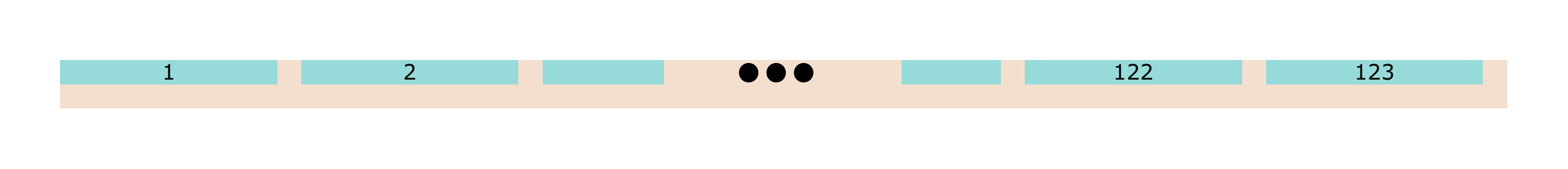}
         \caption{A repetition code made up of surface code patches, with 123 surface patches as in Fig \ref{fig:concat_patch}, $p_x < 1.39\times 10^{-16}$, $p_Z < 2.10\times 10^{-16}$. }
         \label{fig:rc_patch}
     \end{subfigure}
        \caption{An example of concatenation levels for biased repetition code, with one possible configuration of height and physical error rate.}
        \label{fig:biased_diags}
\end{figure*}

As we can engineer rectangular surface code patches with biased logical errors and can perform logical parity measurements between them, we are able to implement a repetition code on top of these biased surface code patches. By then concatenating these logical repetition code qubits with other codes, such as the Steane $\dbracket{7,1,3}$ code, we are able to reach an arbitrary logical error rate. As Hadamard and phase gates are difficult to perform on these rectangular qubits in a fault-tolerant way, compilation above the error-correcting code level is expected to be performed using either gate teleportation or measurement-based techniques~\cite{Litinski2019}. The repetition code level would result in a scheme depicted in Figure~\ref{fig:biased_diags}.

\subsection{Evaluation of results}
To evaluate the performance of this scheme, we must understand the effect of different values of $d_Z$ and $d_X$ on the time required to extract one round of parity measurements in the repetition code. 

We assume that a lattice surgery is used when performing an operation between two repetition codes. To measure the $ZZ$ parity between two repetition codes, we must move the patches so that the qubits of one repetition code are interdigitated with those of the other whilst simultaneously extracting the repetition code stabilizers. The two overlapping syndrome extractions are required when performing a logical CNOT gate or swapping two repetition code qubits.

For bussed extraction, this takes at least $5(d_Z^2 + d_Z + 1)$ cycles in the worst case, where two logical qubits must be moved past each other during a computation. (A computation consists of two parity extractions plus one quantum teleportation~\cite{Erhard2021}). For the non-bussed method, it takes $18 d_Z$ surface code cycles to perform a round of syndrome extraction in the same situation. This is due to the complicated series of moves that must be performed to enable qubits to move past each other while simultaneously performing each of the two repetition codes. A description of the lattice surgery operations can be found in Appendix \ref{sec:non_bus_ls}.

We then multiply the per-time-step rates for each surface code patch during the lattice surgery procedures to determine the expected logical rates for one cycle of syndrome extraction of the repetition code. We call this the effective $p_X$ and $p_Z$ for a given configuration of bussed/non-bussed operation, and code distances $d_X$ and $d_Z$.

We then use these error rates to calculate the probability of an error in the repetition code of distance $d_{rc}$, which is made up of $d_{rc}$ patches. To estimate the logical fidelity of the $X$ errors after the repetition code, we use the CDF of the binomial probability for having more than $\ceil{n/2} - 1$, or more errors within the repetition code block. We do this for each block size until the repetition code has a lower logical $X$ error probability than the $Z$ errors. To estimate the rate of $Z$ errors, we multiply the effective $p_Z$ by $d_{rc}$, as the repetition code is only correcting for $X$ errors. Then, to get an effective error rate per CNOT gate, we have to multiply this by $2d_{rc} + 1$, as it takes that many repetition code cycles to perform a logical CNOT operation. 



\begin{figure*}
     \begin{subfigure}[b]{0.48\textwidth}
         \includegraphics[width=\textwidth]{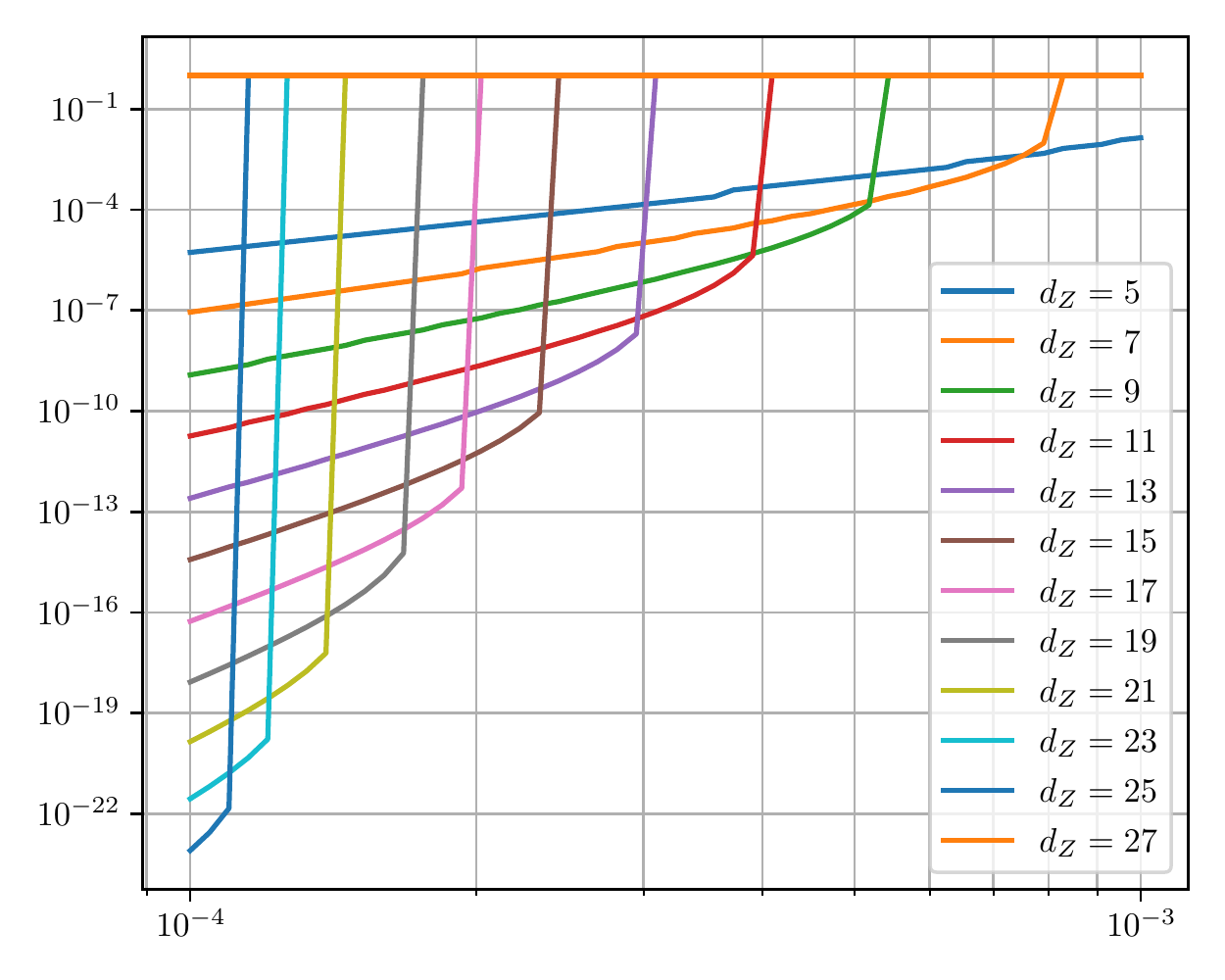}
         \caption{Regularized $p_l$ for $d_X = 3$}
         \label{fig:reg_3_p}
     \end{subfigure}
     \hfill
     \begin{subfigure}[b]{0.48\textwidth}
         \includegraphics[width=\textwidth]{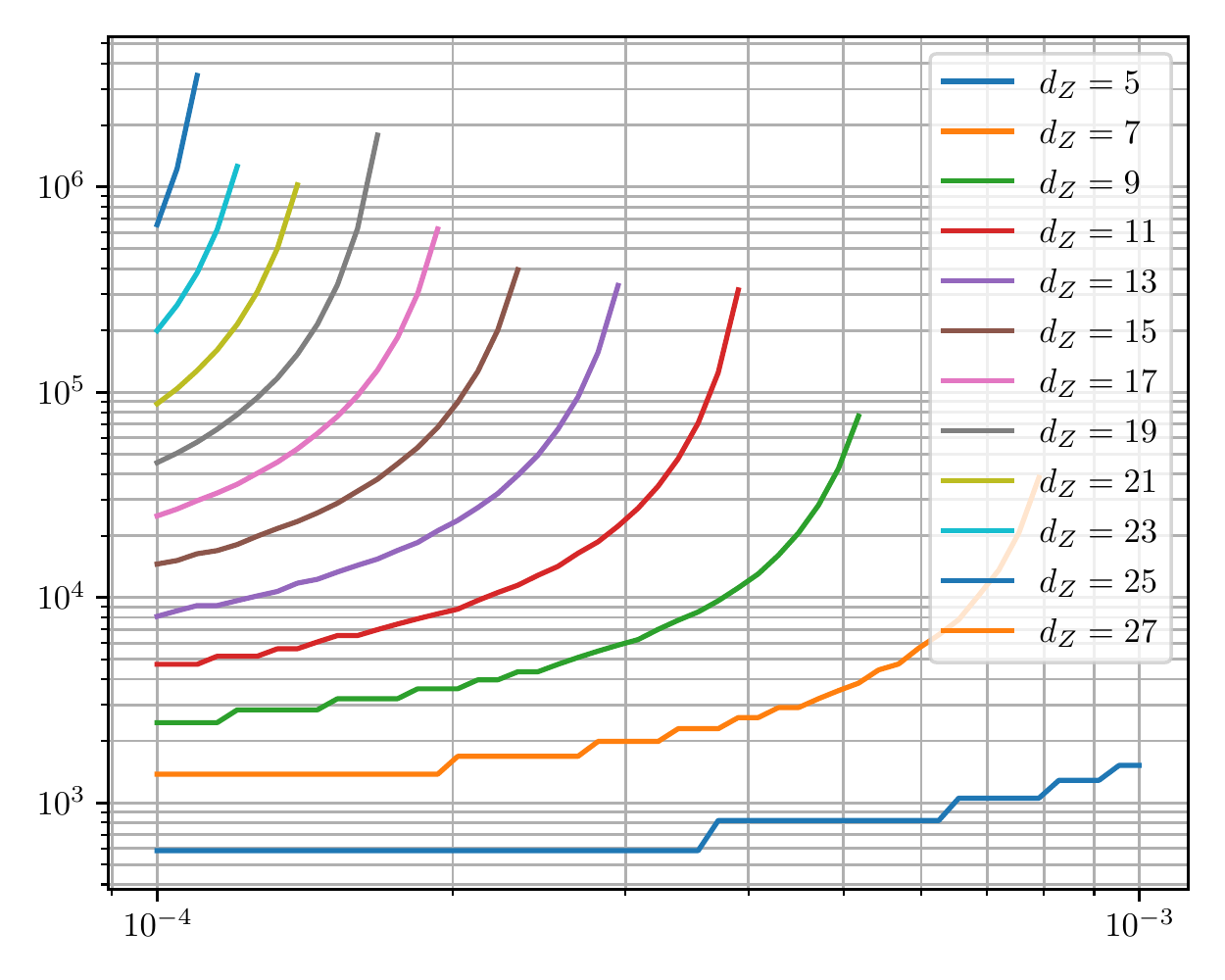}
         \caption{Qubit counts for $d_X = 3$}
         \label{fig:reg_3_q}
     \end{subfigure}
     
     \begin{subfigure}[b]{0.48\textwidth}
         \includegraphics[width=\textwidth]{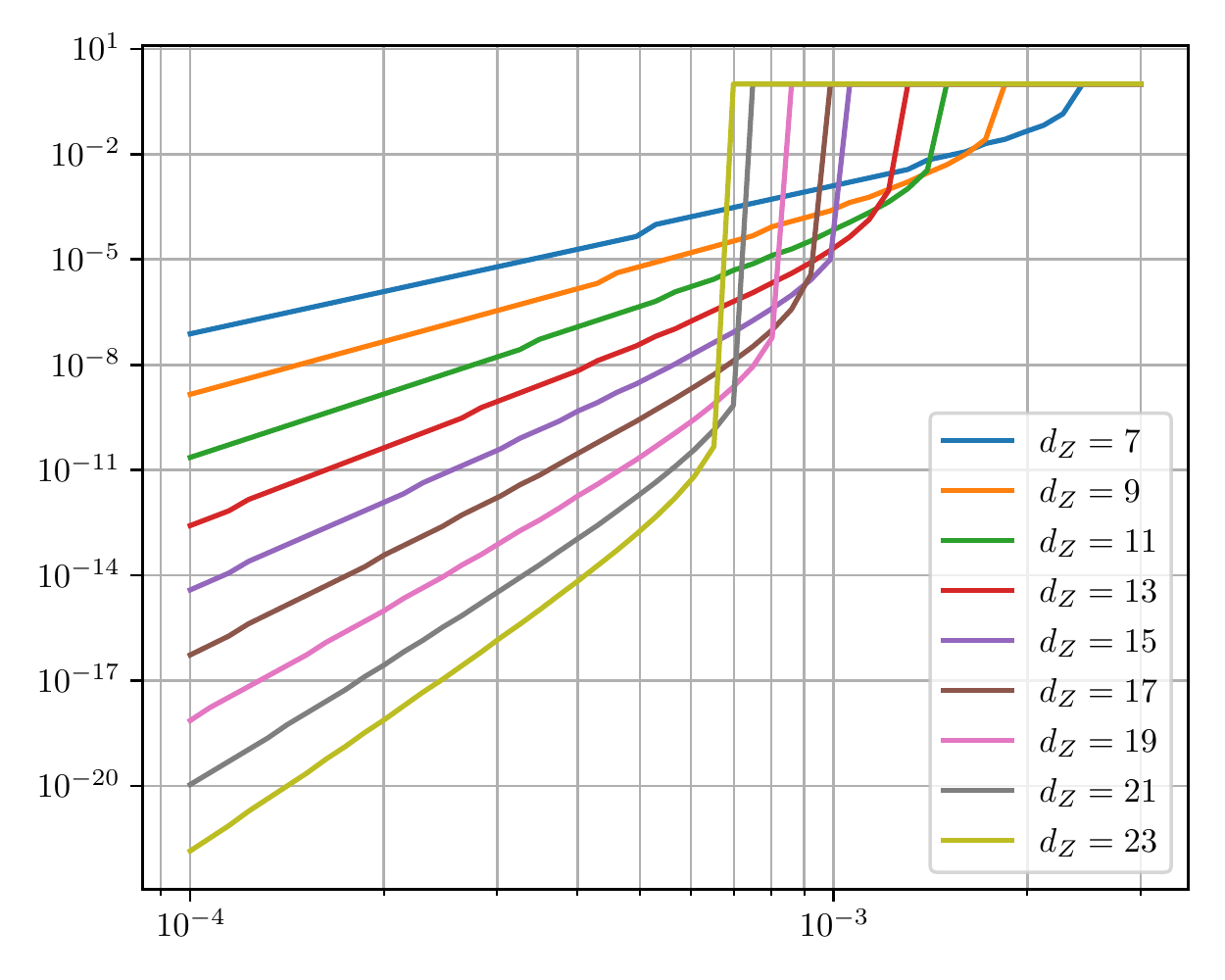}
         \caption{Regularized $p_l$ for $d_X = 5$.}
         \label{fig:reg_5_p}
     \end{subfigure}
     \hfill
     \begin{subfigure}[b]{0.48\textwidth}
         \includegraphics[width=\textwidth]{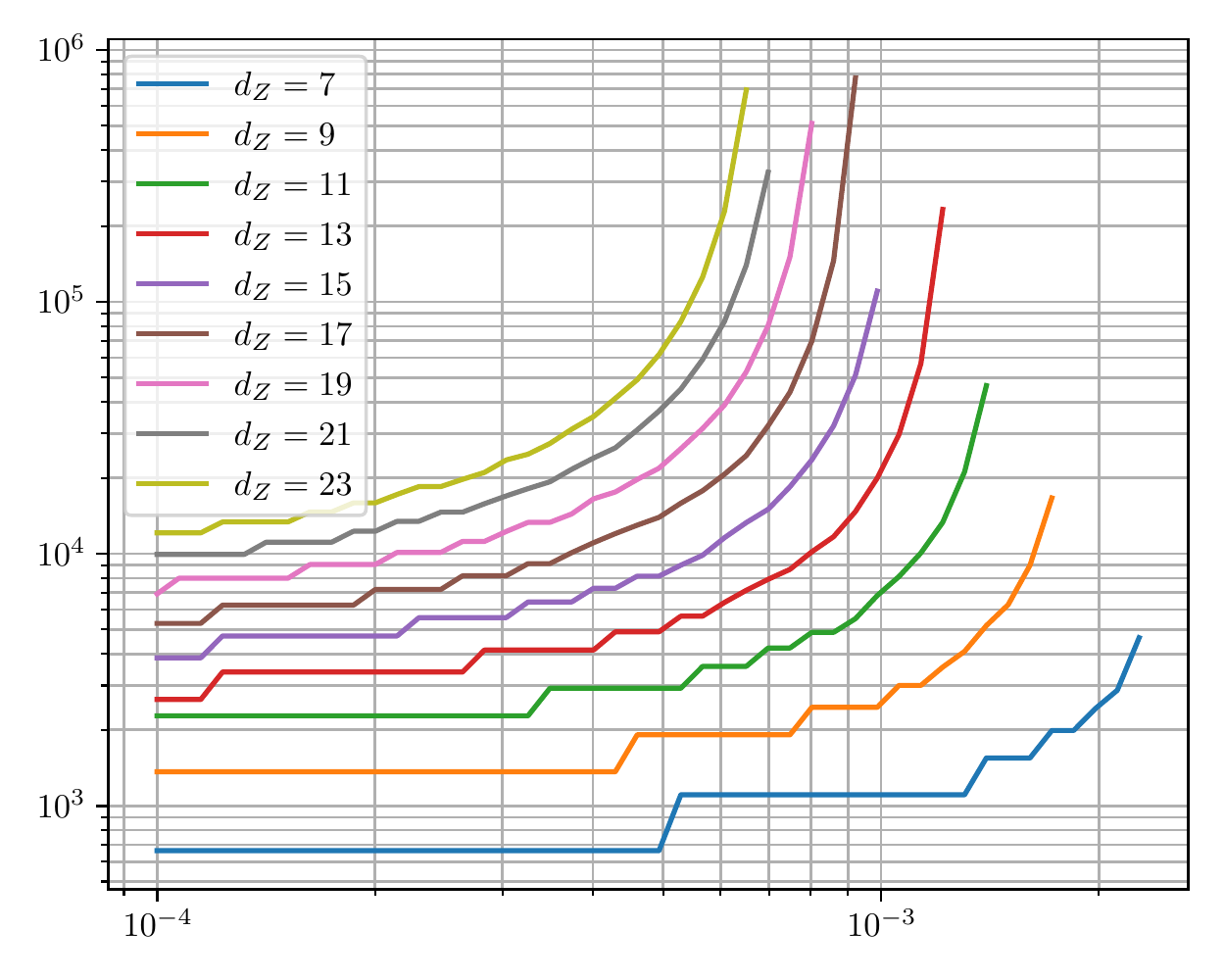}
         \caption{Qubit counts for $d_X = 5$.}
         \label{fig:reg_5_q}
     \end{subfigure}

     \begin{subfigure}[b]{0.48\textwidth}
         \includegraphics[width=\textwidth]{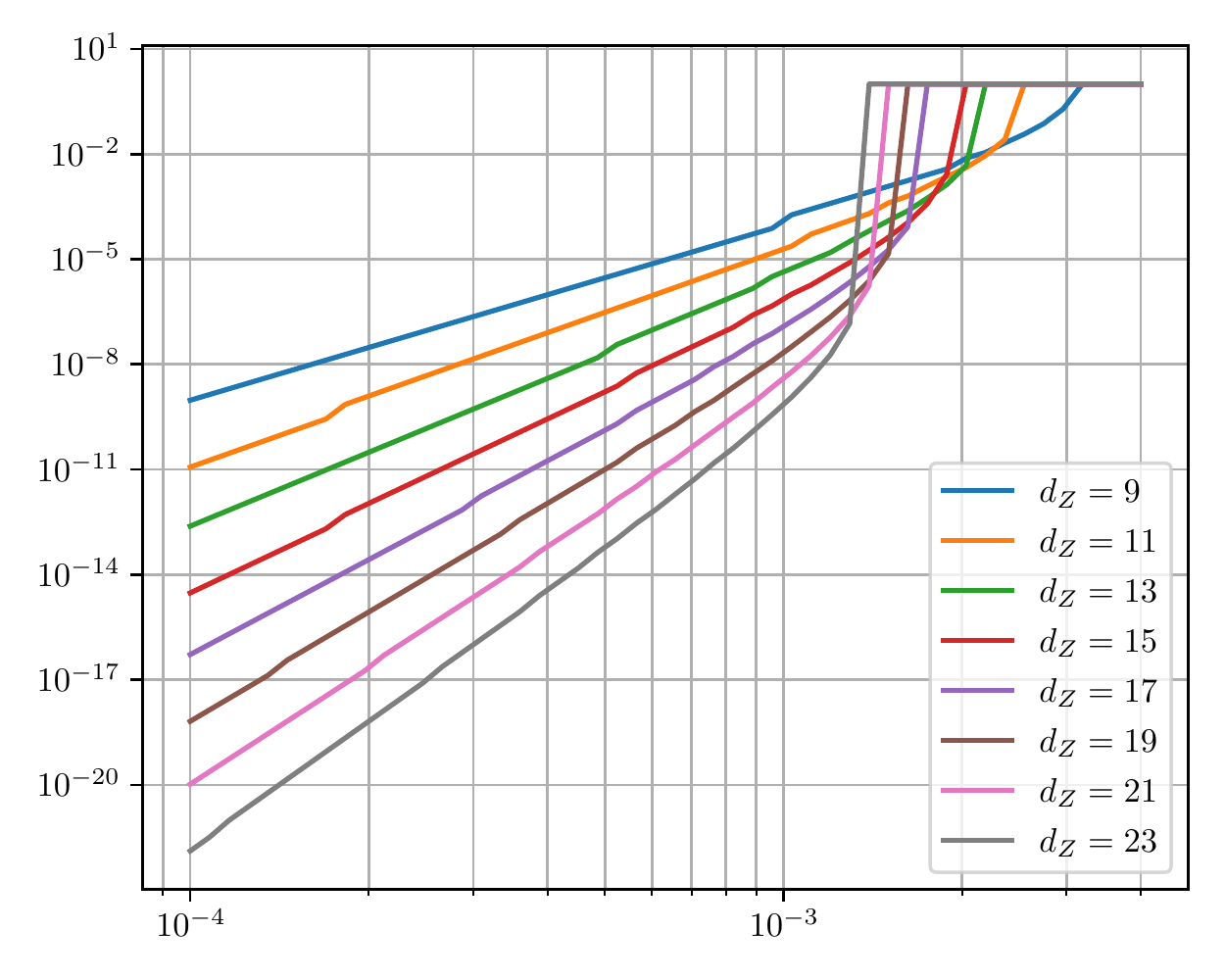}
         \caption{Regularized $p_l$ for $d_X = 7$}
         \label{fig:reg_7_p}
     \end{subfigure}
     \hfill
     \begin{subfigure}[b]{0.48\textwidth}
         \includegraphics[width=\textwidth]{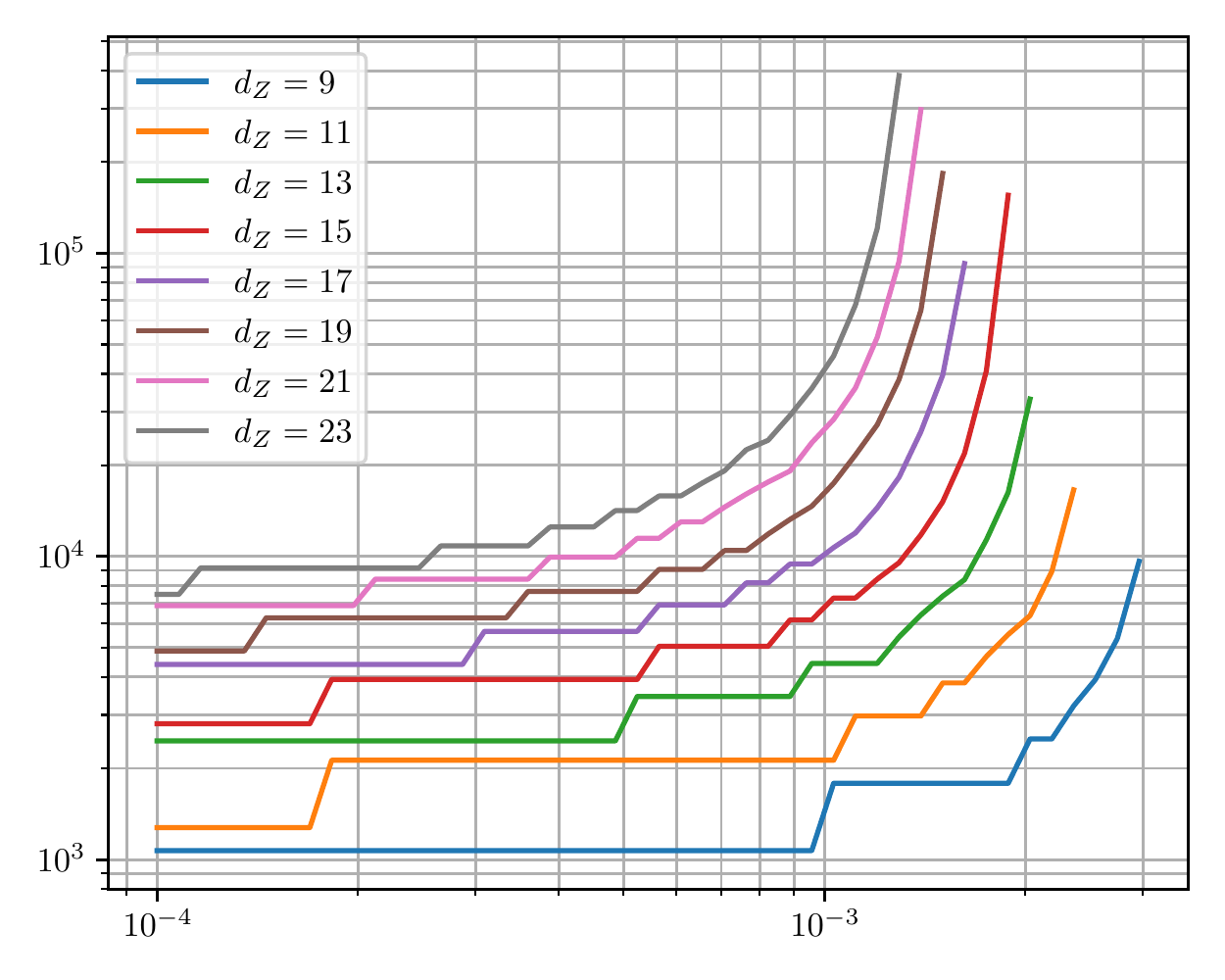}
         \caption{Qubit counts for $d_X = 7$}
         \label{fig:reg_7_q}
     \end{subfigure}
        \caption{Plots of the regularized CNOT fidelity and qubit counts after the repetition code for differing biassed patch distances.}
        \label{fig:reg_figs}
\end{figure*}

If we evaluate the results for each value of $d_X$, we can determine a physical error rate that enables both $X$ and $Z$ errors to be below the threshold for the Steane $\dbracket{7,1,3}$ code presented earlier. You can see the extrapolated logical CNOT error rate for the non-bussed implementation after the repetition code in Figure \ref{fig:reg_figs}.
If we take $p=10^{-6}$ as the threshold on this architecture, then the threshold for $d_X = 3$ is approximately $2\times 10 ^{-4}$; for $d_X = 5$, it is approximately $8\times 10 ^{-4}$, and for $d_X = 7$, it is approximately $1.5 \times 10 ^{-3}$. Architectures for these can be implemented in arrays with widths of 11, 19, and 27 respectively. Whilst the number of qubits required is extremely high at the threshold, it comes
down quickly as you move away from that point. If you compare these values to those estimated for standard CSS codes on square surface code patches presented in Section \ref{sec:concat_perf}, you can see that narrower widths are possible at a given physical error rate and, at certain widths, fewer qubits are required for the same logical error rate. This is traded against the unknown complexity of compiling for this architecture.

\section{Conclusion}
\label{sec:conclusion}
In this work, we have investigated the possibility of reducing interconnect density by restricting the maximum width $w$ of the required qubit array that is needed to accommodate the surface code lattice. Previous attempts to deal with the problem of interconnect density either required more complicated long-distance interactions~\cite{Mohiyaddin2021} or had a much lower code threshold~\cite{JonesFogartyMorelloGyureDzurakLadd2018,Ashley2009} than found in this work. Here we chose to investigate using the surface code as a base layer for other codes. 

In order to evaluate this, we first simulated the performance of rectangular surface code patches, giving a collection of polynomial fits. We then, in section~\ref{sec:bus}, presented a fully fault-tolerant version of a folded surface code bus that enables long-distance qubit parity measurements with much lower overhead than other methods. We used the rectangular simulations of Section~\ref{sec:rectsc} to evaluate the performance of this in differing contexts.

We then determined the performance of both the $\dbracket{15,7,3}$ and Steane $\dbracket{7,1,3}$ codes with the recently described flag-qubit fault-tolerance technique. We simulated these codes with randomised depolarised noise. We then evaluated the results to determine the expected performance of architectures using these codes directly concatenated with the surface code, where long-distance CNOTs between surface code patches are mediated by the folded surface code bus. This gave an array width of 35 when the error rate $p = 10^{-3}$, and 19 when $p=10^{-4}$, compared to widths of 71 and 35 respectively without concatenation. 

Finally, we explored the performance when deliberately biassing the logical error rate and then regularising the logical errors, estimating the width of the minimum qubit performance required to meet the threshold of higher level codes. This showed that at $p < 1.5 \times 10^{-3}$ and $p < 0.8 \times 10^{-3}$ array widths of 27 and 19 respectively should suffice.

We also developed several tools for the evaluation of quantum error correcting codes. While better tools have become available during the progress of this work, especially the excellent stim tool and the increasingly good py-matching decoder, the tools used along with all data we have generated are available from the authors upon request\cite{ToolsData}.

The authors would have liked to evaluate other codes, especially the $\dbracket{23,1,7}$ Golay code, LDPC codes, and concatenated $\dbracket{4,1,2}$ subsystem codes. However, the difficulty of defining efficient flagged fault-tolerant extractions meant that this would have to be done in future work. Similarly, it would be interesting to investigate the compilation and evaluation of the performance of the logical codes above the repetition code above deliberately biased surface codes, as well as the evaluation of the performance of other topological codes in place of the surface code, such as the XZZX code~\cite{BonillaAtaides2021} and the honeycomb code~\cite{Gidney2021faulttolerant}, and their impact on minimum bus width and interconnect density.





\section*{Acknowledgements}
We thank Craig Gidney, Daniel Litinski for their assistance in developing the surface code quantum bus system, specifically the folded design. Alexis Shaw was supported by an Australian Government Research Training Program Scholarship, the ARC Centre of Excellence for Quantum Computation and Communication Technology (CQC2T), project number CE170100012, and a scholarship top-up and extension from the Sydney Quantum Academy. Michael Bremner was supported by the Australian Research Council (ARC) Centre of Excellence for Quantum Computation and Communication Technology (CQC2T), project number CE170100012, as well as by Google Inc. Simon Devitt and Michael Bremner supported were with funding from the Defense Advanced Research Projects Agency [under the Quantum Benchmarking (QB) program under award no. HR00112230007 and HR001121S0026 contracts]. The views, opinions and/or findings expressed are those of the authors and should not be interpreted as representing the official views or policies of the Department of Defense or the U.S. Government. 

\bibliographystyle{aipauth4-2}
\bibliography{cite}

\begin{thebibliography}{63}%
\makeatletter
\providecommand \@ifxundefined [1]{%
 \@ifx{#1\undefined}
}%
\providecommand \@ifnum [1]{%
 \ifnum #1\expandafter \@firstoftwo
 \else \expandafter \@secondoftwo
 \fi
}%
\providecommand \@ifx [1]{%
 \ifx #1\expandafter \@firstoftwo
 \else \expandafter \@secondoftwo
 \fi
}%
\providecommand \natexlab [1]{#1}%
\providecommand \enquote  [1]{``#1''}%
\providecommand \bibnamefont  [1]{#1}%
\providecommand \bibfnamefont [1]{#1}%
\providecommand \citenamefont [1]{#1}%
\providecommand \href@noop [0]{\@secondoftwo}%
\providecommand \href [0]{\begingroup \@sanitize@url \@href}%
\providecommand \@href[1]{\@@startlink{#1}\@@href}%
\providecommand \@@href[1]{\endgroup#1\@@endlink}%
\providecommand \@sanitize@url [0]{\catcode `\\12\catcode `\$12\catcode
  `\&12\catcode `\#12\catcode `\^12\catcode `\_12\catcode `\%12\relax}%
\providecommand \@@startlink[1]{}%
\providecommand \@@endlink[0]{}%
\providecommand \url  [0]{\begingroup\@sanitize@url \@url }%
\providecommand \@url [1]{\endgroup\@href {#1}{\urlprefix }}%
\providecommand \urlprefix  [0]{URL }%
\providecommand \Eprint [0]{\href }%
\providecommand \doibase [0]{https://doi.org/}%
\providecommand \selectlanguage [0]{\@gobble}%
\providecommand \bibinfo  [0]{\@secondoftwo}%
\providecommand \bibfield  [0]{\@secondoftwo}%
\providecommand \translation [1]{[#1]}%
\providecommand \BibitemOpen [0]{}%
\providecommand \bibitemStop [0]{}%
\providecommand \bibitemNoStop [0]{.\EOS\space}%
\providecommand \EOS [0]{\spacefactor3000\relax}%
\providecommand \BibitemShut  [1]{\csname bibitem#1\endcsname}%
\let\auto@bib@innerbib\@empty
\bibitem [{\citenamefont {Aaronson}\ and\ \citenamefont
  {Gottesman}(2004)}]{AaronsonGottesman2004}%
  \BibitemOpen
  \bibfield  {author} {\bibinfo {author} {\bibnamefont {Aaronson},
  \bibfnamefont {S.}}and\ \bibinfo {author} {\bibnamefont {Gottesman},
  \bibfnamefont {D.}},\ }\href {https://doi.org/10.1103/PhysRevA.70.052328}
  {\bibfield  {journal} {\bibinfo  {journal} {Phys. Rev. A}\ }\textbf {\bibinfo
  {volume} {70}},\ \bibinfo {pages} {052328} (\bibinfo {year}
  {2004})}\BibitemShut {NoStop}%
\bibitem [{\citenamefont {Aharonov}\ and\ \citenamefont
  {Ben-Or}(1997)}]{BenOr1997}%
  \BibitemOpen
  \bibfield  {author} {\bibinfo {author} {\bibnamefont {Aharonov},
  \bibfnamefont {D.}}and\ \bibinfo {author} {\bibnamefont {Ben-Or},
  \bibfnamefont {M.}},\ }in\ \href {https://doi.org/10.1145/258533.258579}
  {\emph {\bibinfo {booktitle} {Proceedings of the Twenty-Ninth Annual ACM
  Symposium on Theory of Computing}}},\ \bibinfo {series and number} {STOC
  '97}\ (\bibinfo  {publisher} {Association for Computing Machinery},\ \bibinfo
  {address} {New York, NY, USA},\ \bibinfo {year} {1997})\ p.\ \bibinfo {pages}
  {176–188}\BibitemShut {NoStop}%
\bibitem [{\citenamefont {Ansaloni}\ \emph {et~al.}(2020)\citenamefont
  {Ansaloni}, \citenamefont {Chatterjee}, \citenamefont {Bohuslavskyi},
  \citenamefont {Bertrand}, \citenamefont {Hutin}, \citenamefont {Vinet},\ and\
  \citenamefont {Kuemmeth}}]{Ansaloni2020}%
  \BibitemOpen
  \bibfield  {author} {\bibinfo {author} {\bibnamefont {Ansaloni},
  \bibfnamefont {F.}}, \bibinfo {author} {\bibnamefont {Chatterjee},
  \bibfnamefont {A.}}, \bibinfo {author} {\bibnamefont {Bohuslavskyi},
  \bibfnamefont {H.}}, \bibinfo {author} {\bibnamefont {Bertrand},
  \bibfnamefont {B.}}, \bibinfo {author} {\bibnamefont {Hutin}, \bibfnamefont
  {L.}}, \bibinfo {author} {\bibnamefont {Vinet}, \bibfnamefont {M.}}, and\
  \bibinfo {author} {\bibnamefont {Kuemmeth}, \bibfnamefont {F.}},\ }\href
  {https://doi.org/10.1038/s41467-020-20280-3} {\bibfield  {journal} {\bibinfo
  {journal} {Nature Communications}\ }\textbf {\bibinfo {volume} {11}},\
  \bibinfo {pages} {6399} (\bibinfo {year} {2020})}\BibitemShut {NoStop}%
\bibitem [{\citenamefont {Azad}\ \emph {et~al.}(2022)\citenamefont {Azad},
  \citenamefont {Lipińska}, \citenamefont {Mahato}, \citenamefont {Sachdeva},
  \citenamefont {Bhoumik},\ and\ \citenamefont {Majumdar}}]{Utkarsh}%
  \BibitemOpen
  \bibfield  {author} {\bibinfo {author} {\bibnamefont {Azad}, \bibfnamefont
  {U.}}, \bibinfo {author} {\bibnamefont {Lipińska}, \bibfnamefont {A.}},
  \bibinfo {author} {\bibnamefont {Mahato}, \bibfnamefont {S.}}, \bibinfo
  {author} {\bibnamefont {Sachdeva}, \bibfnamefont {R.}}, \bibinfo {author}
  {\bibnamefont {Bhoumik}, \bibfnamefont {D.}}, and\ \bibinfo {author}
  {\bibnamefont {Majumdar}, \bibfnamefont {R.}},\ }\href
  {https://doi.org/https://doi.org/10.1049/qtc2.12042} {\bibfield  {journal}
  {\bibinfo  {journal} {IET Quantum Communication}\ }\textbf {\bibinfo {volume}
  {3}},\ \bibinfo {pages} {174} (\bibinfo {year} {2022})}\BibitemShut {NoStop}%
\bibitem [{\citenamefont {Backens}(2014)}]{Backens2014}%
  \BibitemOpen
  \bibfield  {author} {\bibinfo {author} {\bibnamefont {Backens}, \bibfnamefont
  {M.}},\ }\href {https://doi.org/10.1088/1367-2630/16/9/093021} {\bibfield
  {journal} {\bibinfo  {journal} {New Journal of Physics}\ }\textbf {\bibinfo
  {volume} {16}},\ \bibinfo {pages} {093021} (\bibinfo {year}
  {2014})}\BibitemShut {NoStop}%
\bibitem [{\citenamefont {de~Beaudrap}\ and\ \citenamefont
  {Horsman}(2020)}]{deBeaudrap2020}%
  \BibitemOpen
  \bibfield  {author} {\bibinfo {author} {\bibnamefont {de~Beaudrap},
  \bibfnamefont {N.}}and\ \bibinfo {author} {\bibnamefont {Horsman},
  \bibfnamefont {D.}},\ }\href {https://doi.org/10.22331/q-2020-01-09-218}
  {\bibfield  {journal} {\bibinfo  {journal} {{Quantum}}\ }\textbf {\bibinfo
  {volume} {4}},\ \bibinfo {pages} {218} (\bibinfo {year} {2020})}\BibitemShut
  {NoStop}%
\bibitem [{\citenamefont {Bonilla~Ataides}\ \emph {et~al.}(2021)\citenamefont
  {Bonilla~Ataides}, \citenamefont {Tuckett}, \citenamefont {Bartlett},
  \citenamefont {Flammia},\ and\ \citenamefont {Brown}}]{BonillaAtaides2021}%
  \BibitemOpen
  \bibfield  {author} {\bibinfo {author} {\bibnamefont {Bonilla~Ataides},
  \bibfnamefont {J.~P.}}, \bibinfo {author} {\bibnamefont {Tuckett},
  \bibfnamefont {D.~K.}}, \bibinfo {author} {\bibnamefont {Bartlett},
  \bibfnamefont {S.~D.}}, \bibinfo {author} {\bibnamefont {Flammia},
  \bibfnamefont {S.~T.}}, and\ \bibinfo {author} {\bibnamefont {Brown},
  \bibfnamefont {B.~J.}},\ }\href {https://doi.org/10.1038/s41467-021-22274-1}
  {\bibfield  {journal} {\bibinfo  {journal} {Nature Communications}\ }\textbf
  {\bibinfo {volume} {12}},\ \bibinfo {pages} {2172} (\bibinfo {year}
  {2021})}\BibitemShut {NoStop}%
\bibitem [{\citenamefont {Bourassa}\ \emph {et~al.}(2021)\citenamefont
  {Bourassa}, \citenamefont {Alexander}, \citenamefont {Vasmer}, \citenamefont
  {Patil}, \citenamefont {Tzitrin}, \citenamefont {Matsuura}, \citenamefont
  {Su}, \citenamefont {Baragiola}, \citenamefont {Guha}, \citenamefont
  {Dauphinais}, \citenamefont {Sabapathy}, \citenamefont {Menicucci},\ and\
  \citenamefont {Dhand}}]{Bourassa2021}%
  \BibitemOpen
  \bibfield  {author} {\bibinfo {author} {\bibnamefont {Bourassa},
  \bibfnamefont {J.~E.}}, \bibinfo {author} {\bibnamefont {Alexander},
  \bibfnamefont {R.~N.}}, \bibinfo {author} {\bibnamefont {Vasmer},
  \bibfnamefont {M.}}, \bibinfo {author} {\bibnamefont {Patil}, \bibfnamefont
  {A.}}, \bibinfo {author} {\bibnamefont {Tzitrin}, \bibfnamefont {I.}},
  \bibinfo {author} {\bibnamefont {Matsuura}, \bibfnamefont {T.}}, \bibinfo
  {author} {\bibnamefont {Su}, \bibfnamefont {D.}}, \bibinfo {author}
  {\bibnamefont {Baragiola}, \bibfnamefont {B.~Q.}}, \bibinfo {author}
  {\bibnamefont {Guha}, \bibfnamefont {S.}}, \bibinfo {author} {\bibnamefont
  {Dauphinais}, \bibfnamefont {G.}}, \bibinfo {author} {\bibnamefont
  {Sabapathy}, \bibfnamefont {K.~K.}}, \bibinfo {author} {\bibnamefont
  {Menicucci}, \bibfnamefont {N.~C.}}, and\ \bibinfo {author} {\bibnamefont
  {Dhand}, \bibfnamefont {I.}},\ }\href
  {https://doi.org/10.22331/q-2021-02-04-392} {\bibfield  {journal} {\bibinfo
  {journal} {{Quantum}}\ }\textbf {\bibinfo {volume} {5}},\ \bibinfo {pages}
  {392} (\bibinfo {year} {2021})}\BibitemShut {NoStop}%
\bibitem [{\citenamefont {Bravyi}\ \emph {et~al.}(2018)\citenamefont {Bravyi},
  \citenamefont {Englbrecht}, \citenamefont {K{\"o}nig},\ and\ \citenamefont
  {Peard}}]{Bravyi2018}%
  \BibitemOpen
  \bibfield  {author} {\bibinfo {author} {\bibnamefont {Bravyi}, \bibfnamefont
  {S.}}, \bibinfo {author} {\bibnamefont {Englbrecht}, \bibfnamefont {M.}},
  \bibinfo {author} {\bibnamefont {K{\"o}nig}, \bibfnamefont {R.}}, and\
  \bibinfo {author} {\bibnamefont {Peard}, \bibfnamefont {N.}},\ }\href
  {https://doi.org/10.1038/s41534-018-0106-y} {\bibfield  {journal} {\bibinfo
  {journal} {npj Quantum Information}\ }\textbf {\bibinfo {volume} {4}},\
  \bibinfo {pages} {55} (\bibinfo {year} {2018})}\BibitemShut {NoStop}%
\bibitem [{\citenamefont {Breuckmann}\ and\ \citenamefont
  {Eberhardt}(2021)}]{Breuckmann2021}%
  \BibitemOpen
  \bibfield  {author} {\bibinfo {author} {\bibnamefont {Breuckmann},
  \bibfnamefont {N.~P.}}and\ \bibinfo {author} {\bibnamefont {Eberhardt},
  \bibfnamefont {J.~N.}},\ }\href {https://doi.org/10.1103/PRXQuantum.2.040101}
  {\bibfield  {journal} {\bibinfo  {journal} {PRX Quantum}\ }\textbf {\bibinfo
  {volume} {2}},\ \bibinfo {pages} {040101} (\bibinfo {year}
  {2021})}\BibitemShut {NoStop}%
\bibitem [{\citenamefont {Cai}\ \emph {et~al.}(2019)\citenamefont {Cai},
  \citenamefont {Fogarty}, \citenamefont {Schaal}, \citenamefont
  {Patom{\"{a}}ki}, \citenamefont {Benjamin},\ and\ \citenamefont
  {Morton}}]{Cai2019}%
  \BibitemOpen
  \bibfield  {author} {\bibinfo {author} {\bibnamefont {Cai}, \bibfnamefont
  {Z.}}, \bibinfo {author} {\bibnamefont {Fogarty}, \bibfnamefont {M.~A.}},
  \bibinfo {author} {\bibnamefont {Schaal}, \bibfnamefont {S.}}, \bibinfo
  {author} {\bibnamefont {Patom{\"{a}}ki}, \bibfnamefont {S.}}, \bibinfo
  {author} {\bibnamefont {Benjamin}, \bibfnamefont {S.~C.}}, and\ \bibinfo
  {author} {\bibnamefont {Morton}, \bibfnamefont {J.~J.~L.}},\ }\href
  {https://doi.org/10.22331/q-2019-12-09-212} {\bibfield  {journal} {\bibinfo
  {journal} {{Quantum}}\ }\textbf {\bibinfo {volume} {3}},\ \bibinfo {pages}
  {212} (\bibinfo {year} {2019})}\BibitemShut {NoStop}%
\bibitem [{\citenamefont {Calderbank}\ and\ \citenamefont
  {Shor}(1996)}]{CalderbankShor1996}%
  \BibitemOpen
  \bibfield  {author} {\bibinfo {author} {\bibnamefont {Calderbank},
  \bibfnamefont {A.~R.}}and\ \bibinfo {author} {\bibnamefont {Shor},
  \bibfnamefont {P.~W.}},\ }\href {https://doi.org/10.1103/PhysRevA.54.1098}
  {\bibfield  {journal} {\bibinfo  {journal} {Phys. Rev. A}\ }\textbf {\bibinfo
  {volume} {54}},\ \bibinfo {pages} {1098} (\bibinfo {year}
  {1996})}\BibitemShut {NoStop}%
\bibitem [{\citenamefont {Chao}\ and\ \citenamefont
  {Reichardt}(2018{\natexlab{a}})}]{ChaoReighardt2018b}%
  \BibitemOpen
  \bibfield  {author} {\bibinfo {author} {\bibnamefont {Chao}, \bibfnamefont
  {R.}}and\ \bibinfo {author} {\bibnamefont {Reichardt}, \bibfnamefont
  {B.~W.}},\ }\href {https://doi.org/10.1038/s41534-018-0085-z} {\bibfield
  {journal} {\bibinfo  {journal} {npj Quantum Information}\ }\textbf {\bibinfo
  {volume} {4}},\ \bibinfo {pages} {42} (\bibinfo {year}
  {2018}{\natexlab{a}})}\BibitemShut {NoStop}%
\bibitem [{\citenamefont {Chao}\ and\ \citenamefont
  {Reichardt}(2018{\natexlab{b}})}]{ChaoReighardt2018}%
  \BibitemOpen
  \bibfield  {author} {\bibinfo {author} {\bibnamefont {Chao}, \bibfnamefont
  {R.}}and\ \bibinfo {author} {\bibnamefont {Reichardt}, \bibfnamefont
  {B.~W.}},\ }\href {https://doi.org/10.1103/PhysRevLett.121.050502} {\bibfield
   {journal} {\bibinfo  {journal} {Phys. Rev. Lett.}\ }\textbf {\bibinfo
  {volume} {121}},\ \bibinfo {pages} {050502} (\bibinfo {year}
  {2018}{\natexlab{b}})}\BibitemShut {NoStop}%
\bibitem [{\citenamefont {Chao}\ and\ \citenamefont
  {Reichardt}(2020)}]{ChaoReighardt2020}%
  \BibitemOpen
  \bibfield  {author} {\bibinfo {author} {\bibnamefont {Chao}, \bibfnamefont
  {R.}}and\ \bibinfo {author} {\bibnamefont {Reichardt}, \bibfnamefont
  {B.~W.}},\ }\href {https://doi.org/10.1103/PRXQuantum.1.010302} {\bibfield
  {journal} {\bibinfo  {journal} {PRX Quantum}\ }\textbf {\bibinfo {volume}
  {1}},\ \bibinfo {pages} {010302} (\bibinfo {year} {2020})}\BibitemShut
  {NoStop}%
\bibitem [{\citenamefont {Coecke}\ and\ \citenamefont
  {Duncan}(2008)}]{CoeckeDuncan2008}%
  \BibitemOpen
  \bibfield  {author} {\bibinfo {author} {\bibnamefont {Coecke}, \bibfnamefont
  {B.}}and\ \bibinfo {author} {\bibnamefont {Duncan}, \bibfnamefont {R.}},\
  }in\ \href@noop {} {\emph {\bibinfo {booktitle} {Automata, Languages and
  Programming}}},\ \bibinfo {editor} {edited by\ \bibinfo {editor}
  {\bibfnamefont {L.}~\bibnamefont {Aceto}}, \bibinfo {editor} {\bibfnamefont
  {I.}~\bibnamefont {Damg{\aa}rd}}, \bibinfo {editor} {\bibfnamefont {L.~A.}\
  \bibnamefont {Goldberg}}, \bibinfo {editor} {\bibfnamefont {M.~M.}\
  \bibnamefont {Halld{\'o}rsson}}, \bibinfo {editor} {\bibfnamefont
  {A.}~\bibnamefont {Ing{\'o}lfsd{\'o}ttir}}, \ and\ \bibinfo {editor}
  {\bibfnamefont {I.}~\bibnamefont {Walukiewicz}}}\ (\bibinfo  {publisher}
  {Springer Berlin Heidelberg},\ \bibinfo {address} {Berlin, Heidelberg},\
  \bibinfo {year} {2008})\ pp.\ \bibinfo {pages} {298--310}\BibitemShut
  {NoStop}%
\bibitem [{\citenamefont {Cross}, \citenamefont {Divincenzo},\ and\
  \citenamefont {Terhal}(2009)}]{CrossDivincenzoTerhal2009}%
  \BibitemOpen
  \bibfield  {author} {\bibinfo {author} {\bibnamefont {Cross}, \bibfnamefont
  {A.~W.}}, \bibinfo {author} {\bibnamefont {Divincenzo}, \bibfnamefont
  {D.~P.}}, and\ \bibinfo {author} {\bibnamefont {Terhal}, \bibfnamefont
  {B.~M.}},\ }\href@noop {} {\bibfield  {journal} {\bibinfo  {journal} {Quantum
  Info. Comput.}\ }\textbf {\bibinfo {volume} {9}},\ \bibinfo {pages}
  {541–572} (\bibinfo {year} {2009})}\BibitemShut {NoStop}%
\bibitem [{\citenamefont {Devitt}\ \emph {et~al.}(2016)\citenamefont {Devitt},
  \citenamefont {Greentree}, \citenamefont {Stephens},\ and\ \citenamefont
  {Van~Meter}}]{Devitt2016}%
  \BibitemOpen
  \bibfield  {author} {\bibinfo {author} {\bibnamefont {Devitt}, \bibfnamefont
  {S.~J.}}, \bibinfo {author} {\bibnamefont {Greentree}, \bibfnamefont
  {A.~D.}}, \bibinfo {author} {\bibnamefont {Stephens}, \bibfnamefont {A.~M.}},
  and\ \bibinfo {author} {\bibnamefont {Van~Meter}, \bibfnamefont {R.}},\
  }\href {https://doi.org/10.1038/srep36163} {\bibfield  {journal} {\bibinfo
  {journal} {Scientific Reports}\ }\textbf {\bibinfo {volume} {6}},\ \bibinfo
  {pages} {36163} (\bibinfo {year} {2016})}\BibitemShut {NoStop}%
\bibitem [{\citenamefont {Erhard}\ \emph {et~al.}(2021)\citenamefont {Erhard},
  \citenamefont {Poulsen~Nautrup}, \citenamefont {Meth}, \citenamefont
  {Postler}, \citenamefont {Stricker}, \citenamefont {Stadler}, \citenamefont
  {Negnevitsky}, \citenamefont {Ringbauer}, \citenamefont {Schindler},
  \citenamefont {Briegel}, \citenamefont {Blatt}, \citenamefont {Friis},\ and\
  \citenamefont {Monz}}]{Erhard2021}%
  \BibitemOpen
  \bibfield  {author} {\bibinfo {author} {\bibnamefont {Erhard}, \bibfnamefont
  {A.}}, \bibinfo {author} {\bibnamefont {Poulsen~Nautrup}, \bibfnamefont
  {H.}}, \bibinfo {author} {\bibnamefont {Meth}, \bibfnamefont {M.}}, \bibinfo
  {author} {\bibnamefont {Postler}, \bibfnamefont {L.}}, \bibinfo {author}
  {\bibnamefont {Stricker}, \bibfnamefont {R.}}, \bibinfo {author}
  {\bibnamefont {Stadler}, \bibfnamefont {M.}}, \bibinfo {author} {\bibnamefont
  {Negnevitsky}, \bibfnamefont {V.}}, \bibinfo {author} {\bibnamefont
  {Ringbauer}, \bibfnamefont {M.}}, \bibinfo {author} {\bibnamefont
  {Schindler}, \bibfnamefont {P.}}, \bibinfo {author} {\bibnamefont {Briegel},
  \bibfnamefont {H.~J.}}, \bibinfo {author} {\bibnamefont {Blatt},
  \bibfnamefont {R.}}, \bibinfo {author} {\bibnamefont {Friis}, \bibfnamefont
  {N.}}, and\ \bibinfo {author} {\bibnamefont {Monz}, \bibfnamefont {T.}},\
  }\href {https://doi.org/10.1038/s41586-020-03079-6} {\bibfield  {journal}
  {\bibinfo  {journal} {Nature}\ }\textbf {\bibinfo {volume} {589}},\ \bibinfo
  {pages} {220} (\bibinfo {year} {2021})}\BibitemShut {NoStop}%
\bibitem [{\citenamefont {Fowler}\ and\ \citenamefont
  {Gidney}(2018)}]{FowlerGidney2018U}%
  \BibitemOpen
  \bibfield  {author} {\bibinfo {author} {\bibnamefont {Fowler}, \bibfnamefont
  {A.~G.}}and\ \bibinfo {author} {\bibnamefont {Gidney}, \bibfnamefont {C.}},\
  }\href {https://doi.org/10.48550/ARXIV.1808.06709} {\enquote {\bibinfo
  {title} {Low overhead quantum computation using lattice surgery},}\ }
  (\bibinfo {year} {2018})\BibitemShut {NoStop}%
\bibitem [{\citenamefont {Fowler}\ \emph
  {et~al.}(2012{\natexlab{a}})\citenamefont {Fowler}, \citenamefont
  {Mariantoni}, \citenamefont {Martinis},\ and\ \citenamefont
  {Cleland}}]{FowlerMariantoniMartinisCleland2012}%
  \BibitemOpen
  \bibfield  {author} {\bibinfo {author} {\bibnamefont {Fowler}, \bibfnamefont
  {A.~G.}}, \bibinfo {author} {\bibnamefont {Mariantoni}, \bibfnamefont {M.}},
  \bibinfo {author} {\bibnamefont {Martinis}, \bibfnamefont {J.~M.}}, and\
  \bibinfo {author} {\bibnamefont {Cleland}, \bibfnamefont {A.~N.}},\ }\href
  {https://doi.org/10.1103/PhysRevA.86.032324} {\bibfield  {journal} {\bibinfo
  {journal} {Phys. Rev. A}\ }\textbf {\bibinfo {volume} {86}},\ \bibinfo
  {pages} {032324} (\bibinfo {year} {2012}{\natexlab{a}})}\BibitemShut
  {NoStop}%
\bibitem [{\citenamefont {Fowler}\ \emph
  {et~al.}(2012{\natexlab{b}})\citenamefont {Fowler}, \citenamefont
  {Whiteside}, \citenamefont {McInnes},\ and\ \citenamefont
  {Rabbani}}]{Fowler2012}%
  \BibitemOpen
  \bibfield  {author} {\bibinfo {author} {\bibnamefont {Fowler}, \bibfnamefont
  {A.~G.}}, \bibinfo {author} {\bibnamefont {Whiteside}, \bibfnamefont
  {A.~C.}}, \bibinfo {author} {\bibnamefont {McInnes}, \bibfnamefont {A.~L.}},
  and\ \bibinfo {author} {\bibnamefont {Rabbani}, \bibfnamefont {A.}},\ }\href
  {https://doi.org/10.1103/PhysRevX.2.041003} {\bibfield  {journal} {\bibinfo
  {journal} {Phys. Rev. X}\ }\textbf {\bibinfo {volume} {2}},\ \bibinfo {pages}
  {041003} (\bibinfo {year} {2012}{\natexlab{b}})}\BibitemShut {NoStop}%
\bibitem [{\citenamefont {Gidney}(2021)}]{Gidney2021}%
  \BibitemOpen
  \bibfield  {author} {\bibinfo {author} {\bibnamefont {Gidney}, \bibfnamefont
  {C.}},\ }\href {https://doi.org/10.22331/q-2021-07-06-497} {\bibfield
  {journal} {\bibinfo  {journal} {{Quantum}}\ }\textbf {\bibinfo {volume}
  {5}},\ \bibinfo {pages} {497} (\bibinfo {year} {2021})}\BibitemShut {NoStop}%
\bibitem [{\citenamefont {Gidney}\ \emph {et~al.}(2021)\citenamefont {Gidney},
  \citenamefont {Newman}, \citenamefont {Fowler},\ and\ \citenamefont
  {Broughton}}]{Gidney2021faulttolerant}%
  \BibitemOpen
  \bibfield  {author} {\bibinfo {author} {\bibnamefont {Gidney}, \bibfnamefont
  {C.}}, \bibinfo {author} {\bibnamefont {Newman}, \bibfnamefont {M.}},
  \bibinfo {author} {\bibnamefont {Fowler}, \bibfnamefont {A.}}, and\ \bibinfo
  {author} {\bibnamefont {Broughton}, \bibfnamefont {M.}},\ }\href
  {https://doi.org/10.22331/q-2021-12-20-605} {\bibfield  {journal} {\bibinfo
  {journal} {{Quantum}}\ }\textbf {\bibinfo {volume} {5}},\ \bibinfo {pages}
  {605} (\bibinfo {year} {2021})}\BibitemShut {NoStop}%
\bibitem [{\citenamefont {Gilbert}\ \emph {et~al.}(2022)\citenamefont
  {Gilbert}, \citenamefont {Tanttu}, \citenamefont {Lim}, \citenamefont {Feng},
  \citenamefont {Huang}, \citenamefont {Cifuentes}, \citenamefont {Serrano},
  \citenamefont {Mai}, \citenamefont {Leon}, \citenamefont {Escott},
  \citenamefont {Itoh}, \citenamefont {Abrosimov}, \citenamefont {Pohl},
  \citenamefont {Thewalt}, \citenamefont {Hudson}, \citenamefont {Morello},
  \citenamefont {Laucht}, \citenamefont {Yang}, \citenamefont {Saraiva},\ and\
  \citenamefont {Dzurak}}]{2201.06679}%
  \BibitemOpen
  \bibfield  {author} {\bibinfo {author} {\bibnamefont {Gilbert}, \bibfnamefont
  {W.}}, \bibinfo {author} {\bibnamefont {Tanttu}, \bibfnamefont {T.}},
  \bibinfo {author} {\bibnamefont {Lim}, \bibfnamefont {W.~H.}}, \bibinfo
  {author} {\bibnamefont {Feng}, \bibfnamefont {M.}}, \bibinfo {author}
  {\bibnamefont {Huang}, \bibfnamefont {J.~Y.}}, \bibinfo {author}
  {\bibnamefont {Cifuentes}, \bibfnamefont {J.~D.}}, \bibinfo {author}
  {\bibnamefont {Serrano}, \bibfnamefont {S.}}, \bibinfo {author} {\bibnamefont
  {Mai}, \bibfnamefont {P.~Y.}}, \bibinfo {author} {\bibnamefont {Leon},
  \bibfnamefont {R.~C.~C.}}, \bibinfo {author} {\bibnamefont {Escott},
  \bibfnamefont {C.~C.}}, \bibinfo {author} {\bibnamefont {Itoh}, \bibfnamefont
  {K.~M.}}, \bibinfo {author} {\bibnamefont {Abrosimov}, \bibfnamefont
  {N.~V.}}, \bibinfo {author} {\bibnamefont {Pohl}, \bibfnamefont {H.-J.}},
  \bibinfo {author} {\bibnamefont {Thewalt}, \bibfnamefont {M.~L.~W.}},
  \bibinfo {author} {\bibnamefont {Hudson}, \bibfnamefont {F.~E.}}, \bibinfo
  {author} {\bibnamefont {Morello}, \bibfnamefont {A.}}, \bibinfo {author}
  {\bibnamefont {Laucht}, \bibfnamefont {A.}}, \bibinfo {author} {\bibnamefont
  {Yang}, \bibfnamefont {C.~H.}}, \bibinfo {author} {\bibnamefont {Saraiva},
  \bibfnamefont {A.}}, and\ \bibinfo {author} {\bibnamefont {Dzurak},
  \bibfnamefont {A.~S.}},\ }\href {https://doi.org/10.48550/ARXIV.2201.06679}
  {\enquote {\bibinfo {title} {On-demand electrical control of spin qubits},}\
  } (\bibinfo {year} {2022})\BibitemShut {NoStop}%
\bibitem [{\citenamefont {Hakkaku}, \citenamefont {Mitarai},\ and\
  \citenamefont {Fujii}(2021)}]{Hakkaku2021}%
  \BibitemOpen
  \bibfield  {author} {\bibinfo {author} {\bibnamefont {Hakkaku}, \bibfnamefont
  {S.}}, \bibinfo {author} {\bibnamefont {Mitarai}, \bibfnamefont {K.}}, and\
  \bibinfo {author} {\bibnamefont {Fujii}, \bibfnamefont {K.}},\ }\href
  {https://doi.org/10.1103/PhysRevResearch.3.043130} {\bibfield  {journal}
  {\bibinfo  {journal} {Phys. Rev. Research}\ }\textbf {\bibinfo {volume}
  {3}},\ \bibinfo {pages} {043130} (\bibinfo {year} {2021})}\BibitemShut
  {NoStop}%
\bibitem [{\citenamefont {He}\ \emph {et~al.}(2019)\citenamefont {He},
  \citenamefont {Gorman}, \citenamefont {Keith}, \citenamefont {Kranz},
  \citenamefont {Keizer},\ and\ \citenamefont {Simmons}}]{He2019}%
  \BibitemOpen
  \bibfield  {author} {\bibinfo {author} {\bibnamefont {He}, \bibfnamefont
  {Y.}}, \bibinfo {author} {\bibnamefont {Gorman}, \bibfnamefont {S.~K.}},
  \bibinfo {author} {\bibnamefont {Keith}, \bibfnamefont {D.}}, \bibinfo
  {author} {\bibnamefont {Kranz}, \bibfnamefont {L.}}, \bibinfo {author}
  {\bibnamefont {Keizer}, \bibfnamefont {J.~G.}}, and\ \bibinfo {author}
  {\bibnamefont {Simmons}, \bibfnamefont {M.~Y.}},\ }\href
  {https://doi.org/10.1038/s41586-019-1381-2} {\bibfield  {journal} {\bibinfo
  {journal} {Nature}\ }\textbf {\bibinfo {volume} {571}},\ \bibinfo {pages}
  {371} (\bibinfo {year} {2019})}\BibitemShut {NoStop}%
\bibitem [{\citenamefont {Herr}\ \emph {et~al.}(2019)\citenamefont {Herr},
  \citenamefont {Paler}, \citenamefont {Devitt},\ and\ \citenamefont
  {Nori}}]{1902.08117}%
  \BibitemOpen
  \bibfield  {author} {\bibinfo {author} {\bibnamefont {Herr}, \bibfnamefont
  {D.}}, \bibinfo {author} {\bibnamefont {Paler}, \bibfnamefont {A.}}, \bibinfo
  {author} {\bibnamefont {Devitt}, \bibfnamefont {S.~J.}}, and\ \bibinfo
  {author} {\bibnamefont {Nori}, \bibfnamefont {F.}},\ }\href
  {https://doi.org/10.48550/ARXIV.1902.08117} {\enquote {\bibinfo {title} {Time
  versus hardware: Reducing qubit counts with a (surface code) data bus},}\ }
  (\bibinfo {year} {2019}),\ \bibinfo {note} {arxiv:1902.08117},\ \Eprint
  {https://arxiv.org/abs/1902.08117} {arXiv:1902.08117 [quant-ph]} \BibitemShut
  {NoStop}%
\bibitem [{\citenamefont {Horsman}\ \emph {et~al.}(2012)\citenamefont
  {Horsman}, \citenamefont {Fowler}, \citenamefont {Devitt},\ and\
  \citenamefont {Meter}}]{Horsman_2012}%
  \BibitemOpen
  \bibfield  {author} {\bibinfo {author} {\bibnamefont {Horsman}, \bibfnamefont
  {C.}}, \bibinfo {author} {\bibnamefont {Fowler}, \bibfnamefont {A.~G.}},
  \bibinfo {author} {\bibnamefont {Devitt}, \bibfnamefont {S.}}, and\ \bibinfo
  {author} {\bibnamefont {Meter}, \bibfnamefont {R.~V.}},\ }\href
  {https://doi.org/10.1088/1367-2630/14/12/123011} {\bibfield  {journal}
  {\bibinfo  {journal} {New Journal of Physics}\ }\textbf {\bibinfo {volume}
  {14}},\ \bibinfo {pages} {123011} (\bibinfo {year} {2012})}\BibitemShut
  {NoStop}%
\bibitem [{\citenamefont {Jeandel}, \citenamefont {Perdrix},\ and\
  \citenamefont {Vilmart}(2018)}]{JeandelPendrixVilmart2018}%
  \BibitemOpen
  \bibfield  {author} {\bibinfo {author} {\bibnamefont {Jeandel}, \bibfnamefont
  {E.}}, \bibinfo {author} {\bibnamefont {Perdrix}, \bibfnamefont {S.}}, and\
  \bibinfo {author} {\bibnamefont {Vilmart}, \bibfnamefont {R.}},\ }in\ \href
  {https://doi.org/10.1145/3209108.3209131} {\emph {\bibinfo {booktitle}
  {Proceedings of the 33rd Annual ACM/IEEE Symposium on Logic in Computer
  Science}}},\ \bibinfo {series and number} {LICS '18}\ (\bibinfo  {publisher}
  {Association for Computing Machinery},\ \bibinfo {address} {New York, NY,
  USA},\ \bibinfo {year} {2018})\ p.\ \bibinfo {pages} {559–568}\BibitemShut
  {NoStop}%
\bibitem [{\citenamefont {Jones}\ \emph {et~al.}(2018)\citenamefont {Jones},
  \citenamefont {Fogarty}, \citenamefont {Morello}, \citenamefont {Gyure},
  \citenamefont {Dzurak},\ and\ \citenamefont
  {Ladd}}]{JonesFogartyMorelloGyureDzurakLadd2018}%
  \BibitemOpen
  \bibfield  {author} {\bibinfo {author} {\bibnamefont {Jones}, \bibfnamefont
  {C.}}, \bibinfo {author} {\bibnamefont {Fogarty}, \bibfnamefont {M.~A.}},
  \bibinfo {author} {\bibnamefont {Morello}, \bibfnamefont {A.}}, \bibinfo
  {author} {\bibnamefont {Gyure}, \bibfnamefont {M.~F.}}, \bibinfo {author}
  {\bibnamefont {Dzurak}, \bibfnamefont {A.~S.}}, and\ \bibinfo {author}
  {\bibnamefont {Ladd}, \bibfnamefont {T.~D.}},\ }\href
  {https://doi.org/10.1103/PhysRevX.8.021058} {\bibfield  {journal} {\bibinfo
  {journal} {Phys. Rev. X}\ }\textbf {\bibinfo {volume} {8}},\ \bibinfo {pages}
  {021058} (\bibinfo {year} {2018})}\BibitemShut {NoStop}%
\bibitem [{\citenamefont {Kempe}\ \emph {et~al.}(2001)\citenamefont {Kempe},
  \citenamefont {Bacon}, \citenamefont {Lidar},\ and\ \citenamefont
  {Whaley}}]{KBLW2001}%
  \BibitemOpen
  \bibfield  {author} {\bibinfo {author} {\bibnamefont {Kempe}, \bibfnamefont
  {J.}}, \bibinfo {author} {\bibnamefont {Bacon}, \bibfnamefont {D.}}, \bibinfo
  {author} {\bibnamefont {Lidar}, \bibfnamefont {D.~A.}}, and\ \bibinfo
  {author} {\bibnamefont {Whaley}, \bibfnamefont {K.~B.}},\ }\href
  {https://doi.org/10.1103/PhysRevA.63.042307} {\bibfield  {journal} {\bibinfo
  {journal} {Phys. Rev. A}\ }\textbf {\bibinfo {volume} {63}},\ \bibinfo
  {pages} {042307} (\bibinfo {year} {2001})}\BibitemShut {NoStop}%
\bibitem [{\citenamefont {Kitaev}(2003)}]{kitaev2003}%
  \BibitemOpen
  \bibfield  {author} {\bibinfo {author} {\bibnamefont {Kitaev}, \bibfnamefont
  {A.}},\ }\href
  {https://doi.org/https://doi.org/10.1016/S0003-4916(02)00018-0} {\bibfield
  {journal} {\bibinfo  {journal} {Annals of Physics}\ }\textbf {\bibinfo
  {volume} {303}},\ \bibinfo {pages} {2} (\bibinfo {year} {2003})}\BibitemShut
  {NoStop}%
\bibitem [{\citenamefont {Kolmogorov}(2009)}]{Kolmogorov2009}%
  \BibitemOpen
  \bibfield  {author} {\bibinfo {author} {\bibnamefont {Kolmogorov},
  \bibfnamefont {V.}},\ }\href {https://doi.org/10.1007/s12532-009-0002-8}
  {\bibfield  {journal} {\bibinfo  {journal} {Mathematical Programming
  Computation}\ }\textbf {\bibinfo {volume} {1}},\ \bibinfo {pages} {43}
  (\bibinfo {year} {2009})}\BibitemShut {NoStop}%
\bibitem [{\citenamefont {Lee}, \citenamefont {Park},\ and\ \citenamefont
  {Heo}(2021)}]{Lee2021}%
  \BibitemOpen
  \bibfield  {author} {\bibinfo {author} {\bibnamefont {Lee}, \bibfnamefont
  {J.}}, \bibinfo {author} {\bibnamefont {Park}, \bibfnamefont {J.}}, and\
  \bibinfo {author} {\bibnamefont {Heo}, \bibfnamefont {J.}},\ }\href
  {https://doi.org/10.1007/s11128-021-03130-z} {\bibfield  {journal} {\bibinfo
  {journal} {Quantum Information Processing}\ }\textbf {\bibinfo {volume}
  {20}},\ \bibinfo {pages} {231} (\bibinfo {year} {2021})}\BibitemShut
  {NoStop}%
\bibitem [{\citenamefont {Lekitsch}\ \emph {et~al.}(2017)\citenamefont
  {Lekitsch}, \citenamefont {Weidt}, \citenamefont {Fowler}, \citenamefont
  {Mølmer}, \citenamefont {Devitt}, \citenamefont {Wunderlich},\ and\
  \citenamefont {Hensinger}}]{Lekitsch2017}%
  \BibitemOpen
  \bibfield  {author} {\bibinfo {author} {\bibnamefont {Lekitsch},
  \bibfnamefont {B.}}, \bibinfo {author} {\bibnamefont {Weidt}, \bibfnamefont
  {S.}}, \bibinfo {author} {\bibnamefont {Fowler}, \bibfnamefont {A.~G.}},
  \bibinfo {author} {\bibnamefont {Mølmer}, \bibfnamefont {K.}}, \bibinfo
  {author} {\bibnamefont {Devitt}, \bibfnamefont {S.~J.}}, \bibinfo {author}
  {\bibnamefont {Wunderlich}, \bibfnamefont {C.}}, and\ \bibinfo {author}
  {\bibnamefont {Hensinger}, \bibfnamefont {W.~K.}},\ }\href
  {https://doi.org/10.1126/sciadv.1601540} {\bibfield  {journal} {\bibinfo
  {journal} {Science Advances}\ }\textbf {\bibinfo {volume} {3}},\ \bibinfo
  {pages} {e1601540} (\bibinfo {year} {2017})}\BibitemShut {NoStop}%
\bibitem [{\citenamefont {Litinski}(2019{\natexlab{a}})}]{Litinski2019}%
  \BibitemOpen
  \bibfield  {author} {\bibinfo {author} {\bibnamefont {Litinski},
  \bibfnamefont {D.}},\ }\href {https://doi.org/10.22331/q-2019-03-05-128}
  {\bibfield  {journal} {\bibinfo  {journal} {{Quantum}}\ }\textbf {\bibinfo
  {volume} {3}},\ \bibinfo {pages} {128} (\bibinfo {year}
  {2019}{\natexlab{a}})}\BibitemShut {NoStop}%
\bibitem [{\citenamefont {Litinski}(2019{\natexlab{b}})}]{Litinski2019b}%
  \BibitemOpen
  \bibfield  {author} {\bibinfo {author} {\bibnamefont {Litinski},
  \bibfnamefont {D.}},\ }\href {https://doi.org/10.22331/q-2019-12-02-205}
  {\bibfield  {journal} {\bibinfo  {journal} {{Quantum}}\ }\textbf {\bibinfo
  {volume} {3}},\ \bibinfo {pages} {205} (\bibinfo {year}
  {2019}{\natexlab{b}})}\BibitemShut {NoStop}%
\bibitem [{\citenamefont {Lloyd}(1993)}]{lloyd1993}%
  \BibitemOpen
  \bibfield  {author} {\bibinfo {author} {\bibnamefont {Lloyd}, \bibfnamefont
  {S.}},\ }\href {https://doi.org/10.1126/science.261.5128.1569} {\bibfield
  {journal} {\bibinfo  {journal} {Science}\ }\textbf {\bibinfo {volume}
  {261}},\ \bibinfo {pages} {1569} (\bibinfo {year} {1993})}\BibitemShut
  {NoStop}%
\bibitem [{\citenamefont {Mohiyaddin}\ \emph {et~al.}(2021)\citenamefont
  {Mohiyaddin}, \citenamefont {Li}, \citenamefont {Brebels}, \citenamefont
  {Simion}, \citenamefont {Dumoulin~Stuyck}, \citenamefont {Godfrin},
  \citenamefont {Shehata}, \citenamefont {Elsayed}, \citenamefont {Gys},
  \citenamefont {Kubicek}, \citenamefont {Jussot}, \citenamefont {Canvel},
  \citenamefont {Massar}, \citenamefont {Weckx}, \citenamefont {Matagne},
  \citenamefont {Mongillo}, \citenamefont {Govoreanu},\ and\ \citenamefont
  {Radu}}]{Mohiyaddin2021}%
  \BibitemOpen
  \bibfield  {author} {\bibinfo {author} {\bibnamefont {Mohiyaddin},
  \bibfnamefont {F.}}, \bibinfo {author} {\bibnamefont {Li}, \bibfnamefont
  {R.}}, \bibinfo {author} {\bibnamefont {Brebels}, \bibfnamefont {S.}},
  \bibinfo {author} {\bibnamefont {Simion}, \bibfnamefont {G.}}, \bibinfo
  {author} {\bibnamefont {Dumoulin~Stuyck}, \bibfnamefont {N.~I.}}, \bibinfo
  {author} {\bibnamefont {Godfrin}, \bibfnamefont {C.}}, \bibinfo {author}
  {\bibnamefont {Shehata}, \bibfnamefont {M.}}, \bibinfo {author} {\bibnamefont
  {Elsayed}, \bibfnamefont {A.}}, \bibinfo {author} {\bibnamefont {Gys},
  \bibfnamefont {B.}}, \bibinfo {author} {\bibnamefont {Kubicek}, \bibfnamefont
  {S.}}, \bibinfo {author} {\bibnamefont {Jussot}, \bibfnamefont {J.}},
  \bibinfo {author} {\bibnamefont {Canvel}, \bibfnamefont {Y.}}, \bibinfo
  {author} {\bibnamefont {Massar}, \bibfnamefont {S.}}, \bibinfo {author}
  {\bibnamefont {Weckx}, \bibfnamefont {P.}}, \bibinfo {author} {\bibnamefont
  {Matagne}, \bibfnamefont {P.}}, \bibinfo {author} {\bibnamefont {Mongillo},
  \bibfnamefont {M.}}, \bibinfo {author} {\bibnamefont {Govoreanu},
  \bibfnamefont {B.}}, and\ \bibinfo {author} {\bibnamefont {Radu},
  \bibfnamefont {I.~P.}},\ }in\ \href
  {https://doi.org/10.1109/IEDM19574.2021.9720606} {\emph {\bibinfo {booktitle}
  {2021 IEEE International Electron Devices Meeting (IEDM)}}}\ (\bibinfo {year}
  {2021})\ pp.\ \bibinfo {pages} {27.5.1--27.5.4}\BibitemShut {NoStop}%
\bibitem [{\citenamefont {Nagayama}\ \emph {et~al.}(2017)\citenamefont
  {Nagayama}, \citenamefont {Fowler}, \citenamefont {Horsman}, \citenamefont
  {Devitt},\ and\ \citenamefont {Meter}}]{Nagayama_2017}%
  \BibitemOpen
  \bibfield  {author} {\bibinfo {author} {\bibnamefont {Nagayama},
  \bibfnamefont {S.}}, \bibinfo {author} {\bibnamefont {Fowler}, \bibfnamefont
  {A.~G.}}, \bibinfo {author} {\bibnamefont {Horsman}, \bibfnamefont {D.}},
  \bibinfo {author} {\bibnamefont {Devitt}, \bibfnamefont {S.~J.}}, and\
  \bibinfo {author} {\bibnamefont {Meter}, \bibfnamefont {R.~V.}},\ }\href
  {https://doi.org/10.1088/1367-2630/aa5918} {\bibfield  {journal} {\bibinfo
  {journal} {New Journal of Physics}\ }\textbf {\bibinfo {volume} {19}},\
  \bibinfo {pages} {023050} (\bibinfo {year} {2017})}\BibitemShut {NoStop}%
\bibitem [{\citenamefont {Nielsen}\ and\ \citenamefont
  {Chuang}(2010)}]{nielsen_chuang_2010}%
  \BibitemOpen
  \bibfield  {author} {\bibinfo {author} {\bibnamefont {Nielsen}, \bibfnamefont
  {M.~A.}}and\ \bibinfo {author} {\bibnamefont {Chuang}, \bibfnamefont
  {I.~L.}},\ }\href {https://doi.org/10.1017/CBO9780511976667} {\emph {\bibinfo
  {title} {Quantum Computation and Quantum Information: 10th Anniversary
  Edition}}}\ (\bibinfo  {publisher} {Cambridge University Press},\ \bibinfo
  {year} {2010})\BibitemShut {NoStop}%
\bibitem [{Note1()}]{Note1}%
  \BibitemOpen
  \bibinfo {note} {This method was determined in discussions involving the
  authors of the Herr paper and Craig Gidney, and is presented here with their
  consent}\BibitemShut {NoStop}%
\bibitem [{\citenamefont {O'Gorman}\ and\ \citenamefont
  {Campbell}(2017)}]{OGorman2017}%
  \BibitemOpen
  \bibfield  {author} {\bibinfo {author} {\bibnamefont {O'Gorman},
  \bibfnamefont {J.}}and\ \bibinfo {author} {\bibnamefont {Campbell},
  \bibfnamefont {E.~T.}},\ }\href {https://doi.org/10.1103/PhysRevA.95.032338}
  {\bibfield  {journal} {\bibinfo  {journal} {Phys. Rev. A}\ }\textbf {\bibinfo
  {volume} {95}},\ \bibinfo {pages} {032338} (\bibinfo {year}
  {2017})}\BibitemShut {NoStop}%
\bibitem [{\citenamefont {Prabhu}\ and\ \citenamefont
  {Reichardt}(2021)}]{PrabhuReichardt2021}%
  \BibitemOpen
  \bibfield  {author} {\bibinfo {author} {\bibnamefont {Prabhu}, \bibfnamefont
  {P.}}and\ \bibinfo {author} {\bibnamefont {Reichardt}, \bibfnamefont
  {B.~W.}},\ }in\ \href {https://doi.org/10.4230/LIPIcs.TQC.2021.5} {\emph
  {\bibinfo {booktitle} {16th Conference on the Theory of Quantum Computation,
  Communication and Cryptography (TQC 2021)}}},\ \bibinfo {series} {Leibniz
  International Proceedings in Informatics (LIPIcs)}, Vol.\ \bibinfo {volume}
  {197},\ \bibinfo {editor} {edited by\ \bibinfo {editor} {\bibfnamefont
  {M.-H.}\ \bibnamefont {Hsieh}}}\ (\bibinfo  {publisher} {Schloss Dagstuhl --
  Leibniz-Zentrum f{\"u}r Informatik},\ \bibinfo {address} {Dagstuhl,
  Germany},\ \bibinfo {year} {2021})\ pp.\ \bibinfo {pages}
  {5:1--5:13}\BibitemShut {NoStop}%
\bibitem [{\citenamefont {Raussendorf}\ and\ \citenamefont
  {Harrington}(2007)}]{Russendorf2007}%
  \BibitemOpen
  \bibfield  {author} {\bibinfo {author} {\bibnamefont {Raussendorf},
  \bibfnamefont {R.}}and\ \bibinfo {author} {\bibnamefont {Harrington},
  \bibfnamefont {J.}},\ }\href {https://doi.org/10.1103/PhysRevLett.98.190504}
  {\bibfield  {journal} {\bibinfo  {journal} {Phys. Rev. Lett.}\ }\textbf
  {\bibinfo {volume} {98}},\ \bibinfo {pages} {190504} (\bibinfo {year}
  {2007})}\BibitemShut {NoStop}%
\bibitem [{\citenamefont {Reichardt}(2020)}]{Reichardt2020}%
  \BibitemOpen
  \bibfield  {author} {\bibinfo {author} {\bibnamefont {Reichardt},
  \bibfnamefont {B.~W.}},\ }\href {https://doi.org/10.1088/2058-9565/abc6f4}
  {\bibfield  {journal} {\bibinfo  {journal} {Quantum Science and Technology}\
  }\textbf {\bibinfo {volume} {6}},\ \bibinfo {pages} {015007} (\bibinfo {year}
  {2020})}\BibitemShut {NoStop}%
\bibitem [{\citenamefont {Roetteler}\ \emph {et~al.}(2017)\citenamefont
  {Roetteler}, \citenamefont {Naehrig}, \citenamefont {Svore},\ and\
  \citenamefont {Lauter}}]{roetteler2017}%
  \BibitemOpen
  \bibfield  {author} {\bibinfo {author} {\bibnamefont {Roetteler},
  \bibfnamefont {M.}}, \bibinfo {author} {\bibnamefont {Naehrig}, \bibfnamefont
  {M.}}, \bibinfo {author} {\bibnamefont {Svore}, \bibfnamefont {K.~M.}}, and\
  \bibinfo {author} {\bibnamefont {Lauter}, \bibfnamefont {K.}},\ }in\
  \href@noop {} {\emph {\bibinfo {booktitle} {Advances in Cryptology --
  ASIACRYPT 2017}}},\ \bibinfo {editor} {edited by\ \bibinfo {editor}
  {\bibfnamefont {T.}~\bibnamefont {Takagi}}\ and\ \bibinfo {editor}
  {\bibfnamefont {T.}~\bibnamefont {Peyrin}}}\ (\bibinfo  {publisher} {Springer
  International Publishing},\ \bibinfo {address} {Cham},\ \bibinfo {year}
  {2017})\ pp.\ \bibinfo {pages} {241--270}\BibitemShut {NoStop}%
\bibitem [{\citenamefont {Shaw}(2022)}]{ToolsData}%
  \BibitemOpen
  \bibfield  {author} {\bibinfo {author} {\bibnamefont {Shaw}, \bibfnamefont
  {A.}},\ }\href@noop {} {\enquote {\bibinfo {title} {Tools and data for the
  paper "quantum computation on a 19-qubit wide 2d nearest neighbour qubit
  array.".}}\ }\bibinfo {howpublished}
  {\url{https://github.com/alexisshaw/RibbonPaper}} (\bibinfo {year}
  {2022})\BibitemShut {NoStop}%
\bibitem [{\citenamefont {Shor}(1996)}]{Shor1996}%
  \BibitemOpen
  \bibfield  {author} {\bibinfo {author} {\bibnamefont {Shor}, \bibfnamefont
  {P.}},\ }in\ \href {https://doi.org/10.1109/SFCS.1996.548464} {\emph
  {\bibinfo {booktitle} {Proceedings of 37th Conference on Foundations of
  Computer Science}}}\ (\bibinfo {year} {1996})\ pp.\ \bibinfo {pages}
  {56--65}\BibitemShut {NoStop}%
\bibitem [{\citenamefont {Shor}(1995)}]{Shor1995}%
  \BibitemOpen
  \bibfield  {author} {\bibinfo {author} {\bibnamefont {Shor}, \bibfnamefont
  {P.~W.}},\ }\href {https://doi.org/10.1103/PhysRevA.52.R2493} {\bibfield
  {journal} {\bibinfo  {journal} {Phys. Rev. A}\ }\textbf {\bibinfo {volume}
  {52}},\ \bibinfo {pages} {R2493} (\bibinfo {year} {1995})}\BibitemShut
  {NoStop}%
\bibitem [{\citenamefont {Steane}(1996)}]{Steane1996}%
  \BibitemOpen
  \bibfield  {author} {\bibinfo {author} {\bibnamefont {Steane}, \bibfnamefont
  {A.~M.}},\ }\href {https://doi.org/10.1103/PhysRevLett.77.793} {\bibfield
  {journal} {\bibinfo  {journal} {Phys. Rev. Lett.}\ }\textbf {\bibinfo
  {volume} {77}},\ \bibinfo {pages} {793} (\bibinfo {year} {1996})}\BibitemShut
  {NoStop}%
\bibitem [{\citenamefont {Stephens}\ and\ \citenamefont
  {Evans}(2009)}]{Ashley2009}%
  \BibitemOpen
  \bibfield  {author} {\bibinfo {author} {\bibnamefont {Stephens},
  \bibfnamefont {A.~M.}}and\ \bibinfo {author} {\bibnamefont {Evans},
  \bibfnamefont {Z.~W.~E.}},\ }\href
  {https://doi.org/10.1103/PhysRevA.80.022313} {\bibfield  {journal} {\bibinfo
  {journal} {Phys. Rev. A}\ }\textbf {\bibinfo {volume} {80}},\ \bibinfo
  {pages} {022313} (\bibinfo {year} {2009})}\BibitemShut {NoStop}%
\bibitem [{\citenamefont {Tomita}\ and\ \citenamefont
  {Svore}(2014)}]{Tomita2014}%
  \BibitemOpen
  \bibfield  {author} {\bibinfo {author} {\bibnamefont {Tomita}, \bibfnamefont
  {Y.}}and\ \bibinfo {author} {\bibnamefont {Svore}, \bibfnamefont {K.~M.}},\
  }\href {https://doi.org/10.1103/PhysRevA.90.062320} {\bibfield  {journal}
  {\bibinfo  {journal} {Phys. Rev. A}\ }\textbf {\bibinfo {volume} {90}},\
  \bibinfo {pages} {062320} (\bibinfo {year} {2014})}\BibitemShut {NoStop}%
\bibitem [{\citenamefont {Tuckett}, \citenamefont {Bartlett},\ and\
  \citenamefont {Flammia}(2018)}]{TuckettBartlettFlamia2018}%
  \BibitemOpen
  \bibfield  {author} {\bibinfo {author} {\bibnamefont {Tuckett}, \bibfnamefont
  {D.~K.}}, \bibinfo {author} {\bibnamefont {Bartlett}, \bibfnamefont {S.~D.}},
  and\ \bibinfo {author} {\bibnamefont {Flammia}, \bibfnamefont {S.~T.}},\
  }\href {https://doi.org/10.1103/PhysRevLett.120.050505} {\bibfield  {journal}
  {\bibinfo  {journal} {Phys. Rev. Lett.}\ }\textbf {\bibinfo {volume} {120}},\
  \bibinfo {pages} {050505} (\bibinfo {year} {2018})}\BibitemShut {NoStop}%
\bibitem [{\citenamefont {Tuckett}\ \emph {et~al.}(2019)\citenamefont
  {Tuckett}, \citenamefont {Darmawan}, \citenamefont {Chubb}, \citenamefont
  {Bravyi}, \citenamefont {Bartlett},\ and\ \citenamefont
  {Flammia}}]{Tuckett2019}%
  \BibitemOpen
  \bibfield  {author} {\bibinfo {author} {\bibnamefont {Tuckett}, \bibfnamefont
  {D.~K.}}, \bibinfo {author} {\bibnamefont {Darmawan}, \bibfnamefont {A.~S.}},
  \bibinfo {author} {\bibnamefont {Chubb}, \bibfnamefont {C.~T.}}, \bibinfo
  {author} {\bibnamefont {Bravyi}, \bibfnamefont {S.}}, \bibinfo {author}
  {\bibnamefont {Bartlett}, \bibfnamefont {S.~D.}}, and\ \bibinfo {author}
  {\bibnamefont {Flammia}, \bibfnamefont {S.~T.}},\ }\href
  {https://doi.org/10.1103/PhysRevX.9.041031} {\bibfield  {journal} {\bibinfo
  {journal} {Phys. Rev. X}\ }\textbf {\bibinfo {volume} {9}},\ \bibinfo {pages}
  {041031} (\bibinfo {year} {2019})}\BibitemShut {NoStop}%
\bibitem [{\citenamefont {Vandersypen}\ \emph {et~al.}(2017)\citenamefont
  {Vandersypen}, \citenamefont {Bluhm}, \citenamefont {Clarke}, \citenamefont
  {Dzurak}, \citenamefont {Ishihara}, \citenamefont {Morello}, \citenamefont
  {Reilly}, \citenamefont {Schreiber},\ and\ \citenamefont
  {Veldhorst}}]{Vandersypen2017}%
  \BibitemOpen
  \bibfield  {author} {\bibinfo {author} {\bibnamefont {Vandersypen},
  \bibfnamefont {L.~M.~K.}}, \bibinfo {author} {\bibnamefont {Bluhm},
  \bibfnamefont {H.}}, \bibinfo {author} {\bibnamefont {Clarke}, \bibfnamefont
  {J.~S.}}, \bibinfo {author} {\bibnamefont {Dzurak}, \bibfnamefont {A.~S.}},
  \bibinfo {author} {\bibnamefont {Ishihara}, \bibfnamefont {R.}}, \bibinfo
  {author} {\bibnamefont {Morello}, \bibfnamefont {A.}}, \bibinfo {author}
  {\bibnamefont {Reilly}, \bibfnamefont {D.~J.}}, \bibinfo {author}
  {\bibnamefont {Schreiber}, \bibfnamefont {L.~R.}}, and\ \bibinfo {author}
  {\bibnamefont {Veldhorst}, \bibfnamefont {M.}},\ }\href
  {https://doi.org/10.1038/s41534-017-0038-y} {\bibfield  {journal} {\bibinfo
  {journal} {npj Quantum Information}\ }\textbf {\bibinfo {volume} {3}},\
  \bibinfo {pages} {34} (\bibinfo {year} {2017})}\BibitemShut {NoStop}%
\bibitem [{\citenamefont {Veldhorst}\ \emph {et~al.}(2017)\citenamefont
  {Veldhorst}, \citenamefont {Eenink}, \citenamefont {Yang},\ and\
  \citenamefont {Dzurak}}]{Veldhorst2017}%
  \BibitemOpen
  \bibfield  {author} {\bibinfo {author} {\bibnamefont {Veldhorst},
  \bibfnamefont {M.}}, \bibinfo {author} {\bibnamefont {Eenink}, \bibfnamefont
  {H.~G.~J.}}, \bibinfo {author} {\bibnamefont {Yang}, \bibfnamefont {C.~H.}},
  and\ \bibinfo {author} {\bibnamefont {Dzurak}, \bibfnamefont {A.~S.}},\
  }\href {https://doi.org/10.1038/s41467-017-01905-6} {\bibfield  {journal}
  {\bibinfo  {journal} {Nature Communications}\ }\textbf {\bibinfo {volume}
  {8}},\ \bibinfo {pages} {1766} (\bibinfo {year} {2017})}\BibitemShut
  {NoStop}%
\bibitem [{\citenamefont {Wang}, \citenamefont {Fowler},\ and\ \citenamefont
  {Hollenberg}(2011)}]{WangFowlerHollenberg2011}%
  \BibitemOpen
  \bibfield  {author} {\bibinfo {author} {\bibnamefont {Wang}, \bibfnamefont
  {D.~S.}}, \bibinfo {author} {\bibnamefont {Fowler}, \bibfnamefont {A.~G.}},
  and\ \bibinfo {author} {\bibnamefont {Hollenberg}, \bibfnamefont
  {L.~C.~L.}},\ }\href {https://doi.org/10.1103/PhysRevA.83.020302} {\bibfield
  {journal} {\bibinfo  {journal} {Phys. Rev. A}\ }\textbf {\bibinfo {volume}
  {83}},\ \bibinfo {pages} {020302} (\bibinfo {year} {2011})}\BibitemShut
  {NoStop}%
\bibitem [{\citenamefont {van~de Wetering}(2020)}]{2012.13966}%
  \BibitemOpen
  \bibfield  {author} {\bibinfo {author} {\bibnamefont {van~de Wetering},
  \bibfnamefont {J.}},\ }\href {https://doi.org/10.48550/ARXIV.2012.13966}
  {\enquote {\bibinfo {title} {Zx-calculus for the working quantum computer
  scientist},}\ } (\bibinfo {year} {2020}),\ \bibinfo {note}
  {arxiv:2012.13966},\ \Eprint {https://arxiv.org/abs/2012.13966}
  {arXiv:2012.13966 [quant-ph]} \BibitemShut {NoStop}%
\bibitem [{\citenamefont {Xue}\ \emph {et~al.}(2022)\citenamefont {Xue},
  \citenamefont {Russ}, \citenamefont {Samkharadze}, \citenamefont {Undseth},
  \citenamefont {Sammak}, \citenamefont {Scappucci},\ and\ \citenamefont
  {Vandersypen}}]{Xue2022}%
  \BibitemOpen
  \bibfield  {author} {\bibinfo {author} {\bibnamefont {Xue}, \bibfnamefont
  {X.}}, \bibinfo {author} {\bibnamefont {Russ}, \bibfnamefont {M.}}, \bibinfo
  {author} {\bibnamefont {Samkharadze}, \bibfnamefont {N.}}, \bibinfo {author}
  {\bibnamefont {Undseth}, \bibfnamefont {B.}}, \bibinfo {author} {\bibnamefont
  {Sammak}, \bibfnamefont {A.}}, \bibinfo {author} {\bibnamefont {Scappucci},
  \bibfnamefont {G.}}, and\ \bibinfo {author} {\bibnamefont {Vandersypen},
  \bibfnamefont {L.~M.~K.}},\ }\href
  {https://doi.org/10.1038/s41586-021-04273-w} {\bibfield  {journal} {\bibinfo
  {journal} {Nature}\ }\textbf {\bibinfo {volume} {601}},\ \bibinfo {pages}
  {343} (\bibinfo {year} {2022})}\BibitemShut {NoStop}%
\bibitem [{\citenamefont {Yang}\ \emph {et~al.}(2020)\citenamefont {Yang},
  \citenamefont {Leon}, \citenamefont {Hwang}, \citenamefont {Saraiva},
  \citenamefont {Tanttu}, \citenamefont {Huang}, \citenamefont
  {Camirand~Lemyre}, \citenamefont {Chan}, \citenamefont {Tan}, \citenamefont
  {Hudson}, \citenamefont {Itoh}, \citenamefont {Morello}, \citenamefont
  {Pioro-Ladri{\`e}re}, \citenamefont {Laucht},\ and\ \citenamefont
  {Dzurak}}]{Yang2020}%
  \BibitemOpen
  \bibfield  {author} {\bibinfo {author} {\bibnamefont {Yang}, \bibfnamefont
  {C.~H.}}, \bibinfo {author} {\bibnamefont {Leon}, \bibfnamefont {R.~C.~C.}},
  \bibinfo {author} {\bibnamefont {Hwang}, \bibfnamefont {J.~C.~C.}}, \bibinfo
  {author} {\bibnamefont {Saraiva}, \bibfnamefont {A.}}, \bibinfo {author}
  {\bibnamefont {Tanttu}, \bibfnamefont {T.}}, \bibinfo {author} {\bibnamefont
  {Huang}, \bibfnamefont {W.}}, \bibinfo {author} {\bibnamefont
  {Camirand~Lemyre}, \bibfnamefont {J.}}, \bibinfo {author} {\bibnamefont
  {Chan}, \bibfnamefont {K.~W.}}, \bibinfo {author} {\bibnamefont {Tan},
  \bibfnamefont {K.~Y.}}, \bibinfo {author} {\bibnamefont {Hudson},
  \bibfnamefont {F.~E.}}, \bibinfo {author} {\bibnamefont {Itoh}, \bibfnamefont
  {K.~M.}}, \bibinfo {author} {\bibnamefont {Morello}, \bibfnamefont {A.}},
  \bibinfo {author} {\bibnamefont {Pioro-Ladri{\`e}re}, \bibfnamefont {M.}},
  \bibinfo {author} {\bibnamefont {Laucht}, \bibfnamefont {A.}}, and\ \bibinfo
  {author} {\bibnamefont {Dzurak}, \bibfnamefont {A.~S.}},\ }\href
  {https://doi.org/10.1038/s41586-020-2171-6} {\bibfield  {journal} {\bibinfo
  {journal} {Nature}\ }\textbf {\bibinfo {volume} {580}},\ \bibinfo {pages}
  {350} (\bibinfo {year} {2020})}\BibitemShut {NoStop}%
\bibitem [{\citenamefont {Zwerver}\ \emph {et~al.}(2022)\citenamefont
  {Zwerver}, \citenamefont {Kr{\"a}henmann}, \citenamefont {Watson},
  \citenamefont {Lampert}, \citenamefont {George}, \citenamefont
  {Pillarisetty}, \citenamefont {Bojarski}, \citenamefont {Amin}, \citenamefont
  {Amitonov}, \citenamefont {Boter}, \citenamefont {Caudillo}, \citenamefont
  {Correas-Serrano}, \citenamefont {Dehollain}, \citenamefont {Droulers},
  \citenamefont {Henry}, \citenamefont {Kotlyar}, \citenamefont {Lodari},
  \citenamefont {L{\"u}thi}, \citenamefont {Michalak}, \citenamefont {Mueller},
  \citenamefont {Neyens}, \citenamefont {Roberts}, \citenamefont {Samkharadze},
  \citenamefont {Zheng}, \citenamefont {Zietz}, \citenamefont {Scappucci},
  \citenamefont {Veldhorst}, \citenamefont {Vandersypen},\ and\ \citenamefont
  {Clarke}}]{Zwerver2022}%
  \BibitemOpen
  \bibfield  {author} {\bibinfo {author} {\bibnamefont {Zwerver}, \bibfnamefont
  {A.~M.~J.}}, \bibinfo {author} {\bibnamefont {Kr{\"a}henmann}, \bibfnamefont
  {T.}}, \bibinfo {author} {\bibnamefont {Watson}, \bibfnamefont {T.~F.}},
  \bibinfo {author} {\bibnamefont {Lampert}, \bibfnamefont {L.}}, \bibinfo
  {author} {\bibnamefont {George}, \bibfnamefont {H.~C.}}, \bibinfo {author}
  {\bibnamefont {Pillarisetty}, \bibfnamefont {R.}}, \bibinfo {author}
  {\bibnamefont {Bojarski}, \bibfnamefont {S.~A.}}, \bibinfo {author}
  {\bibnamefont {Amin}, \bibfnamefont {P.}}, \bibinfo {author} {\bibnamefont
  {Amitonov}, \bibfnamefont {S.~V.}}, \bibinfo {author} {\bibnamefont {Boter},
  \bibfnamefont {J.~M.}}, \bibinfo {author} {\bibnamefont {Caudillo},
  \bibfnamefont {R.}}, \bibinfo {author} {\bibnamefont {Correas-Serrano},
  \bibfnamefont {D.}}, \bibinfo {author} {\bibnamefont {Dehollain},
  \bibfnamefont {J.~P.}}, \bibinfo {author} {\bibnamefont {Droulers},
  \bibfnamefont {G.}}, \bibinfo {author} {\bibnamefont {Henry}, \bibfnamefont
  {E.~M.}}, \bibinfo {author} {\bibnamefont {Kotlyar}, \bibfnamefont {R.}},
  \bibinfo {author} {\bibnamefont {Lodari}, \bibfnamefont {M.}}, \bibinfo
  {author} {\bibnamefont {L{\"u}thi}, \bibfnamefont {F.}}, \bibinfo {author}
  {\bibnamefont {Michalak}, \bibfnamefont {D.~J.}}, \bibinfo {author}
  {\bibnamefont {Mueller}, \bibfnamefont {B.~K.}}, \bibinfo {author}
  {\bibnamefont {Neyens}, \bibfnamefont {S.}}, \bibinfo {author} {\bibnamefont
  {Roberts}, \bibfnamefont {J.}}, \bibinfo {author} {\bibnamefont
  {Samkharadze}, \bibfnamefont {N.}}, \bibinfo {author} {\bibnamefont {Zheng},
  \bibfnamefont {G.}}, \bibinfo {author} {\bibnamefont {Zietz}, \bibfnamefont
  {O.~K.}}, \bibinfo {author} {\bibnamefont {Scappucci}, \bibfnamefont {G.}},
  \bibinfo {author} {\bibnamefont {Veldhorst}, \bibfnamefont {M.}}, \bibinfo
  {author} {\bibnamefont {Vandersypen}, \bibfnamefont {L.~M.~K.}}, and\
  \bibinfo {author} {\bibnamefont {Clarke}, \bibfnamefont {J.~S.}},\ }\href
  {https://doi.org/10.1038/s41928-022-00727-9} {\bibfield  {journal} {\bibinfo
  {journal} {Nature Electronics}\ }\textbf {\bibinfo {volume} {5}},\ \bibinfo
  {pages} {184} (\bibinfo {year} {2022})}\BibitemShut {NoStop}%
\end{thebibliography}%

\appendix

\section{A lattice surgery construction for non-bus repetiton codes on rect patches.}
\label{sec:non_bus_ls}
In order to perform non-bussed extraction of the repetition code in approximately $9 d_Z + 4$ time steps $\times$ qubit area per qubit patches , one can use the series of lattice surgery in the figure below, one logical qubit is indexed by roman numerals, the other with letters. The sequence has the effect of shifting the repetition code qubits to the right every extraction. By taking the mirror
image of the sequence a shift to the left could be achieved. To have no-net movement extractions should alternate between these two sets. The sequence also only extracts the repetition code for one of the two sets of qubits, to perform the extraction for both one should perform the sequence once for each of the repetition code qubits
this then takes $18 d_Z + 8$.

To perform a cnot one then inter-digitates two repetition codes using the lattice surgery procedure, and performs either an $XX$ or $ZZ$ parity measurement between each of the individual sub-patches in the lattice.

\begin{figure*}
         \includegraphics[width=\textwidth]{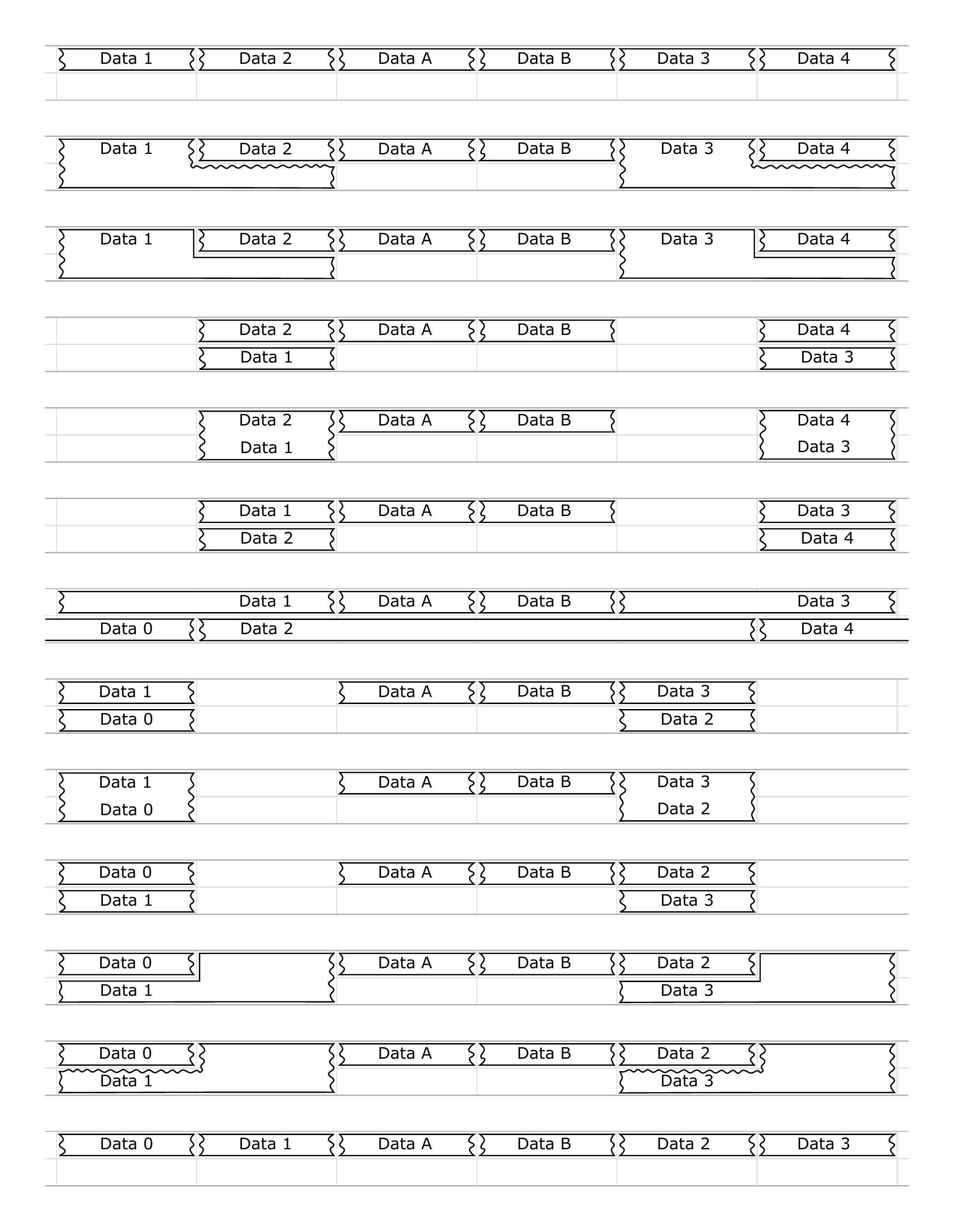}
         \caption{Lattice surgery instructions pt 1}
         \label{fig:rect_sc_1}
\end{figure*}

\end{document}